\documentclass[12pt,a4paper,openright,oneside]{book} 

\usepackage{fancyheadings}

\usepackage{thesis}
\usepackage{setspace}
\usepackage{epsfig}
\usepackage{graphics}
\usepackage{psfrag}

\usepackage{titlesec}
\titleformat{\chapter}[display]{\sffamily
\Huge}{\thechapter}{2ex}{}[\vspace{2ex} \titlerule]
\renewcommand{\chaptername}{Chapter}

\usepackage{caption}
\usepackage{makeidx}

\makeindex
\usepackage{eurosans} 

\usepackage{afterpage}

\reversemarginpar

\newcommand{\ket}[1]{|#1\rangle}
\newcommand{\bra}[1]{\langle#1|}

\begin{document}

\pagestyle{empty}

\begin{center}
  {\sc{University of London}}\\[0.5cm]
  Imperial College of
  Science, Technology and Medicine\\ The
  Blackett Laboratory\\ Quantum Optics \& Laser Science Group\\[1.5cm]
  
  {\Huge Quantum Information Processing}\\[0.5cm]
  {\Huge  with Single Photons }\\[0.5cm]
 
  {\Large{Yuan Liang Lim}}\\[1.5cm]

  Thesis submitted in partial fulfilment of the \\
  requirements for the degree of\\
  Doctor of Philosophy\\
  of the University of London\\
  and the Diploma of Membership of Imperial College.\\[.8cm]
  \vfill\today
\end{center}


\clearpage \pagestyle{empty}


\begin{center}
  {\Huge Abstract}\\[1cm]
\end{center}
Photons are natural carriers of quantum information due to their ease of distribution and long lifetime. This thesis concerns various 
related aspects of quantum information processing with single photons. 
Firstly, we demonstrate  $N$-photon entanglement 
generation through a generalised $N \times N$ symmetric beam splitter 
known as the Bell multiport. A wide variety of 4-photon entangled states as well as the $N$-photon 
W-state can be generated with an unexpected non-monotonic decreasing probability of success with $N$. We also show how the same setup can 
be used to generate multiatom entanglement. A further study of multiports also leads us to a multiparticle generalisation of the 
Hong-Ou-Mandel dip 
which holds for all Bell multiports of even number of input ports.

Next, we demonstrate a generalised linear optics based photon  filter that has a constant success 
probability regardless of the number of photons involved. This filter has the 
highest reported success probability and 
is interferometrically robust. Finally, we  demonstrate 
how repeat-until-success quantum computing can be performed with two distant nodes 
with unit success probability using only linear optics resource. We further show that using non-identical photon sources, robustness can 
still be achieved, an illustration of the nature and advantages of measurement-based quantum computation.   A 
direct application to the same setup leads naturally to arbitrary multiphoton
state 
generation on demand. Finally, we demonstrate how  
polarisation entanglement of photons can be detected from the emission of two atoms in a Young's double-slit type experiment without 
linear optics, resulting in both atoms being also maximally entangled. 


\begin{publications}
\\
1. Y. L. Lim and A. Beige, {\em Photon polarisation entanglement from distant 
dipole sources}, J. Phys. A {\bf 38}, L7 (2005), quant-ph/0308095  \\
2. Y. L. Lim and A. Beige, {\em Push button generation of multiphoton 
entanglement}, Proc. SPIE {\bf 5436}, 118 (2004), quant-ph/0403125 \\
3. Y. L. Lim and A. Beige, {\em An efficient quantum filter for multiphoton 
states}, J. Mod. Opt. {\bf 52}, 1073 (2005), quant-ph/0406008 \\
4. Y. L. Lim and A. Beige, {\em Multiphoton entanglement through a Bell 
multiport beam splitter}, Phys. Rev. A {\bf 71}, 062311 (2005), quant-ph/0406047 \\
5. Y. L. Lim and A. Beige and L. C. Kwek, {\em Repeat-Until-Success linear optics quantum computing}, Phys. Rev. Lett. {\bf 95}, 030505 
(2005), quant-ph/0408043 \\
6. Y. L. Lim and A. Beige, {\em Generalised Hong-Ou-Mandel Experiments with 
Bosons and Fermions},  New J. Phys {\bf 7}, 155 (2005), quant-ph/0505034 \\
7. Y. L. Lim, S. Barrett, A. Beige, P. Kok and L. C. Kwek, {\em Repeat-until-success distributed quantum computing with 
stationary and flying qubits},(submitted to Phys. Rev. A), quant-ph/0508218 \\
\end{publications}  

\begin{acknowledgements}
I thank both Dr Almut Beige and Prof Sir Peter Knight for being my PhD supervisors. Almut has patiently guided me from being a novice to 
the stage where I can hopefully say interesting and new things about physics. A valuable lesson which I have learnt from her is the 
importance of being dogged and persevering in research. She has also taught me to believe in the impossible. Had it not been for her encouragement, I might have given up too easily on a difficult problem! Peter has been very encouraging and supportive to my education and have helped to ensure that I get the support and opportunities to attend 
summer schools and conferences, all of which have proven to be crucial to this PhD experience.  I also thank both of them for guiding me in my thesis and giving me many 
valuable comments. Furthermore, many of the results in this thesis have been inspired from my interactions with them.

I thank Dan Browne for kindly ploughing through my thesis and offering many valuable scientific feedback, as well as suggesting improvements to the English. I also thank him for our many discussions that has greatly enriched me, and his patience in answering my many curious questions. In addition, I also thank Shash for reading my introductory chapter and also patiently explaining things like stabilizers and POVMs to me. I also thank Terry Rudolph for providing me with many inspiring and provocative thoughts that has helped to shape my research. 

I thank Jim Franson for many stimulating discussions and encouragement on many of the work in this thesis, when he was visiting Imperial College. I also thank Geoff Pryde and Marek {\. Z}ukowski  for their encouragement and interest in my work. I also appreciate Martin Plenio for his encouragement.

I thank Jesus Roger-Salazar for kindly giving me the latex template for this thesis. I also thank Jens Eisert for his kind help with Appendix A. 

I thank my collaborators Sean Barrett and Pieter Kok for their friendship and sharing their valuable experience and knowledge in research 
and physics. I also thank another collaborator Kwek Leong Chuan from Quantum LAH for funding a few of my visits to my homeland Singapore to facilitate research collaborations, 
and also helping me to open up opportunities for future research. My appreciation goes also to Christian Kurtsiefer for getting me fired up with excitement on things that can be done back in Singapore after my PhD.

Special thanks goes to Hugo Cable for being the best of buddies to me in QOLS. We have  both "grown up" much together through this PhD experience and I wish him all the best after his PhD.

I also thank Jae, Rachele, Jeremy and Adele for their friendship.

I want to specially thank Huang Sen and Emily for such a special and dear friendship that cannot be expressed with words.

I thank a special brother, Bae Joon Woo. By divine circumstances, our paths had crossed twice - once in IQING 2002 and the other time in Cargese 2004, and we have since maintained close contact. He has been a special encouragement and inspiration to me and I thank him for his friendship.

I would like to thank David Oliver, Ros, Carmel, Ladi and Iyiola of Riverpark Church who had so warmly welcomed me when I first arrived in London and eased my settling in. Also many thanks to all members of the church for their love and prayer support during my three years in London. 

I thank David Ong, Adrian Ying, Patrick Wong as well as Melvin Tan for their friendship and encouragement, and for always remembering me in their prayers even when I was away from Singapore.

I also thank Kin Seng and Kien Boon who have been responsible for encouraging me to apply for the DSO postgraduate scholarship program and facilitating the application process. I thank Kin Seng in particular for his encouragement and belief in me. I also thank Yuqing and Ernest for their prayers and support. I thank Poh Boon for helping me to get into the Applied Physics Lab in DSO, where my love for physics was rekindled. Of course, many thanks to DSO for granting me the funding to complete the PhD program in Imperial College!

I thank everyone in my family, especially Daddy, Mummy, Granny, Aunty Lucia, Aunty Letitia, my dear sister Yuan Ping,   brother Yuan Sing, as well as my parents-in-law for all their love, support and encouragement. My love of physics was first ignited when my dad bought me the book, {\em At the opening of the atomic era}, when I was only a child. Ever since, I had always wanted to be a physicist!

I thank my wife Puay-Sze for her neverending patience, support and love which have been so crucial to me. I also want to make special mention of my soon-to-be-born son, Isaac, for the joy and additional motivation that he brings to me during the last lap of my thesis-writing!

I thank God who has loved me, always guided me and taken care of me.

\end{acknowledgements}

\tableofcontents


\pagestyle{fancyplain} 

\renewcommand{\chaptername}{Chapter} 

\chapter{Introduction}\label{overview}

Photons are natural carriers of quantum information owing to their long 
lifetimes and ease of distribution and this  constitutes the main motivation 
for this thesis.  In quantum information processing, entanglement plays its role in 
diverse applications such as quantum cryptography, implementation of universal 
quantum gates, tests of non-locality, and is prevalent in 
all known quantum algorithms that provide an exponential speedup compared to 
classical algorithms. Entanglement, a still elusive concept, is strictly defined as the situation where a quantum state cannot be 
decomposed into a  convex sum of tensor product density matrix states. The ability to generate or manipulate entanglement is  thus a
key ingredient to 
quantum information processing. In this thesis, we focus primarily on 
various aspects of entangled state generation, detection, manipulation  and 
exploitation for quantum information processing using single 
photons.  We 
consider novel means of how single photon sources can be manipulated through the 
photons they emit and vice-versa the way photons can be 
manipulated with the aid of single photon sources. Furthermore, as will be 
seen, these two apparently different tasks can often be 
exploited for each other. Therefore, an alternative title to this thesis could well be ``Photon 
assisted quantum computation". We start here by giving a 
brief overview to quantum information processing bringing single photons into 
a general context. A more detailed survey of the 
research done in this thesis can be found in the relevant chapters.

\section{Brief Introduction to quantum information processing and single 
photons}

Quantum information processing is a remarkably diverse  and interdisciplinary 
field.  In the words of  Knill and Nielsen \cite{Knill02}, it is  ``The science 
of  the theoretical, experimental and technological areas covering the use of 
quantum mechanics for communication and computation." It includes quantum 
information theory, quantum communication, quantum computation, quantum  
algorithms and their complexity and quantum control. In general,  these 
fields are not mutually exclusive and often have substantial overlaps. It is therefore somewhat artificial to attempt to classify them as 
separate subfields.

Early ideas of quantum information processing began with Feynman, who considered
the question of efficient 
simulation of a quantum 
system  \cite{Feynman82}. He speculated that the only efficient 
simulation of a quantum system that 
could be achieved  would come from another quantum 
system. Following that, Deutsch and Jozsa \cite{Deutsch85,Deutsch92} 
demonstrated 
the existence of quantum algorithms that are more efficient than classical 
algorithms. Later, Shor \cite{Shor94}, building on the work of Simon 
\cite{Simon94} as well as Deutsch and Jozsa,  demonstrated a quantum 
algorithm for prime factorisation that is exponentially faster than any known 
classical algorithm. Both the  Shor and Deutsch-Jozsa algorithm, as well as the 
Simon algorithm are actually special cases of the more general algorithm for the 
problem known as the Hidden Subgroup Problem(HSP). In fact, all known algorithms 
belonging to HSP class, at least for the case of finite Abelian groups,  are exponentially more efficient than the best known 
corresponding classical algorithms. Finally, Grover \cite{Grover96} demonstrated a 
fundamentally different algorithm that is $\sqrt{N}$ faster than the best known 
classical algorithm for  an $N$ element database search without any partial 
information.  Entanglement appears to be  necessary for 
quantum algorithms that yield exponential speedups compared to classical 
algorithms \cite{Linden01,Harrow03}.

The basic logical unit in each of these algorithms are so-called qubits,
which hold the quantum information. Each of these qubits can be in any 
superposition between two  orthogonal logical states, denoted $\ket{0}$ and 
$\ket{1}$, constituting the computation basis states, 
in analogy to classical bits `0' and `1' in classical computing. For example, a qubit $\ket{\psi}$ can be written as a state vector in a 2-dimensional 
Hilbert space given by $\ket{\psi}=a\ket{0}+b\ket{1}$ where $a$ and $b$ are arbitrary normalised  complex coefficients. In contrast, a 
classical bit can only be in the state `0' or `1'. In addition, there are many possible physical realisations of a qubit. For 
example, the computational basis states can be the Zeeman ground states of atoms, the direction of the spin of electrons or polarisation 
states of photons.  The coherent evolution of many qubits which can be in an
arbitrary superposition\footnote{Note that with $N$ qubits, the Hilbert space spans an exponentially large dimension given by $2^N$.} can be 
thought as a mechanism that enables massive 
parallelism in the computation, hence leading to a possible exponential speedup 
compared to classical 
computation.  At the same time, any $N$-qubit unitary 
operation can be decomposed to two-qubit unitary operations \cite{Barenco95}. It is important to 
note that there exist two-qubit  gates, which together with arbitrary single qubit 
operations, can simulate any two-qubit unitary operation 
\cite{DiVincenzo95}. 
Any two-qubit gate fufilling the above universality criterion  is known as a universal two-qubit gate. Notable examples of such gates are
CNOT 
and CZ gates, which perform a controlled non-trivial single qubit unitary 
operation on a target qubit 
dependent on the state of the control qubit. Specifically, given  two input qubits, a control and target one, the CNOT operation flips 
the target qubit only if the control qubit is in the logical state $\ket{1}_{\rm c}$. Similarly, the CZ gate yields a negative sign to 
the target qubit $\ket{1}_{\rm t}$ only if the control qubit is also in the state $\ket{1}_{\rm c}$.

\begin{figure}
\begin{minipage}{\columnwidth}
\begin{center}
\vspace*{0cm}
\psfrag{Psi}{$\ket{\psi}$}
\psfrag{Esi}{$\ket{\Psi^+}$}
\psfrag{HPsi}{$H\ket{\psi}$}
\psfrag{H}{$H$}
\psfrag{HPH}{$HPH$}
\resizebox{\columnwidth}{!}{\rotatebox{0}{\includegraphics{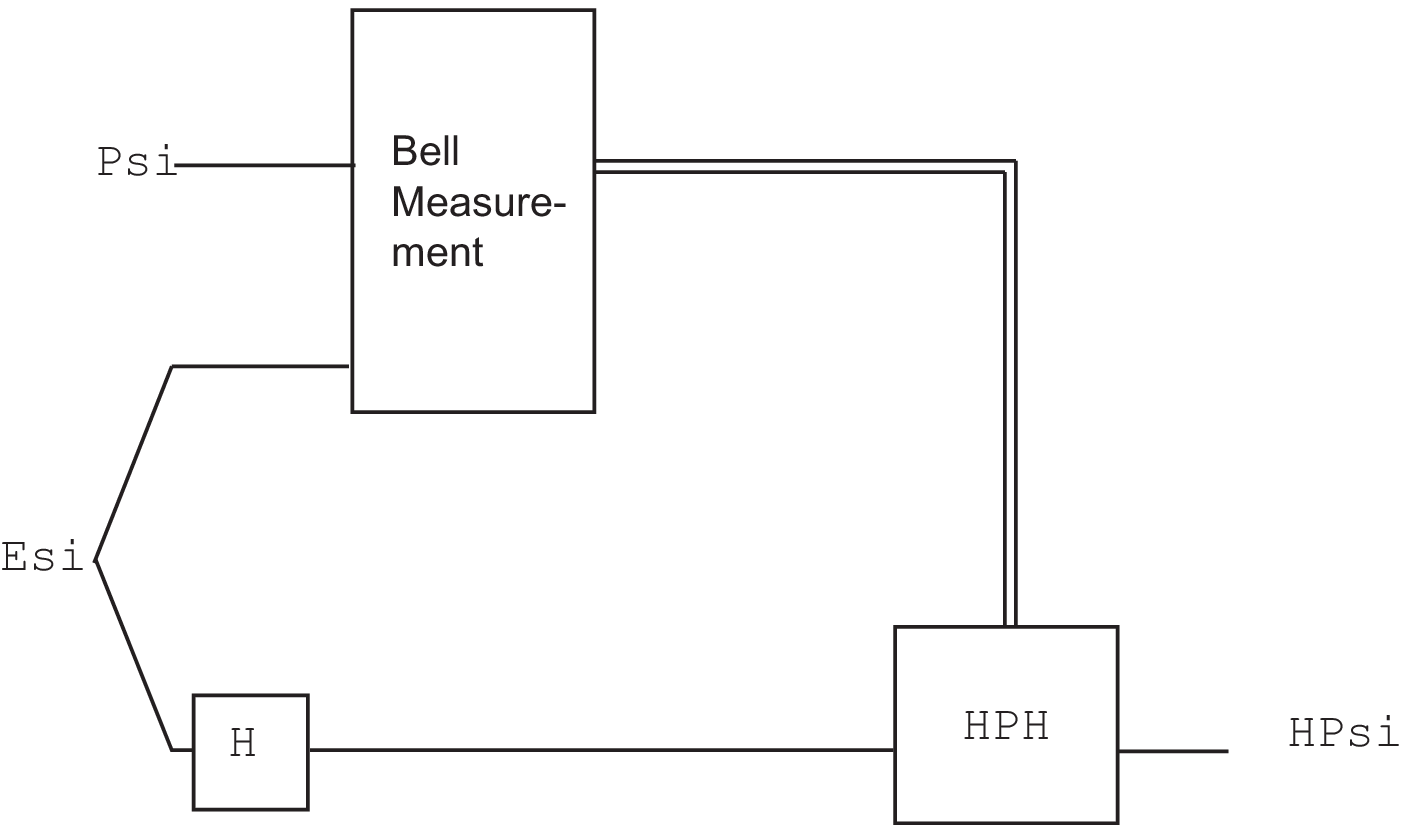}}}  
\end{center}
\vspace*{0cm}
\caption{Teleportation of a unitary operation such as a Hadamard gate $H$ over the input state $\ket{\psi}$. The action of a Hadamard 
gate is defined by $H\ket{0}=\ket{+}=\frac{1}{\sqrt{2}}(\ket{0}+\ket{1})$ and $H\ket{1}=\ket{-}=\frac{1}{\sqrt{2}}(\ket{0}-\ket{1})$. The entangled ancilla $\frac{1}{\sqrt{2}}(\ket{0+}+\ket{1-})$ is given by 
acting $H$ on $\ket{\Psi^+}=\frac{1}{\sqrt{2}}(\ket{00}+\ket{11})$. A Bell measurement between one of the qubits of the entangled 
ancilla and the input state $\ket{\psi}$ is performed. $P$ consists of a local unitary operation that depends on the measurement syndrome 
deriving from the Bell measurement. After measurement and the operation $HPH$, the teleported state becomes $H\ket{\psi}$. Note that 
ordinary teleportation is given by replacing the gate $H$ with an identity operator.} \label{tele}
\end{minipage}
\end{figure}

In addition, application of these two-qubit gates to a suitable product state can entangle 
the input state and we shall now make a short detour to illustrate some basic properties of an entangled state.  For example, the state 
$\frac{1}{\sqrt{2}}(\ket{10}-\ket{11})$ is not entangled because it can be written as a tensor product of two qubit states given by  
$\frac{1}{\sqrt{2}}(\ket{0}-\ket{1})\otimes\ket{1}$. We now treat this as an input state where the left qubit is treated as the control 
and the right qubit  is treated as the target.  The application of a CNOT gate to this separable state yields a 
particularly interesting maximally entangled bipartite state, known as the singlet state $\ket{\Phi^-}$ given by  
$\ket{\Phi^-}=\frac{1}{\sqrt{2}}(\ket{01}-\ket{10})$ where  the control and target subscripts have been dropped for clarity. This state is 
invariant under any qubit rotation applied equally to the two qubits. This means that in the basis given by $\ket{0'}$ and $\ket{1'}$, 
the state $\ket{\Phi^-}$ is again given by $\frac{1}{\sqrt{2}}(\ket{0'1'}-\ket{1'0'})$.  The two parties each holding a qubit will 
always measure different basis states, no matter what common basis states they share. We now try to construct a two party non-entangled 
state that might yield a similar measurement syndrome. For example, the non-entangled mixed-state $\rho_{\rm no 
ent}=\frac{1}{2}(\ket{01}\bra{01}+\ket{10}\bra{10})$ will no doubt yield positive correlations of different states in the measurement basis 
$\ket{0}$ and $\ket{1}$. Unfortunately, this will no longer be true in another measurement basis, say 
$\ket{\pm}=\frac{1}{\sqrt{2}}(\ket{0}\pm\ket{1})$. This shows that an entangled state can have stronger correlations than is possible 
compared to a non-entangled state. Due to the existence of these special correlations, a bipartite entangled state for example, cannot be thought of 
as two separate parties. We give briefly an example of the exploitation of this correlation. Suppose we have an input state given by 
$\ket{\psi}=\alpha\ket{0}+\beta\ket{1}$ together with an ancilla $\ket{\Phi^-}$. We can then write the tripartite state 
$\ket{\psi}\ket{\Phi^-}$ as $\alpha\ket{001}-\alpha\ket{010}+\beta\ket{101}-\beta\ket{110}$ omitting the normalisation factor. If a projective 
measurement is performed on the first two parties such that a maximally entangled state say 
$\ket{\Psi^+}=\frac{1}{\sqrt{2}}(\ket{00}+\ket{11})$ is detected (an example of a Bell measurement), this immediately projects the third 
qubit to the state $\alpha\ket{1}-\beta\ket{0}$, which is local unitary equivalent to the original input qubit $\ket{\psi}$. We have 
therefore transferred the input state by measurement to one of the qubits in the entangled ancilla using the correlation found in $\ket{\Phi^-}$ as well in $\ket{\Psi^+}$. 
This is also known as teleportation \cite{Bennett93}.

We have earlier defined what entanglement is by saying what it is not!  Although 
we have already some limited success on entanglement measures (see Ref. \cite{Plenio05} and references therein) and criteria 
\cite{Horodecki96,Peres96} to help us establish whether a state is entangled or not, the full understanding of what entanglement really is remains 
elusive\footnote{This is the author's personal perception.}.

Returning to the discussion on universal gates, one might assume that such gate 
operations should be accomplished by coherent means, for example, with a controlled evolution of the Hamiltonian governing interactions 
between qubits possibly with an external control agent, such as a laser beam, with the Cirac-Zoller gate for trapped ions 
\cite{Cirac95} as a famous example.
This is however too restrictive  and it is worth commenting briefly on 
approaches which use entangled 
resources to simulate universal gates with a measurement-based approach instead 
of using purely coherent 
evolutions. These approaches may be important for a future scalable quantum 
computing implementation. Notable examples 
are teleportation-based \cite{Gottesman99},  and cluster 
state \cite{Briegel01,Raussendorf01,Raussendorf03}  approaches. Both approaches require the preparation of a highly entangled ancilla 
which 
subsequently acts as a useful resource for quantum computation. The basic philosophy of the measurement-based approach is to bury all the 
``difficult" quantum operations in the offline preparation of the entangled ancilla. Quantum computation then proceeds by measurement, 
which is hopefully an easier operation. Generally, to simulate any $N$-qubit operation by these approaches, we require at least two-qubit 
interactions or gates for the preparation of the entangled ancilla.  In particular, cluster state approaches allow for 
universal 
quantum computation  without the need of coherent qubit to qubit interaction once the cluster state\footnote{A  cluster state is prepared 
for example by first initialising  a lattice arrangement of qubits in  the state$\frac{1}{\sqrt{2}}(\ket{0}+\ket{1})$. A CZ gate is then 
performed between each nearest neighbour to form the cluster state.} has been 
prepared. Appropriate single qubit measurements in a cluster state allows for any quantum algorithm to be simulated. This was proven by 
Raussendorf {\em et al.}  \cite{Raussendorf03} by exploiting the correlations found in the cluster state, which is a highly entangled one. These measurements destroy the entanglement of the 
cluster state and hence, the cluster state is not reusable. Therefore, the term ``one-way quantum computing" is used interchangeably with 
cluster state quantum computing.  In the 
so-called teleportation based approach, the desired unitary operation is ``teleported"  onto an output state with the help of a suitably 
prepared entangled resource and Bell  measurements (see (\ref{completeBellintro})). Refer to Fig.~\ref{tele} for an example of  the teleportation of a  single 
qubit unitary operation. Note that this can be extended to any multiqubit unitary operation. For a  general discussion of the measurement-based 
approach, see Ref. \cite{Aliferis04,Childs05}.

There is however another kind of approach that seems to share  properties of both the coherent and measurement-based approaches. Examples 
are given in \cite{Beige00a,Franson04} where a Zeno-type  measurement induces a coherent evolution. A Zeno effect can 
be understood as the process of halting an evolution based on continuous strong measurements. This is a very useful tool to freeze 
undesired evolution. Applied to cavity QED \cite{Beige00a}, an environment induced Zeno-type effect suppresses the cavity decay, that would usually decohere the system. Applied to photons, \cite{Franson04} the Zeno effect can prevent the undesired 2-photon occupation, associated 
with a failure event, in a doped fiber with a very large 2-photon absorption cross-section and with negligible  1-photon absorption cross-section. Therefore, it 
is a special kind of ``deterministic" postselection.

In parallel to these developments, came the invention of quantum error correction codes by Calderbank, Shor and Steane (CSS) 
\cite{Shor95,Calderbank96, Steane96}.  It was initially thought that this was impossible due to the notion that quantum states are 
fragile, characterised with a continuous degree of freedom and generally subjected to  noise of continuous nature which leads to 
decoherence. 
Furthermore, the quantum no-cloning theorem \cite{Wootters82} ruled out the naive method of state copy to combat against noise, as often 
used in classical communication and computation. CSS however showed that quantum error correction was possible with the help of encoding 
operations and the measurement of error syndromes. This important result led to the concept of fault-tolerant quantum computation where 
one 
can asymptopically approach error-free computation with suitable encodings and error corrections provided that the error probability of 
gates do not exceed a certain threshold \cite{Gottesman98}.

Therefore, a lot of effort both experimentally and theoretically, has been 
focussed on the physical implementation of universal two-qubit gates.  
General criteria 
for a scalable quantum computing system were formulated by DiVincenzo 
\cite{DiVincenzo00}. Note that this criteria, based on the conventional gate model for quantum computation, have been formulated before  
the recent development of new paradigms of quantum computation, such as measurement-based approaches to quantum computation or even 
hybrid models. A relook of this criteria may be timely. To date, 
gate implementation has been implemented using NMR techniques on a molecule (perfluorobutadienyl iron complex) \cite{Vandersypen01} where 
a seven-qubit 
Shor's algorithm for the prime factorisation of the number 15 was demonstrated. 
In 
trapped ions, the  Cirac-Zoller gate \cite{Schmidt-Kaler03}, a geometric two-ion phase gate \cite{Leibfried03} the Deutsch-Jozsa 
algorithm \cite{Gulde03}, determinstic teleportation of ions \cite{Barrett04a,Riebe04}, quantum error correction \cite{Chiaverini04} as 
well as a semi-classical quantum Fourier transform \cite{Chiaverini05} has been demonstrated. These systems consist of qubits which are 
stationary with a possibly long decoherence time which makes them suitable as quantum memories. On 
the other hand, disadvantages of using stationary qubits alone include 
 the requirement for precise coherent control. Furthermore, interaction with 
remote 
stationary qubits is difficult.

Alternatively, single photons, generally loosely thought  of as a single excitation in the electromagnetic field, are natural flying 
qubits with 
long decoherence time (compared to gate operations) and are useful for the distribution of quantum 
information. At optical frequencies, the background photon count rate is virtually zero. Furthermore, photons are bosons and they obey 
the following 
commutation rules,
\begin{equation}
\left[a_i, a_j^\dagger \right]=\delta_{ij}\, \, , \left[a_i, a_j 
\right]=\left[a_i^\dagger, a_j^\dagger \right]=0
\end{equation}where $a_i$ $(a_i^\dagger)$ is the photon annihilation (creation) 
operator for a certain  mode $i$, 
$\left[\hat{a},\hat{b}\right]=\hat{a}\hat{b}-\hat{b}\hat{a}$ and $\delta_{ij}=1$ 
for 
$i=j$ or $0$ otherwise. Photons can in general be described in various encodings 
or degree of freedom, 
such as polarisation, spatial 
or frequency, or even angular momentum. For example, in polarisation 
encoding, one can assign the logical 
qubit $\ket{0}_L$ and $\ket{1}_L$ to any two orthogonal polarisations, such as 
the horizontal and vertical 
polarisations.
Single qubit operations for photons 
are extremely easy \cite{James01,Englert01} to implement with waveplates, polarisation rotators etc. However, there exists practically no coupling 
between photons 
in vacuum and hence 
a two-qubit gate implementation between photons is difficult, which is one of the 
reasons why photons are so stable. Indeed, an early proposal \cite{Milburn89} of a photonic universal three-qubit conditional SWAP gate, known as the Fredkin gate, requires 
Kerr nonlinearity to produce  intensity-dependent phase shifts. The Kerr nonlinearity required is extremely huge\footnote{See Ref. \cite{Turchette95} for a proof-of-principle demonstration with cavity QED.} if operation at the single photon level is required, which pose a severe experimental challenge. One of the early explorations of how quantum logic can be simulated (inefficiently and requiring exponential 
resources) with linear optical elements alone is found in the paper by 
Cerf {\em et al.} \cite{Cerf98}. The word ``linear optics''\footnote{This definition would certainly include squeezing which is not part 
of the standard linear optical quantum information processing toolbox. We do not have to include squeezing in this thesis, although weak 
squeezing with photon detectors can result in a heralded single photon source.  The linear optical quantum information processing toolbox 
we consider consist only of photon sources, detectors, beam splitters and phase plates.  } is defined in the sense in 
which the 
Hamiltonian that describe the photon transformation has only at most quadratic terms in photon creation or destruction operators. In this 
way, the resulting Heisenberg equations of motion are linear in terms of photon creation or destruction operators. Cerf {\em et al.}'s 
scheme is however not generally applicable to quantum computation with different photons as it operates on a Hilbert space of two degrees 
of freedom(polarisation and momentum) on the same photon instead of different photons. 
Following that, a very important no-go theorem by L{\"u}tkenhaus {\em et al.} \cite{Lutkenhaus99} showed that complete Bell state 
measurement with unit efficiency is impossible with linear optics resource alone, despite having ancillas and conditional measurements as 
resources. Note that their work covers the case where the Bell state is defined with two photons regardless of the type of encoding, 
which applies generally to quantum 
computation with different photons. Further work 
\cite{Calsamiglia01} (see also related work by Vaidman and Yoran \cite{Vaidman99}) in this direction led to the result that given no 
ancillas 
as resources, linear optics-based Bell measurement yields a success probability of at most $50 \%$ (see Chapter \ref{minsk} where such an 
example is given).
We define, without loss of generality, the basis 
states of a complete Bell measurement as
\begin{eqnarray} \label{completeBellintro}
&&\ket{\Phi^\pm}=\frac{1}{\sqrt{2}}(a_{1,h}^\dagger a_{2,v}^\dagger \pm 
a_{1,v}^\dagger a_{2,h}^\dagger ) \ket{0}_{\rm vac} \, , \nonumber \\
&&\ket{\Psi^\pm}=\frac{1}{\sqrt{2}}(a_{1,h}^\dagger a_{2,h}^\dagger \pm 
a_{1,v}^\dagger a_{2,v}^\dagger ) \ket{0}_{\rm vac} \, .
\end{eqnarray} Here, $a_{i,\lambda}^\dagger$ refers to a photon creation 
operator for spatial mode $i$ with polarisation mode $\lambda$. Bell states, which are maximally entangled two-qubit states, provide 
quantum correlations which feature as a crucial ingredient in many aspects of quantum information processing such as teleportation, 
entanglement swapping etc.. 

Later, the seminal 
paper by Knill {\em et al.} \cite{Knill01} demonstrated  that quantum 
computing can be implemented efficiently (i.e. with polynomial resource) with 
photons and linear optics elements 
if one has deterministic single photon sources with perfect photon-resolving 
detectors. They proposed a photon nonlinear gate operation based on photon interference in a linear optics setup together with 
postselection. Their scheme also makes use of a teleportation based approach \cite{Gottesman98} with the help of Bell state measurements. 
They  managed to 
approach near $100 \%$ efficient Bell measurement with the aid of asymtopically 
large number of highly entangled photons  without contradicting the no-go theorem of L{\"u}tkenhaus {\em et al.} Franson {\em et al.} 
\cite{Franson02} have subsequently improved this scaling tremendously, with feedforward corrections, from the failure rate of  $1/n$ to $1/n^2$ where $n$ is the number 
of ancilla photons. Probabilistic gate operations, based on Ref. \cite{Knill01} with some clever improvements, 
between photons have since been  
demonstrated 
experimentally \cite{Brien03,Pittman03,Gasparoni04,Zhao05} and serve as a testbed for 
quantum computation.  

Unfortunately, approaches using purely photon and 
linear optics alone seem to require huge practical resources for scaling even if 
they are polynomial \cite{Scheel03,Scheel04a,Eisert05}. In principle, this can be alleviated through a photonic 
cluster state computation model in which the cluster state  
can be built in an efficient manner  \cite{Browne05,Nielsen04}.  The cluster state than serves as a universal palette for any quantum 
computation that should proceed by measurement with unit efficiency in 
principle. A recent working demonstration of a postselected 4-photon cluster state quantum computation is found in Ref. \cite{Walther05}. 
It is however still
necessary to implement photon memory and this is currently still a great experimental challenge. 

Going by a different thread from the usual linear optics quantum computation, it has been recently shown that
relatively weak, but non-zero Kerr nonlinearity 
\cite{Munro05,Nemoto04} is sufficient for implementing universal gates between photons with unit efficiency. The surprising thing is that 
one does not really need strong Kerr nonlinearity for this. The trick is to use a homodyne measurement with an intense coherent state 
source to compensate for the weak nonlinearity. This promising approach has many applications useful to photonic based quantum 
computation. Besides implementing photonic gates with unit efficiency,  it can be used as a photon counting  non-demolition 
measurement or to turn a weak coherent pulse into a heralded single photon source.

One might envision a hybrid approach using the best properties of both 
stationary and flying qubits (photons) 
which is a key feature in this thesis. Motivations of such hybrid 
approaches have been first considered by Van Enk {\em et al.} \cite{Enk97,Enk98}  in quantum 
networking, where information can be 
sent to distant nodes via flying qubits between stationary nodes consisting of stationary qubits. The stationary qubit (for example, atoms 
or ions) function as a qubit with long decoherence time as well as acting as a quantum memory. Such an approach opens the possibility 
of 
distributed quantum computing. 

The basic 
component of such a network requires stationary qubit to flying qubit interfaces 
which is commonly found in cavity QED and atomic ensemble implementations. 

Parallel developments in the field of quantum communication which essentially 
involves the exchange of classical or quantum information through classical and 
quantum channels includes quantum cryptography, teleportation, distributed quantum computation 
etc. In the field of quantum cryptography, also widely known as quantum key 
distribution, protocols such as BB84, Ekert \cite{Bennett84,Ekert91} show the 
possibility of two parties establishing a secret key with no possibility of an 
eavesdropper being able to share any part of the secret key. The main principles 
used are the quantum no-cloning theorem  and the fact that a 
measurement of a state 
generally disturbs the original state. The eavesdropper attempting to learn anything of the secret key necessarily reduces the measured 
correlation observed between the two rightful parties, Alice and Bob. Such an observation signals the presence of a possible eavesdropper 
if the correlation is below a certain bound. Again, due to their nature of being 
flying qubits,  all experiments to-date 
implementing quantum key distributions 
involve photons \cite{Peng05,Kurtsiefer02,Gisin02}. Particularly, Ekert's 
protocol requires the preparation of an 
entangled pair of photons. Related to Ekert's protocol is the so-called Bell's 
inequality violation test \cite{Bell65,Clauser69}. This is a deep test for  
ruling out a local hidden variable theory that can make predictions similar to 
quantum mechanics. Such a test involves the repeated preparation of an entangled pair of particles followed by independent measurements 
on each of the qubits to obtain a statistical correlation function. All local hidden-variable theories will yield a bound in the 
correlation function. According to quantum mechanics, this bound can be violated.
The violation has been widely demonstrated\footnote{There exist two loopholes applying to experiments demonstrating the violation 
of the Bell's inequality. One is the lightcone loophole that would still allow a possible local realistic interpretation. The other is 
the detection loophole where the whole ensemble may not violate the Bell's inequality although the detected subensemble is perceived to 
violate it. To date, there has  been no experiments that closes both loopholes.}  for the case where at least one of the particles is a 
photon. For that of two photons, the violation has been observed from atomic cascade emission \cite{Aspect82} as well from spontaneous 
parametric downconversion \cite{Ou88}. Particularly interesting, 
the 
experiment performed by Blinov {\em et al.} \cite{Blinov04,Moehring04} demonstrated 
entanglement between an ion and a photon or in other words, a stationary and a 
flying qubit. They also demonstrated for the first time, a Bell inequality violation between particles of different species, namely an 
atom and a photon. This provides a building block to 
distributed quantum computation between distant ions assisted by photons. 
Teleportation also plays an especially important role in quantum communication. Augmented with quantum repeaters \cite{Briegel98} based 
on entanglement purification 
\cite{Bennett96a}, states can be transferred with high fidelity through teleportation with a robustly created perfect entangled ancilla. 
Experiments 
with photons over long distances have also been performed 
\cite{Ursin04,Riedmatten04} further illustrating the use of photons as an 
information 
carrier. 

As attractive as it is to use single photons in quantum information processing, five major sources of decoherence and errors are 
relevant. 
They are interferometric stability, mode matching (spatial and temporal), photon loss as well as detector accuracy and efficiency. 
Various aspects of these issues will be addressed in this thesis although we do not claim to fully resolve all these issues.

We have also seen in this section how important entanglement generation and manipulation of single 
photons is to the field of quantum information processing. 
This short introduction, in which we have not discussed those aspects of quantum information theory which are out of the scope of this 
thesis, obviously 
does not do justice 
to the wide field of quantum 
information processing. The interested reader is invited to refer to the 
book by Nielsen and Chuang 
\cite{Nielsen00} for an excellent exposition.

\section{Thesis Overview}

The central theme of this thesis 
is the manipulation and preparation of  qubits (be it stationary or flying qubits) with single photons. The bulk of the 
research work based on this theme is 
described from Chapters \ref{firework} to 
\ref{photon} and a brief overview is given as follows.

In Chapter \ref{firework}, we show that a wide range of highly entangled 
multiphoton states, including {\em W}-states, can be prepared 
by interfering  {\em single} photons inside a Bell multiport beam splitter and 
using postselection. Multiphoton entanglement being an 
important resource for linear optics quantum computing motivates the work in  
this chapter. The results that we obtain is photon encoding independent and thus 
 have wide applicability. We perform further studies on the multiport in the next 
chapter for a different application.

In Chapter \ref{fusion}, we study an important aspect of multiphoton 
interference, namely, the generalised Hong-Ou-Mandel(HOM) effect 
that plays a crucial role to many aspects of linear optics based quantum 
computation with photons. The famous HOM dip for two photons, 
where two identical photons entering  separate input arms of a 50:50 
beamsplitter never exit in separate output arms, plays an important role 
in quantum information processing such as the characterisation of single photon 
sources, Bell measurements etc. Here, we present a new generalisation of 
the HOM dip for multiparticle scattering through a multiport.
 
In Chapter \ref{hummingbird}, we propose a scheme for implementing a 
multipartite quantum filter that uses entangled photons as a 
resource. Such filters have applications in the building of cluster states  and 
are shown 
to be universal. The scheme that we propose is highly efficient and uses the 
least resources of all comparable current schemes.

In Chapter \ref{minsk}, we describe  an architecture of distributed quantum 
computing that can be realised with single photon sources 
without the need of highly entangled ancilla states. The ability to perform 
gate operations between arbitrary qubits, and not only 
between next neighbours, yields a significant improvement of the scalability of 
quantum computing architectures. This can be achieved 
with the help of distributed quantum computing, where the information of 
stationary qubits is encoded in the states of flying 
qubits (i.e. single photons), which then allow to establish a communication  
between  distant sources. We describe the implementation 
of an eventually deterministic universal two-qubit gate operation between single photon 
sources, despite  the restriction of the no-go theorem 
on deterministic Bell measurements with linear optics. This is a novel demonstration of an efficient repeat-until-success architecture to 
quantum computation.

In Chapters 
\ref{firework} and \ref{photon}, the entangled photons are shown to be generated 
postselectively or at best preselectively. Ideally, one would like to generate these entangled photons on demand. Interestingly, 
by combining ideas of photon interaction with their sources together with 
measurements from Chapters \ref{firework} and \ref{minsk}, we 
show in Chapter \ref{demand} that distributed photon entanglement  can be 
generated on demand. This can then serve as a useful tool for 
the diverse applications already mentioned.

So far, linear optics has played a crucial component in the preceding chapters. 
 Penultimately, in Chapter \ref{photon}, we do not 
consider any linear optics manipulation of light at all. Indeed, we recall the  Young's 
double-slit experiment in the 
context of two distant dipole sources in free space without cavities. 
Experiments have shown that interference fringes can be observed by coherent 
light scattered by the dipole sources.  Taking a step 
further, we show  that polarisation entanglement can also be produced by 
initially unentangled {\em distant} single photon sources in 
free space which at the same time also results in entanglement between the 
sources. This adds new perspectives to common notions where it 
is widely 
thought that photon polarisation entanglement can only be obtained via pair 
creation within the {\em same} source or via postselective measurements on 
photons that overlapped within their coherence time inside a linear optics 
setup. 

Finally, we close in Chapter \ref{outlook} with a summary and give 
limitations and an outlook of the work of 
this thesis.

\chapter{Multiphoton Entanglement through a Bell Multiport Beam Splitter using 
Independent Photons} \label{firework}

\section{Introduction} 

In this chapter, we are concerned with the practical generation of multiphoton entanglement. It is 
not possible to 
create a direct interaction between photons and hence they are difficult to 
entangle as already highlighted in Chapter \ref{overview}. One way to overcome this problem is to create polarisation or time-bin 
entanglement via photon pair creation within the same source as in atomic 
cascade and parametric down-conversion experiments. This has already been 
demonstrated experimentally by many groups 
\cite{Aspect82,Kwiat95,Brendel99,Thew02,Riedmatten04}. Other, still theoretical 
proposals employ certain features of the combined level structure of atom-cavity 
systems \cite{Gheri98,Lange00,Schon05}, photon emission from atoms in free space (described in Chapter \ref{photon})
or suitably initialised distant single photon sources (to be demonstrated in Chapter \ref{demand}).

Alternatively, highly entangled multiphoton states can be prepared using 
independently generated single photons with no entanglement in the initial 
state, linear optics and postselection. This method shall be the main focus of this chapter. In general, the photons should enter the 
linear optics network such that all information about the origin of each photon 
is erased. Afterwards postselective measurements are performed in the output 
ports of the network \cite{Lapaire03}. Using this approach, Shih and Alley 
verified the generation of maximally entangled photon pairs in 1988 by passing 
two photons simultaneously through a 50:50 beam splitter and detecting them in 
different output ports of the setup \cite{Shih88}. For a recent experiment based 
on this idea using quantum dot technology, see Ref.~\cite{Fattal04}. 

Currently, many groups experimenting with single photons favour parametric down 
conversion because of the quality of the output states produced. However, these 
experiments cannot be scaled up easily, since they do not provide efficient 
control over the arrival times of the emitted photons. It is therefore 
experimentally challenging to interfere more than two photons successfully. 
Interesting experiments involving up to five photons have nevertheless been performed 
\cite{Eibl03,Bourennane04,Zhao03,Zhao04,Zhao05} but going to higher photon numbers 
might require different technologies. To find alternatives to parametric down 
conversion, a lot of effort has been made over the last years to propose 
experimentally realisable sources for the generation of single photons on demand 
\cite{Law97,Kuhn99,Duan03,Jeffrey04}. Following these proposals, a variety of 
experiments has already been performed, demonstrating the feasibility and 
characterising the quality of  sources based on atom-cavity systems 
\cite{Kuhn00,Kuhn02,Lange04,Mckeever04}, quantum dots \cite{Benson00,Pelton02} 
and NV color centres \cite{Kurtsiefer00,Beveratos02}. Before we proceed further, it is appropriate to give a more detailed survey of the 
above mentioned single photon sources.

\subsection{Photon sources}

Photon sources can be generally subdivided into sources that give strictly antibunched 
photons, (i.e. the normalised intensity time correlation, also known as $g^{(2)}(\tau)$, is smaller than unity 
for zero time separation) or sources that yield otherwise.  True 
single photon sources  yield only antibunched photons with $g^{(2)}(0)=0$. Furthermore, an ideal turnstile 
single photon source should consistently yield exactly one photon in the same 
pure quantum state whenever required. Particularly for applications 
\cite{Knill01} relying on Hong-Ou-Mandel two-photon type interference, it is 
important for photons to be indistinguishable and of high purity.   An 
example of such a candidate source is an atom-like system which includes quantum 
dots, diamond NV-color centers and atom-cavity systems. These systems 
also afford push-button photon generation, which is an ideal requirement for 
experiments requiring single photons such as quantum cryptography or linear 
optics based quantum computing. When a photon is 
required, the source can be triggered to yield a photon.  There also exist 
approximate single photon sources that cannot be directly triggered on 
demand. In principle, even these sources can simulate an on-demand single photon 
source with the help of photon memory and non-demolition 
measurement, a currently challenging experimental requirement that has undergone 
much interest and development. Two prominent examples of 
pseudo single photon sources are a weak coherent laser pulse and the parametric 
downconversion source. We review below a selection of single photon sources that 
are currently in use. 

\subsection {Weak Coherent laser pulse}

A laser pulse can be modelled to a good approximation as a equal weighted 
mixture of coherent states of the same amplitude $\alpha$ but different phase 
$\phi$ \cite{vanEnk02,Molmer97}. This is equivalent to a mixture\footnote{In fact, the relative and not the absolute phase in quantum optics experiments turns out to be the crucial parameter. So, it is equally valid and may be more useful to model the laser pulse as an effective pure coherent state instead of  a mixture of coherent states as in (\ref{fallacy}). One should however be careful to ascribe  realism to such an interpretation. This issue has been a source of hot debate. See Ref. \cite{Bartlett05} and references therein for further discussion.} of photon Fock states weighted with a 
Poissonian distribution,
\begin{equation}\label{fallacy}
\rho_{\rm laser}=\int d\phi \ket{\alpha {\rm e}^{{\rm i} \phi}}\bra{\alpha {\rm 
e}^{{\rm i} \phi}}=\sum_n \frac{{\rm  e}^{-\alpha^2} \alpha^{2 
n}}{n!}\ket{n}\bra{n}.
\end{equation} The probability weight of an $n$-photon Fock state is thus given by  
$P(n)=\frac{{\rm  e}^{-\alpha^2} \alpha^{2 n}}{n!}$. When $\alpha \ll 1$, then $P(0) 
\gg P(1) \gg P(2)$. This implies that a weak laser pulse can indeed function as a 
pseudo single photon source. This necessarily implies low count 
rate for single photons which is due to the necessity to use a weak pulse to suppress any 
multiphoton component weighted by Poissonian statistics. Furthermore, any single 
photon pulse generated must be detected postselectively and cannot be heralded(except with the help of a photon non-demolition 
measurement) since $P(0) \gg P(1)$ therefore implying a necessarily 
large vacuum component. The weak coherent laser pulse finds its application in 
quantum key distribution (QKD) for example. It was once thought that photon-number splitting attack would be a strong impediment to 
achieve a high key rate in the presence of channel loss.
However, in the light of some recent advancement of secure QKD protocols robust against photon-number splitting attack, such as the 
decoy-state \cite{Hwang03} and strong phase-reference pulse \cite{Koashi04} protocol, the weak coherent laser pulse is likely to remain 
an important tool for QKD. 

\subsection{Parametric downconversion}

A useful photon source arises from the process of spontaneous parametric downconversion(SPDC). 
Such a 
source is used widely in a large number of quantum optics experiments such as the famous Hong-Ou-Mandel effect \cite{Hong87}. Similar to 
the coherent laser pulse 
source, it is also not a true single photon source. It is, however 
able 
to yield  a wide variety of multiphoton states postselectively. If only a single 
photon is desired, it can act as a heralded source where 
a trigger allows one to infer the emission of a photon in a certain mode. On the other hand, it is widely used 
to generate entangled photon pairs \cite{Kwiat95,Tittel98,Brendel99,Thew02,Riedmatten04} in various encodings such as polarisation,energy-time, time-bin etc. and a wide variety of experiments ranging from fundamental test of quantum 
mechanics to linear optics quantum computation have been performed with it.
SPDC can generally yield quite a high count rate of entangled photon pairs, for example about $10^5-10^6 s^{-1}$ \cite{Kurtsiefer01,Kumar04}. However, 
experiments for multiphoton interferometery typically yield, 
for example for $N=4$ photons, a coincidence count rate of $10^{-2} s^{-1}$\cite{Pan98}.  This low 
count rate is partially due to  both the random nature of photon emission as well as the need for frequency filters to erase the 
time-stamp of 
the generated photons 
for experiments such as entanglement swapping with Bell measurements \cite{Zukowski93}. The reason is due to the strong temporal correlation of the signal and idler photons emitted.
Due to this, only 
quantum optics experiment in the few photons level utilising the above states 
($N\leq 5$) are viable. 

Although there exist quasi-deterministic schemes, for 
example in Ref. \cite{Pittman02a,Jeffrey04} for photon 
generation, they require photon recycling circuits or photon memories, both still
experimentally challenging. On the other hand, parametric downconversion is 
useful 
for generating squeezed states (see for example Ref.~\cite{Wu86}), which are useful for applications in continuous 
variable quantum information processing.
The SPDC process can be roughly understood in terms of a higher 
energy 
photon being converted by an energy conserving process to two lower energy photons, traditionally known as the 
signal and idler photons. If the signal and idler photons are of the same polarisation, this is known as a 
Type-I process. If their polarisation are mutually orthogonal, this is known as a Type-II process. To generate 
a photon pair, a birefringent noncentrasymmetric nonlinear crystal is pumped by a laser, either in cw mode or 
pulsed mode. Phase matching conditions determine the direction and frequencies of the signal and idler photon pair 
generated.
 
We now denote the photon creation operator with  frequency $\omega$, polarisation $\lambda$ and direction 
$\hat{\bf k}$  as $a^\dagger_{\hat{\bf k},\lambda}(\omega)$. We denote the emitted directions(polarisation) of 
a signal and idler photon as $\hat{\bf k}_s$($\lambda_s$) and $\hat{\bf k}_i$($\lambda_i$) respectively and 
we assume a type-II process. As in Kwiat {\em et al.}\cite{Kwiat95}, we assume  the presence of two photon 
collection directions, $\hat{\bf k}_A$ and $\hat{\bf k}_B$ such that when $\hat{\bf k}_A=\hat{\bf k}_i$ or 
$\hat{\bf k}_s$, $\hat{\bf k}_B=\hat{\bf k}_s$ or $\hat{\bf k}_i$ respectively. In these two directions, 
together with frequency filters,  the postselected 2-photon state $\ket{\psi}$  generated by SPDC 
\cite{Zukowski95} is then  given by
\begin{eqnarray}
\ket{\psi} &=&
\int d\omega_p \int d\omega_i \int d\omega_s F(\omega_p,\omega_i,\omega_s) \nonumber \\
&&[a^\dagger_{\hat{\bf k}_A,\lambda_i}(\omega_i) a^\dagger_{\hat{\bf k}_B,\lambda_s}(\omega_s)+
 a^\dagger_{\hat{\bf k}_A,\lambda_s}(\omega_s) a^\dagger_{\hat{\bf k}_B,\lambda_i}(\omega_i)]\ket{\rm vac}
\end{eqnarray} where $\omega_p$ is the pump frequency and  $F(\omega_p,\omega_i,\omega_s)$ is a function 
dependent on the phase matching condition as well as the frequency envelope of the pump and the frequency 
filters. Under suitable phase matching condition, spatial pin-hole filtering, and or frequency filtering, 
$F(\omega_p,\omega_i,\omega_s)$ can be highly peaked at $F(\omega_p,\omega_p/2,\omega_p/2)$ and  $\ket{\psi}$ 
therefore reduces approximately to a polarisation Bell state \cite{Kwiat95,Zukowski95} that is widely used in 
quantum optics experiments.

\subsection{Atom-like systems for the generation of single photons on demand}

Candidate systems that could yield single photons on demand  include 
mainly atom-like systems such as atoms, quantum dots, NV (Nitrogen-Vacancy) color centers or even 
molecules. These proposals are mainly based on the ability to excite the photon source which 
then decays back to a ground state as a result yielding a photon. Due to the 
fact that every photon generated by this method requires an excitation time 
overhead, this results in naturally  antibunched photon production. These systems are more recent developments, compared to SPDC sources 
and weak coherent laser pulses. They benefit from  recent technological advancements such as semiconductor processing, laser cooling 
and trapping etc. and are still an exciting development avenue. Quantum information processing has further served as an important 
motivating factor, as is investigated in this thesis, for the continual development of these sources.

The quantum dot single photon source is operated by performing
a sharp laser pulse excitation to an excited level representing the creation of a so-called excited 
exciton which rapidly decays non-radiatively to the lowest excited 
state of the exciton. A subsequent slower decay back to the ground state yields 
a photon. In practice, biexcitonic excitation is usually preferable, due to the ability to spectrally isolate the 
last but one photon \cite{Santori00}. With the quantum dot integrated in monolithic cavity structures, the 
spontaneous emission rate can be increased substantially with the 
emission mainly into the cavity mode which results in directed photon emission. 
Due to the non-radiative decay in the excitation process, there is a slight 
uncertainty in the photon emission time. Even with this and all other effects 
contributing to decoherence, photon pulses of sufficient purity and 
indistinguishability can be generated consistently to observe a Hong-Ou-Mandel 
2-photon interference\cite{Santori02} at low temperatures. If spectral purity is not needed, single photon 
generation can still be performed at room temperature \cite{Michler00}. The quantum dot also allows for 
coherent 
Raman excitation schemes \cite{Kiraz04} and may lead to an attractive solid-state 
alternative to photon guns based on atom-cavity systems. A good review on the 
physics of photon generation through quantum dots is found in Santori {\em et 
al.} \cite{Santori04}. It is worth mentioning also the quantum dot can  be excited electronically via a 
Coloumb blockade and Pauli effect \cite{Kim99} leading to an electronic turnstile single photon device.

NV color center, an optically active defect inside a diamond nanocrystal, is an alternative atom-like system 
for photon generation. Unlike the quantum dot, for applications in quantum 
cryptography where the purity of photons generated is not too important, NV 
color centers can be maintained at room temperature during operation. Their key advantage lies in the fact that 
they boast of extremely stable operation even at room temperature. The 
excitation  of an NV color center to generate a single photon is similar to that 
of the quantum dot. To date, photon antibunching has been observed with this 
method \cite{Kurtsiefer00,Beveratos02}. The demonstration of Hong-Ou-Mandel 
2-photon interference would probably require  cyrogenic temperature operation. 
Before moving to atom-cavity systems, we mention that single molecules are yet another attractive atom-like system capable of yielding 
single photons. In fact, the most recent experiment with  a TDI (Terrylenediimide) molecule at cryogenic temperature have yielded photons 
demonstrating the Hong-Ou-Mandel 2-photon interference \cite{Kiraz05}.

The atom-cavity system consists of an atom ideally trapped in a high-finesse 
cavity. A laser and cavity-driven Raman process which is described in more 
detail in Chapter \ref{minsk} transfers a photon in the cavity which subsequently 
leaks out.  This has been experimentally demonstrated 
\cite{Kuhn02,Mckeever04,Lange04} and photons generated from such systems have a 
sufficient purity and consistency to observe the Hong-Ou-Mandel 2-photon 
interference \cite{Legero04}. In principle, barring imperfections such as photon absorption, weak 
cavity coupling, the photon generation probability approaches unity. The 
atom-cavity system allows also for generation of entangled photons on demand \cite{Schon05,Gheri98}. 
More generally, it also allows for the state of an atom to be redundantly 
encoded to the photon it generates which is described in Chapter \ref{minsk}.

Current experimental achievements of all these sources have admittedly not yet achieved photon production on demand. The best reported 
photon production efficiency \cite{Mckeever04} is still less than $70 \%$ although there is no limit in principle to achieving near unit 
efficiency. Compared to SPDC, these  sources generally demand greater experimental complexity at the present. Moreover, these sources are generally not as wavelength tunable as SPDC sources, although this need not be a real disadvantage.  However, with 
strong motivations for scalability in linear optics quantum computation and distributed quantum computing as well as quantum cryptography, much effort to the development of a robust true 
photon on demand source is ongoing.

\subsection{Single photons and multiport} \label{photonmultiport}

Motivated by the above recent developments, several authors studied the creation of 
multiphoton entanglement by passing photons generated by a single photon source 
through a linear optics network 
\cite{Zukowski97,Lee02,Fiurasek02,Zou02,Wang03,Sagi03,Pryde03,Mikami04,Shi05}. A variety 
of 
setups has been considered. Zukowski {\em et al.} showed that the $N \times N$ 
Bell multiport beam splitter (see below) can be used to produce higher dimensional EPR (Einstein-Podolsky-Rosen)
correlations \cite{Zukowski97}. Shi and Tomita \cite{Shi05} for example studied 3 and 4-photon {\em W}-state preparation with  multiports 
which led to high generation efficency. They also conjectured but did not prove that a symmetric $N \times N$  multiport may be used to 
generate an $N$-photon {\em W}-state. Mikami {\em et al.}  studied the generation of $N$-photon states through parametric 
downconversion, coherent laser states and multiports with photon number-resolving detectors.   Such multiports  have an important 
application in boosting the success probability of linear optics teleportation to  
near unity \cite{Knill01} using a special highly entangled multiphoton ancilla.  

Special attention has been paid to the optimisation of schemes for the 
generation of the so-called NOON state with special applications in lithography 
\cite{Lee02,Fiurasek02,Zou02,Pryde03}. Wang studied the event-ready generation 
of maximally entangled photon pairs without photon number-resolving detectors 
\cite{Wang03} and Sagi proposed a scheme for the generation of $N$-photon 
polarisation entangled GHZ states \cite{Sagi03}. It is also possible to prepare 
arbitrary multiphoton states \cite{Fiurasek03} using for example probabilistic 
but universal linear optics quantum gates, like the one described in 
Refs.~\cite{Pittman02,Knill01} or using a large enough optical cluster state 
\cite{Yoran03,Nielsen04,Browne05} which still remains an experimental challenge. 
However, these approaches are not always the most favourable and often require a 
large number of entangled photon ancillas. 

\begin{figure}
\begin{minipage}{\columnwidth}
\begin{center}
\resizebox{\columnwidth}{!}{\rotatebox{0}{\includegraphics{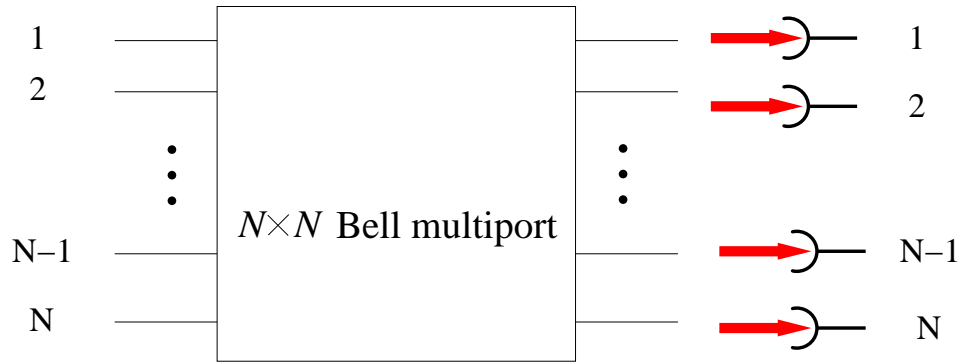}}}  
\end{center}
\caption{Experimental setup for the generation of multiphoton entanglement by 
passing $N$ single photons through an $N \times N$ Bell multiport beam splitter. 
The state preparation is considered successful under the condition of the 
collection of one photon per output.} \label{scheme}
\end{minipage}
\end{figure}

Here we are interested in the generation of highly entangled qubit states of $N$ 
photons using only a single photon source and a symmetric $N \times N$ Bell 
multiport beam splitter, which can be realised by combining single beam 
splitters into a symmetric linear optics network with $N$ input and $N$ output 
ports \cite{Zukowski97,Torma95}.  In the two-photon case, the 
described scheme 
simplifies to the experiment by Shih and Alley \cite{Shih98}. To entangle $N$ 
photons, every input port $i$ of the Bell multiport should be entered by a 
single photon prepared in a state $|\lambda_i \rangle$. The photons then 
interfere with each other before leaving the setup (see Fig.~\ref{scheme}). We 
consider the state preparation as successful under the condition of the 
collection of one photon per output port, which can be relatively easily 
distinguished from cases with at least one empty output port. In general, this can be done with photon number-resolving non-demolition 
detectors \cite{Munro05,Nemoto04} and we obtain preselected multiphoton entanglement. Otherwise, the entangled state is postselected  
without the need of photon number-resolving detectors.  Postselected photon state preparation is nevertheless useful if one can 
accomplish a non-trivial task. Examples 
are teleportation \cite{Joo03}, quantum secret sharing, secure quantum key distribution \cite{Chen05}, testing entanglement with 
witnesses and observing a violation of Bell's inequality \cite{Bourennane04a,Toth05} .

One advantage of using a Bell multiport beam splitter for the generation of 
multiphoton entanglement is that it redirects the photons without changing their 
inner degrees of freedom, like polarisation, arrival time and frequency. The 
described setup can therefore be used to generate polarisation, time-bin and 
frequency entanglement. Especially, time-bin entanglement can be very robust 
against decoherence and has, for example, applications in long-distance fibre 
communication \cite{Gisin02}. Moreover, the preparation of the input product 
state does not require control over the relative phases of the incoming photons, 
since the phase factor of each photon contributes at most to a global phase of 
the combined state with no physical consequences. 

It is the purpose of this chapter to explore some novel properties of a Bell 
multiport beam splitter. This chapter is organised as follows. In Section 
\ref{firework:scatter} we introduce the notation for the description of photon scattering 
through a linear optics setup. Section \ref{fourphoton} shows that a wide range 
of highly entangled photon states can be obtained for $N=4$, including the {\em 
W}-state, the GHZ-state and a double singlet state. Afterwards we discuss the 
generation of $W$-states for arbitrary photon numbers $N$ and calculate the 
corresponding probabilities for a successful state preparation. We observe an 
interesting non-monotonic decreasing trend in the success probability as $N$ 
increases owing to quantum interference. Finally we 
conclude our results in Section \ref{firework:conclusions}.

\section{Photon scattering through a linear optics setup} \label{firework:scatter}

Let us first introduce the notation for the description of the transition of the 
photons through the $N \times N$  multiport beam splitter. In the following, 
$|+ \rangle$ and $|- \rangle$ denote the state of a photon with  polarisation 
``$+$'' and ``$-$'' respectively.  Alternatively, $|+ \rangle$ could describe a 
single photon with an earlier arrival time or a higher frequency than a photon 
prepared in $|- \rangle$. As long as the states $|\pm \rangle$ are orthogonal 
and the incoming photons are in the same state with respect to all other degrees 
of freedom, except of course their input spatial positions, the calculations presented in this paper apply throughout. Moreover, 
we assume that each input port $i$ is entered by one independently generated 
photon prepared in $|\lambda_i \rangle = \alpha_{i+} |+\rangle_i + \alpha_{i-} 
|-\rangle_i$, where $\alpha_{i \pm}$ are complex coefficients with $|\alpha_{i+} 
|^2 + |\alpha_{i-}|^2 = 1$. If $a_{i \mu}^\dagger$ denotes the creation operator 
for one photon with  mode $\mu$ in input port $i$, the $N$-photon input 
state can be written as 
\begin{eqnarray} \label{firework:in}
|\phi_{\rm in} \rangle &=& \prod_{i=1}^N \Big( \sum_{\mu=+,-} 
\alpha_{i \mu} \, a_{i \mu}^\dagger \Big) \, |0 \rangle 
\end{eqnarray}
with $|0 \rangle$ being the vacuum state with no photons in the setup. 

\begin{figure}\label{pyramid}
\begin{center}
\includegraphics[scale=0.75]{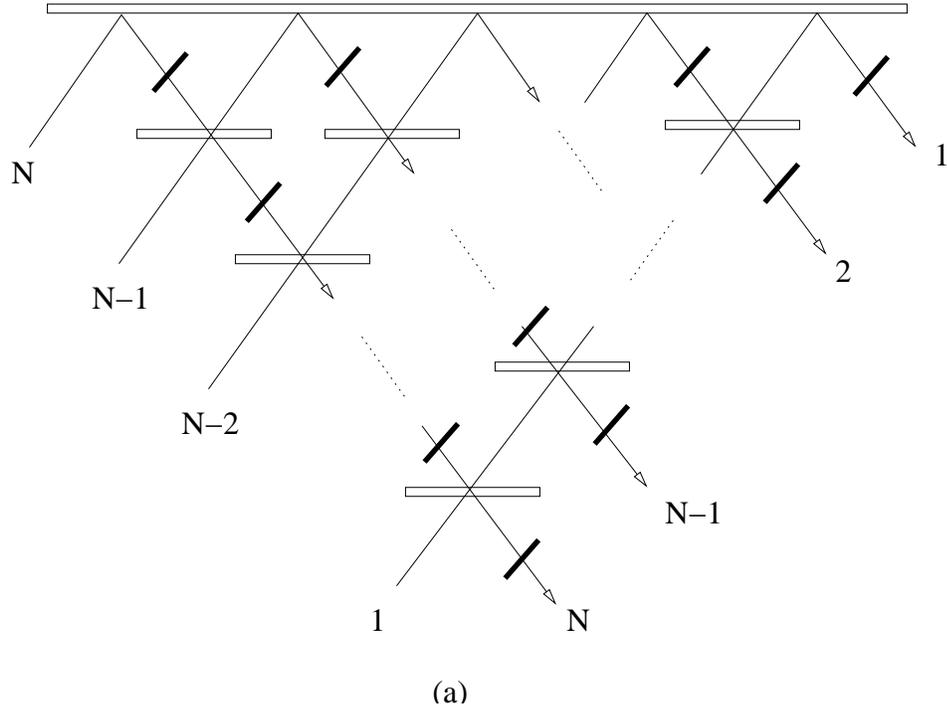} 
\end{center}
\caption{ Pyramidal construction of a $N \times N$ multiport consisting of 
beamsplitters and phase plates } \label{construction}
\end{figure}

Let us now introduce the unitary $N \times N$-multiport transformation operator, 
namely the scattering matrix $S$, that relates the input state of the system to 
the corresponding output state
\begin{eqnarray} \label{fin}
|\phi_{\rm out} \rangle &=& S \, |\phi_{\rm in} \rangle \, .
\end{eqnarray}
Using Eq.~(\ref{firework:in}) and the relation $S^\dagger S = I \!\! I$ therefore yields
\begin{eqnarray} \label{fin2}
|\phi_{\rm out} \rangle &=& S \, \Big( \sum_{\mu=+,-} \alpha_{1 \mu} \, 
a_{1 \mu}^\dagger \Big) 
\, S^\dagger S \, \Big( \sum_{\mu=+,-} \alpha_{2 \mu} \, a_{2 
\mu}^\dagger \Big) \cdot \, . \, . \, . \, \cdot S^\dagger S \, \Big( 
\sum_{\mu=+,-} \alpha_{N \mu} \, 
a_{N \mu}^\dagger \Big) \, S^\dagger S  \, |0 \rangle \nonumber \\
&=& \prod_{i=1}^N \, \Big( \, \sum_{\mu=+,-} \alpha_{i \mu} \, S \, a_{i 
\mu}^\dagger \, S^\dagger \, \Big) \, |0 \rangle \, .
\end{eqnarray}
In the following, the matrix elements $U_{ji}$ of the unitary transformation 
matrix $U$ denote the amplitudes for the  redirection of a photon in input $i$ 
to output $j$. Generally speaking, an $N \times N$ multiport described by any arbitrary transfer matrix $U$ 
may be constructed by a pyramidal arrangement of beamsplitters and phase plates 
as shown in Fig.~\ref{construction}.

The most familiar example of a multiport is the  $2 \times 2$ beamsplitter that has 
2 input and 2 output ports. It can be described by a unitary $2 \times 2$ matrix $B(R,\phi)$ 
given by
\begin{equation}
B(R,\phi)=\left( \begin{array}{cc} \sqrt{T} & {\rm e}^{{\rm i}\phi}\sqrt{R} \\ 
\sqrt{R} & 
-{\rm e}^{{\rm i}\phi} \sqrt{T} \end{array} \right),
\end{equation} where the $R$ denotes the reflectivity and $T=1-R$ denotes the 
transmittivity of the beamsplitter. The phase $\phi$ is obtained by placing a phase 
shifter at one of the input ports.

Reck \cite{Reck94,Reck96} (see also \cite{Sun01}) has shown this using similar methods as used in Gaussian 
elimination. The key to his proof is to factorize the matrix $U$ into a product 
of block matrices describing only $2 \times 2$ beam splitter matrices together with phase 
shifts. We begin by defining the $N \times N$ matrix $T_{pq}(R_{pq}, \phi_{pq})$ 
which is essentially an identity matrix except for possibly four elements indexed by $pp$, $pq$, $qp$ and $qq$ 
which denote  effectively a $2 \times 2$ unitary matrix. Note 
that $T_{pq}(R_{pq}, \phi_{pq})$ represents a $2 \times 2$ beamsplitter with 
matrix $B(R_{pq},\phi_{pq})$  where the two input ports are $p$ and $q$ input 
ports of the $N \times N$ multiport.

For the inverse matrix of $U$ denoted as $U^{-1}$, it is possible to find 
appropriate  $T_{pq}(R_{pq}, \phi_{pq})$ such that $U^{-1} \prod^{1}_{i=N-1} 
T_{Ni}(\omega_{Ni}, \phi_{Ni})$ is another unitary matrix where the last row and 
column contains only zeros except on the diagonal element which contains only a 
phase factor. Defining $\prod^{1}_{i=N-1} T_{Ni}(\omega_{Ni}, \phi_{Ni})=L(N)$, 
we can systematically reduce $U^{-1}$ to a diagonal matrix $D^{-1}$ that 
contains only phase factors in the diagonal elements by the following operation,
\begin{equation}
U^{-1} L(N)L(N-1)...L(2)=D^{-1}.
\end{equation}
It is clear that $U=L(N)L(N-1)...L(2)D$ can be built by a series of $2 \times 2$ 
beamplitters with phase shifters in each of the $N$ output ports corresponding 
to the diagonal elements of $D^{-1}$. Indeed, the pyramidal construction shown in 
Fig.~\ref{pyramid} corresponds precisely to such a decomposition operated in reverse.  From this 
construction, the maximum number of $2 \times 2$ beamsplitters needed for the 
construction for a $N \times N$ multiport is given by $N(N-1)/2$.
 
Since the multiport beam splitter does not contain any elements 
that change the inner degrees of freedom of the incoming photons, the transition 
matrix $U$ does not depend on $\mu$. Denoting the creation operator for a single 
photon with parameter $\mu$ in output port $j$ by $b_{j \mu}^\dagger$ therefore 
yields 
\begin{eqnarray} \label{tran}
S \, a_{i \mu}^\dagger \, S^\dagger = \sum_j U_{ji} \, b_{j \mu}^\dagger \, .
\end{eqnarray}
Inserting this into Eq.~(\ref{fin}) we can now calculate the output state of the 
$N \times N$ multiport given the initial state (\ref{firework:in}) and obtain
\begin{eqnarray} \label{firework:output1}
|\phi_{\rm out} \rangle &=& \prod_{i=1}^N \, \Bigg[ \, \sum_{j=1}^N \, U_{ji} \, 
\Big( \sum_{\mu=+,-} \alpha_{i \mu} \, b_{j \mu}^\dagger \Big) \, 
\Bigg] \, |0 \rangle \, . ~
\end{eqnarray}
This equation describes the independent redirection of all photons to their 
potential output ports. Conservation of the norm of the state vector is 
guaranteed by the unitarity of the transition matrix $U$. It is also important to note that any multiplication of phase factors in any of 
the input or output ports as well as any relabelling of the input or output ports constitutes a multiport which is essentially equivalent 
to the original multiport. In this sense, the original multiport is defined up to an equivalence class.

The state preparation is considered successful under the condition of the 
collection of one photon per output port. To calculate the final state, we apply 
the corresponding projector to the output state (\ref{firework:output1}) and find that 
the thereby postselected $N$-photon state equals, up to normalisation, 
\begin{equation} \label{firework:output2}
|\phi_{\rm pro} \rangle=\sum_{\sigma} \Bigg[ \prod_{i=1}^N U_{\sigma(i) i} \Big( 
\sum_{\mu=+,-} \alpha_{i \mu} b_{\sigma (i) \mu}^{\dagger} \Big) 
\Bigg]  \, |0 \rangle \, .~
\end{equation}
Here $\sigma$ are the $N!$ possible permutations of the $N$ items $\{1,\, 2, \, 
..., \, N\}$. Note that the bosonic statistics of photons has been taken into account inherently in the formulation. A further 
elaboration on this is due in Chapter \ref{fusion}. Moreover, the norm of the state (\ref{firework:output2}) squared, namely
\begin{equation} \label{firework:suc}
P_{\rm suc} = \| \, |\phi_{\rm pro} \rangle \, \|^2 \, ,
\end{equation}
is the success rate of the scheme and probability for the collection of one 
photon in each output $j$.

\subsection{The Bell multiport beam splitter}

Motivated by a great variety of applications, we are particularly interested in 
the generation of highly entangled photon states of a high symmetry, an example 
being {\em W}-states. This suggests to consider symmetric multiports as described in \ref{photonmultiport}, which 
redirect each incoming photon with equal probability to all potential output 
ports. A special example for such an $N \times N$ multiport is the Bell 
multiport beam splitter. Its transformation matrix
\begin{eqnarray}\label{fourier}
U_{ji} &=& {\textstyle{1 \over \sqrt{N}} \, \omega_N^{(j-1)(i-1)}}
\end{eqnarray} 
is also known as a discrete Fourier transform matrix and has been widely 
considered in the literature \cite{Zukowski97,Torma95,Torma98}. Indeed, the Bell multiport is a linear optical realisation of a quantum Fourier transform. Here $\omega_N$ 
denotes the $N$-th root of unity,
\begin{equation} \label{root}
\omega_N \equiv \exp \left( 2{\rm i} \pi /N \right) \, .
\end{equation} 
Proceeding as in Section II.D of Ref.~\cite{Zukowski97}, it can easily be 
verified that $U$ is unitary as well as symmetric. Especially for $N=2$, the 
transition matrix (\ref{fourier}) describes a single 50:50 beam splitter.

\section{The generation of 4-photon states} \label{fourphoton}

Before we discuss the $N$-photon case, we investigate the possibility of preparing 
highly entangled 4-photon states using specially prepared photons and a $4 
\times 4$ Bell multiport beam splitter. For $N=4$, the transition matrix 
(\ref{fourier}) becomes 
\begin{eqnarray} \label{firework:matrix4}
U &=& {\textstyle {1 \over 2}} \left( \begin{array}{rrrr} 
1 & 1 & 1 & 1\\
1 & \omega_4 & \omega_4^2 & \omega_4^3\\
1 & \omega_4^2 & \omega_4^4 & \omega_4^6\\
1 & \omega_4^3 & \omega_4^6 & \omega_4^9 \end{array} \right)  
= {\textstyle {1 \over 2}} \left( \begin{array}{rrrr}
1 & 1 & 1 & 1\\
1 & {\rm i} & -1 & -{\rm i}\\
1 & -1 & 1 & -1\\
1 & -{\rm i} & -1 & {\rm i} \end{array} \right) \, .~  \nonumber \\ &&
\end{eqnarray}
The following analysis illustrates the richness of the problem as well as  
motivating possible generalisations for the case of arbitrary photon numbers. 

\subsection{Impossible output states} \label{pose}

Let us first look at the seemingly trivial situation, where every input port of 
the multiport beamspliter is entered by one photon in the same state, let us say 
in $|+ \rangle$, so that
\begin{equation} \label{inni}
|\phi_{\rm in} \rangle = a_{1 +}^{\dagger}a_{2 +}^{\dagger}a_{3 +}^{\dagger}a_{4 
+}^{\dagger} \, |0 \rangle \, .
\end{equation}
Using Eqs.~(\ref{firework:output2}) and (\ref{firework:matrix4}), we then find that the collection 
of one photon per output port prepares the system in the postselected state 
\begin{equation}
|\phi_{\rm pro} \rangle = \sum_{\sigma} \Bigg[ \prod_{i=1}^4 U_{\sigma (i)i} \, 
b_{i +}^{\dagger} \Bigg] \, 
|0 \rangle = 0 \, .
\end{equation} 
This means, that it is impossible to pass four photons in the same state through 
the considered setup with each of them leaving the multiport in a different 
output port. More generally speaking, the state with four photons in the same 
state does not contribute to the event of collecting one photon per output port. 
It is therefore impossible to prepare any superposition containing the states 
$b_{1 +}^{\dagger}b_{2 +}^{\dagger}b_{3 
+}^{\dagger}b_{4 +}^{\dagger} \, |0 \rangle$ or $b_{1 -}^{\dagger}b_{2 
-}^{\dagger}b_{3 -}^{\dagger}b_{4 -}^{\dagger} \, |0 \rangle$, respectively. The 
reason is destructive interference of certain photon states within the linear 
optics setup, which plays a crucial role for the generation of multiphoton 
entanglement via postselection. This effect is further studied and generalised in Chapter \ref{fusion}.

\subsection{The 4-photon {\em W}-state} \label{W4}

We now focus our attention on the case, where input port 1 is entered by a 
photon prepared in $|+ \rangle$ while all other input ports are entered by a 
photon in $|- \rangle$, i.e.
\begin{equation} \label{Win}
|\phi^W_{\rm in} \rangle = a_{1 +}^{\dagger}a_{2 -}^{\dagger}a_{3 
-}^{\dagger}a_{4 -}^{\dagger} \, |0 \rangle \, .
\end{equation}
Using again Eqs.~(\ref{firework:output2}) and (\ref{firework:matrix4}), we find that the 
collection of one photon per output port
corresponds to the postselected 4-photon state 
\begin{eqnarray} \label{proni}
|\phi_{\rm pro}^W \rangle &=& \sum_{j=1}^4 U_{j1} \, b_{j +}^{\dagger} \, 
\sum_{\sigma_j}  \Bigg[  
\prod_{i=2}^4 U_{\sigma_j(i) i} \, b_{\sigma_j (i) -}^{\dagger} \Bigg] \, 
|0 \rangle \, , \nonumber \\&&
\end{eqnarray} 
where the $\sigma_j$ are the $3!$ permutations that map the list $\{2, \, 3, \,  
4\}$ onto the list $\{1, \, ..., \, (j-1), \, (j+1), \, ..., \, 4 \}$. If 
$|j_{\rm out} \rangle$ denotes the state with one photon in $|+ \rangle$ in 
output port $j$ and one photon in $|-\rangle$ everywhere else, 
\begin{equation} \label{z}
|j_{\rm out} \rangle \equiv b_{N -}^{\dagger} \, . \, . \, . \, b_{(j+1) 
-}^{\dagger} b_{j +}^{\dagger} b_{(j-1) -}^{\dagger} \, . \, . \, . \, b_{1 
-}^{\dagger} \, |0\rangle \, ,
\end{equation}
and $\beta_j$ is a complex probability amplitude, then the output state 
(\ref{proni}) can be written as
\begin{eqnarray} \label{pro3}
|\phi_{\rm pro}^W \rangle &=& \sum_{j=1}^4 \beta_j  \, |j_{\rm out} \rangle \, .
\end{eqnarray}
Furthermore, we introduce the reduced transition matrices $U_{\rm red}^{(j)}$, 
which are obtained by deleting the first column and the $j$-th row of the 
transition matrix $U$. Then one can express each $\beta_j$ as the permanent\footnote{Note that the permanent of matrix $U$ is  ${\rm perm}(U)=\sum_{\sigma}\prod_{i=1}^N U_{i\sigma(i)}.$} 
\cite{Horn85,Scheel04,Minc78} of a matrix,
\begin{equation} \label{c}
\beta_j = U_{j 1} \sum_{\sigma_j} \prod_{i=2}^4 U_{\sigma_j (i)i} 
= U_{j 1} \, {\rm perm} \, \left( U_{\rm red}^{(j) {\rm T}} \right) \, .
\end{equation}
The output state (\ref{pro3}) equals a {\em W}-state, if the coefficients 
$\beta_j$ are all of the same size and differ from each other at most by a phase 
factor. 

To show that this is indeed the case, we calculate the reduced matrices $U_{\rm 
red}^{(j)}$ explicitly\footnote{The reason for not simplifying these matrices 
is that the following equations 
provide the motivation for the proof of the general case in Section 
\ref{doubleW}.} and obtain 
\begin{eqnarray} \label{sing}
&& \hspace*{-0.5cm} U_{\rm red}^{(1)} = {\textstyle {1 \over 2}} 
\left( \begin{array}{rrrr} \omega_4 & \omega_4^2 & \omega_4^3 \\ \omega_4^2 & 
\omega_4^4 & 
\omega_4^6 \\ \omega_4^3 & \omega_4^6 & \omega_4^9 \end{array} \right) \, , ~
U_{\rm red}^{(2)} =  {\textstyle {1 \over 2}} 
\left( \begin{array}{rrrr} 1 & 1 & 1 \\ \omega_4^2 & \omega_4^4 & \omega_4^6 \\ 
\omega_4^3  
& \omega_4^6 & \omega_4^9 \end{array} \right) \, ,  \nonumber \\
&& \hspace*{-0.5cm} U_{\rm red}^{(3)} = {\textstyle {1 \over 2}} 
\left( \begin{array}{rrrr} 1 & 1 & 1 \\ \omega_4 & \omega_4^2 & \omega_4^3 \\ 
\omega_4^3 & 
\omega_4^6 & \omega_4^9 \end{array} \right) \, , ~
U_{\rm red}^{(4)} =  {\textstyle {1 \over 2}} 
\left( \begin{array}{rrrr} 1 & 1 & 1 \\ \omega_4 & \omega_4^2 & \omega_4^3 \\ 
\omega_4^2 & 
\omega_4^4 & \omega_4^6 \end{array} \right) \, . \nonumber \\
\end{eqnarray}
The coefficients $\beta_j$ differ at most by a phase factor, if the norm of the 
permanents of the transpose of these reduced matrices is for all $j$ the same. 
To show that this is the case, we now define the vector 
\begin{equation}
{\bf v} = (\omega_4, \omega_4^2, \omega_4^3) \, ,
\end{equation}
multiply each row of the matrix $U_{\rm red}^{(1)}$ exactly $(j-1)$ times with 
${\bf v}$ and obtain the new matrices   
\begin{eqnarray}
&&  \hspace*{-0.5cm} \tilde{U}_{\rm red}^{(1)} = U_{\rm red}^{(1)} \, , ~
\tilde{U}_{\rm red}^{(2)} =  {\textstyle {1 \over 2}}
\left( \begin{array}{rrrr} \omega_4^2 & \omega_4^4 & \omega_4^6 \\ \omega_4^3 & 
\omega_4^6 & \omega_4^9 \\ 1 & 1 & 1 \end{array} \right) \, , ~ \nonumber \\
&& \hspace*{-0.5cm} \tilde{U}_{\rm red}^{(3)} =  {\textstyle {1 \over 2}}
\left( \begin{array}{rrrr} \omega_4^3 & \omega_4^6 & \omega_4^9 \\ 1 & 1 & 1 \\ 
\omega_4 &
\omega_4^2 & \omega_4^3 \end{array} \right) \, , ~ 
\tilde{U}_{\rm red}^{(4)} = U_{\rm red}^{(4)}  \, . 
\end{eqnarray}
The above described multiplication amounts physically to the multiplication of 
the photon input state with an overall phase factor and 
\begin{equation}
\left| \, {\rm perm} \left( U_{\rm red}^{(1) {\rm T}} \right) \, \right| = 
\left| \, {\rm perm} \left( \tilde{U}_{\rm red}^{(j) {\rm T}} \right) \, \right| 
\, . 
\end{equation}
Moreover, using the cyclic symmetry of permanents \cite{Horn85}, we see that 
\begin{equation}
{\rm perm} \left( U_{\rm red}^{(j) {\rm T}} \right) = {\rm perm} \left( 
\tilde{U}_{\rm red}^{(j) {\rm T}} \right) \, .
\end{equation}
This implies together with Eq.~(\ref{c}) that the norm of the coefficients 
$\beta_j$ is indeed the same for all $j$. Furthermore, using the above argument 
based on the multiplication of phase factors to the photon input state, one can 
show that
\begin{equation} \label{phase}
\beta_j= \beta_1 \left( \prod_{k=0}^3 \omega_4^k \right)^{j-1} \, .
\end{equation}
Inserting this into Eq.~(\ref{proni}), we find that the postselected state with 
one photon per output port equals, after normalisation\footnote{In the 
following we denote normalised states by marking them with the $\hat{~}$ 
symbol.}, the {\em W}-state 
\begin{equation} \label{beta}
|\hat{\phi}_{\rm pro}^W \rangle 
= {\textstyle {1 \over 2}} \, \big[ \, b_{1 +}^{\dagger}b_{2 -}^{\dagger}b_{3 
-}^{\dagger}b_{4 -}^{\dagger}-b_{1 -}^{\dagger}b_{2 
+}^{\dagger}b_{3 -}^{\dagger}b_{4 -}^{\dagger} 
+ b_{1 -}^{\dagger} b_{2 -}^{\dagger} b_{3 +}^{\dagger} b_{4 -}^{\dagger} 
-b_{1 -}^{\dagger} b_{2 -}^{\dagger} b_{3 -}^{\dagger} b_{4  
+}^{\dagger} \, \big] \, |0 \rangle \, .
\end{equation} 
In analogy, we conclude that an input state with one photon in $|-\rangle$ in 
input port 1 and a photon in $|+ \rangle$ in each of the other input ports, 
results in the preparation of the {\em W}-state 
\begin{equation} \label{beta2}
|\hat{\phi}_{\rm pro}^{W'} \rangle 
= {\textstyle {1 \over 2}} \, \big[ \, b_{1 -}^{\dagger}b_{2 +}^{\dagger}b_{3 
+}^{\dagger}b_{4 +}^{\dagger}-b_{1 +}^{\dagger}b_{2 -}^{\dagger}b_{3 
+}^{\dagger}b_{4 +}^{\dagger} 
+b_{1 +}^{\dagger}b_{2 +}^{\dagger}b_{3 -}^{\dagger}b_{4 +}^{\dagger}-b_{1 
+}^{\dagger}b_{2 +}^{\dagger}b_{3 +}^{\dagger}b_{4 -}^{\dagger} \, \big] \, 
|0 \rangle 
\end{equation}
under the condition of the collection of one photon per output port. Both 
states, (\ref{beta}) and (\ref{beta2}), can be generated with probability 
\begin{equation} \label{s}
P_{\rm suc}={\textstyle {1 \over 16}} \, .
\end{equation}
Transforming them into the usual form of a $W$-state with equal coefficients of 
all amplitudes \cite{Dur00} only requires further implementation of a Pauli 
$\sigma_z$ operation (i.e.~a state dependent sign flip) on either the first and 
the third or the second and the fourth output photon, respectively.

Although symmetry considerations may suggest that one can obtain a {\em W}-state given the described input, it is not obvious from a 
rigorous point of view that this is the case. We have therefore performed explicit calculations to obtain the output state. In fact, 
naive application of a symmetry argument may lead to an incorrect predicted state  which we will show in the next 2 subsections. 

\subsection{The 4-photon GHZ-state} \label{fun}

Besides generating {\em W}-states, the proposed setup can also be used to 
prepare 4-photon GHZ-states. This requires, feeding each of the input ports 1 
and 3 with one photon in $|+ \rangle$ while the input ports 2 and 4 should each 
be entered by a photon in $|- \rangle$ such that
\begin{equation}
|\phi_{\rm in}^{\rm GHZ} \rangle = a_{1 +}^{\dagger}a_{2 -}^{\dagger}a_{3 
+}^{\dagger}a_{4 -}^{\dagger} \, |0 \rangle \, .
\end{equation}
Calculating again the output state under the condition of collecting one photon 
per output port, we obtain 
\begin{equation} 
|\phi_{\rm pro}^{\rm GHZ} \rangle 
= \sum_{\sigma}U_{\sigma (1)1}U_{\sigma (2)2}U_{\sigma (3)3}U_{\sigma (4)4} 
b_{\sigma (1)+}^{\dagger}b_{\sigma (2) -}^{\dagger}b_{\sigma (3) 
+}^{\dagger}b_{\sigma (4) -}^{\dagger} \, |0 \rangle \ ,
\end{equation} 
where the $\sigma$ are the $4!$ permutations that map the list $\{1, \, 2, \, 3, 
\, 4 \}$ onto itself. On simplification, one finds that there are only two 
constituent states with non-zero coefficients and $|\phi_{\rm pro}^{\rm GHZ} 
\rangle$ becomes after normalisation
\begin{equation} \label{outGHZ}
|\hat{\phi}_{\rm pro}^{\rm GHZ} \rangle = {\textstyle {1 \over \sqrt{2}}} \, 
\big[ \, b_{1 +}^{\dagger}b_{2 
-}^{\dagger}b_{3 +}^{\dagger}b_{4 -}^{\dagger}-b_{1 
-}^{\dagger}b_{2+}^{\dagger}b_{3 -}^{\dagger}b_{4 +}^{\dagger} \, \big] \, |0 
\rangle 
\end{equation} 
which equals the GHZ-state up to local operations. Transforming (\ref{outGHZ}) 
into the usual form of the GHZ-state requires changing the state of two of the 
photons, for example, from $|+ \rangle$ into $|- \rangle$. This can be realised 
by applying a Pauli $\sigma_x$ operation to the first output port as well as a 
$\sigma_y$ operation to the third output. 

Finally, we remark that the probability for the creation of the GHZ-state 
(\ref{ss}) is twice as high as the probability for the generation of a {\em 
W}-state (\ref{s}), 
\begin{equation} \label{ss}
P_{\rm suc} = {\textstyle {1 \over 8}} \, .
\end{equation} 
Unfortunately, the experimental setup shown in Fig.~\ref{scheme} does not allow 
for the preparation of GHZ-states for arbitrary photon numbers $N$. For a 
detailed description of polarisation entangled GHZ states using a different 
network of $50:50$ and polarising beam splitters, see Ref.~\cite{Sagi03}.

\subsection{The 4-photon double singlet state}

For completeness, we now ask for the output of the proposed state preparation 
scheme, given that the input state equals
\begin{equation}
|\phi_{\rm in} \rangle = a_{1 +}^{\dagger} a_{2 +}^{\dagger} a_{3 -}^{\dagger} 
a_{4 -}^{\dagger} |0 \rangle \, .
\end{equation}
Proceeding as above, we find that this results in the preparation of the state
\begin{equation}
|\phi_{\rm pro}^{\rm DS} \rangle
= \sum_{\sigma}U_{\sigma (1)1}U_{\sigma (2)2}U_{\sigma (3)3}U_{\sigma (4)4} 
b_{\sigma (1)+}^{\dagger}b_{\sigma (2) +}^{\dagger}b_{\sigma 
(3)-}^{\dagger}b_{\sigma (4)-}^{\dagger}\, |0 \rangle 
\end{equation} 
under the condition of the collection of one photon per output port. Here the 
permutation operators $\sigma$ are defined as in Section \ref{fun}, which yields
\begin{eqnarray} \label{DS}
|\hat{\phi}_{\rm pro}^{\rm DS}  \rangle 
&=& {\textstyle {1 \over 2}} \, \big[ \, b_{1 +}^{\dagger}b_{2 +}^{\dagger}b_{3 
-}^{\dagger}b_{4 -}^{\dagger}+b_{1 -}^{\dagger}b_{2 
-}^{\dagger}b_{3 +}^{\dagger}b_{4 +}^{\dagger} 
-b_{1 +}^{\dagger}b_{2 -}^{\dagger}b_{3 -}^{\dagger}b_{4 +}^{\dagger}-b_{1 
-}^{\dagger}b_{2 +}^{\dagger}b_{3 +}^{\dagger}b_{4 -}^{\dagger} \, ] \, |0 
\rangle \nonumber \\
&=& \frac{1}{\sqrt{2}}[ b_{1 +}^{\dagger}b_{3 -}^{\dagger}-b_{3 +}^{\dagger}b_{1 -}^{\dagger}] \otimes \frac{1}{\sqrt{2}}[ b_{2 
+}^{\dagger}b_{4 -}^{\dagger}-b_{2 +}^{\dagger}b_{4 -}^{\dagger}]\ket{0} \, .
\end{eqnarray} 
This state can be prepared with probability
\begin{equation}
P_{\rm suc} = {\textstyle {1 \over 16}} \, . 
\end{equation}
The state (\ref{DS}) is a double singlet state, i.e.~a tensor product of two 
2-photon singlet states, with a high robustness against decoherence 
\cite{Eibl03,Bourennane04}. In this 2 subsections, naive symmetry considerations may suggest  a state with equal 
superpositions of all permutations of the given input state as the output. We have seen here clearly that this is not the case.

\subsection{The general 4-photon case}

Finally, we consider the situation where the input state is of the general form 
(\ref{firework:in}). Calculating Eq.~(\ref{firework:output2}), we find that the unnormalised 
output state under the condition of one photon per output port equals in this 
case
\begin{eqnarray} \label{final}
|\phi_{\rm pro} \rangle &=&
{\textstyle {{\rm i} \over 4}} \big( \, \gamma_1+\gamma_2-\gamma_3-\gamma_4 \, 
\big) \, |\hat{\phi}_{\rm pro}^{\rm DS} \rangle 
+ {\textstyle {1 \over 2 \sqrt{2}}} \big( \, \gamma_6-\gamma_5\, \big) \, 
|\hat{\phi}_{\rm pro}^{\rm GHZ} \rangle \nonumber \\
&& + {\textstyle {1 \over 4}} \big( \, \gamma_8 +\gamma_{10}-\gamma_7-\gamma_9 
\, \big) \, |\hat{\phi}_{\rm pro}^{W} \rangle 
+ {\textstyle{1 \over 4}} \big( \, 
\gamma_{12}+\gamma_{14}-\gamma_{11}-\gamma_{13} \, \big) \, |\hat{\phi}_{\rm 
pro}^{W'} \rangle ~~~
\end{eqnarray} 
with the coefficients
\begin{eqnarray} \label{coefficients}
&\gamma_{1}=\alpha_{1+}\alpha_{2+}\alpha_{3-}\alpha_{4-} \, , ~
& \gamma_{2}=\alpha_{1-}\alpha_{2-}\alpha_{3+}\alpha_{4+}  \, ,\nonumber \\
& \gamma_{3}=\alpha_{1-}\alpha_{2+}\alpha_{3+}\alpha_{4-} \, , ~
& \gamma_{4}=\alpha_{1+}\alpha_{2-}\alpha_{3-}\alpha_{4+}  \, ,\nonumber \\
& \gamma_{5}=\alpha_{1+}\alpha_{2-}\alpha_{3+}\alpha_{4-} \, , ~
& \gamma_{6}=\alpha_{1-}\alpha_{2+}\alpha_{3-}\alpha_{4+}  \, , \nonumber \\
& \gamma_{7}=\alpha_{1+}\alpha_{2-}\alpha_{3-}\alpha_{4-} \, , ~
& \gamma_{8}=\alpha_{1-}\alpha_{2+}\alpha_{3-}\alpha_{4-}  \, , \nonumber \\
& \gamma_{9}=\alpha_{1-}\alpha_{2-}\alpha_{3+}\alpha_{4-} \, , ~
& \gamma_{10}=\alpha_{1-}\alpha_{2-}\alpha_{3-}\alpha_{4+}  \, , \nonumber \\
& \gamma_{11}=\alpha_{1-}\alpha_{2+}\alpha_{3+}\alpha_{4+} \, , ~
& \gamma_{12}=\alpha_{1+}\alpha_{2-}\alpha_{3+}\alpha_{4+}  \, , \nonumber \\
& \gamma_{13}=\alpha_{1+}\alpha_{2+}\alpha_{3-}\alpha_{4+} \, , ~
& \gamma_{14}=\alpha_{1+}\alpha_{2+}\alpha_{3+}\alpha_{4-} \, .~~~~~~
\end{eqnarray} 
The form of the coefficients (\ref{coefficients}) reflects the full symmetry of 
the transformation of the input state. Each of the entangled states 
$|\hat{\phi}_{\rm pro}^{\rm DS} \rangle$, $|\hat{\phi}_{\rm pro}^{\rm GHZ} 
\rangle$, $|\hat{\phi}_{\rm pro}^{W} \rangle$ and $|\hat{\phi}_{\rm pro}^{W'} 
\rangle$ are generated independently from the different constituent parts of the 
input (\ref{firework:in}). Besides, Eq.~(\ref{final}) shows that the output state is 
constrained to be of a certain symmetry, namely the symmetry introduced by the $N \times N$ Bell multiport and the postselection criteria of 
finding one photon per output port.

\section{The generation of $N$-photon {\em W}-states} \label{doubleW} 

Using the same arguments as in Section \ref{W4}, we now show that the $N \times 
N$ Bell multiport beam splitter can be used for the generation of {\em W}-states 
for arbitrary photon numbers $N$. Like Bell states, {\em W}-states are highly 
entangled but their entanglement is more robust \cite{Dur00}. Moreover, as $N$ 
increases, {\em W}-states perform better than the corresponding GHZ states 
against noise admixture in experiments to violate local realism \cite{Sen(De)03} 
and are important for optimal cloning protocols \cite{Buzek96}. 

In analogy to Eq.~(\ref{Win}), we assume that the initial state contains one 
photon in $|+ \rangle$ in the first input port, while every other input port is 
entered by a photon prepared in $|- \rangle$ so that
\begin{eqnarray} 
|\phi_{\rm in} \rangle &=& a_{1 +}^{\dagger} \prod_{i=2}^N a_{i -}^{\dagger} \, 
|0 \rangle \, . 
\end{eqnarray} 
Using Eq.~(\ref{firework:output2}), we find that the state of the system under the 
condition of the collection of one photon per output port equals
\begin{eqnarray} \label{output3} 
|\phi_{\rm pro} \rangle &=& \sum_{j=1}^N U_{j1} \, b_{j +}^{\dagger} \, 
\sum_{\sigma_j}  \Bigg[  
\prod_{i=2}^N U_{\sigma_j(i) i} \, b_{\sigma_j (i) -}^{\dagger} \Bigg] \, 
|0 \rangle \, , \nonumber \\&& 
\end{eqnarray} 
where the $\sigma_j$ are the $(N-1)!$ permutations that map the list $\{2, \, 3, 
\, ..., \, N\}$ onto the list $\{1,\, 2, \, ..., \, (j-1), \, (j+1), \, ..., \, 
N \}$. As expected, the output is a superposition of all states with one photon 
in $|+ \rangle$ and all other photons prepared in $|- \rangle$. 

To prove that Eq.~(\ref{output3}) describes indeed a {\em W}-state, we use again 
the notation introduced in Eqs.~(\ref{z}) and (\ref{pro3}) and write 
\begin{eqnarray} \label{pro4} 
|\phi_{\rm pro} \rangle & \equiv& \sum_j \beta_j  \, |j_{\rm out} \rangle \, .
\end{eqnarray} 
To show that the coefficients $\beta_j$ differ from $\beta_1$ at most by a phase 
factor, we express the amplitudes $\beta_j$ as in Eq.~(\ref{c}) using the 
permanents of the reduced transition matrices and find
\begin{equation} \label{ccc} 
\beta_j = U_{j 1} \sum_{\sigma_j} \prod_{i=2}^N U_{\sigma_j (i)i} = U_{j 1} \, 
{\rm perm} \, \left( U_{\rm red}^{\rm T}(j) \right) \, . 
\end{equation} 
Inserting the concrete form of the transition matrix $U$, this yields 
\begin{equation} 
\beta_j = \frac{1}{\sqrt{N^N}}\sum_{\sigma_j} \prod_{i=2}^N \omega_N^{(\sigma_j 
(i)-1)(i-1)} \, .~~ 
\end{equation} 
Proceeding as in Section \ref{W4}, we now multiply $\beta_j$ with the phase 
factor
\begin{equation} 
v_j \equiv \left( \prod_{k=0}^{N-1}  \omega_N^k \right)^{-(j-1)} 
\end{equation} 
and obtain
\begin{eqnarray} \label{generalproof} 
v_j \, \beta_j &=& {\textstyle {1 \over \sqrt{N^N}}} \sum_{\sigma_j} 
\prod_{i=2}^N \omega_N^{(\sigma_j (i)-j)(i-1)} \nonumber \\
&=&  {\textstyle {1 \over \sqrt{N^N}}} \sum_{\sigma_j} \prod_{i=2}^N 
\omega_N^{\big({\rm mod}_N (\sigma_j (i)-j) \big)(i-1)} \, .~~
\end{eqnarray} 
The expression ${\rm mod}_N(\sigma_j (i)-j)+1$ represents a set of $(N-1)!$ 
permutations that map $\{2, \, 3, \, ..., \, N\}$ onto the list $\{2, \, 3, \, 
..., \, N\}$. It is therefore equivalent to the permutations $\sigma_1(i)$, 
which allows us to simplify Eq.~(\ref{generalproof}) even further and to show 
that
\begin{equation} \label{close} 
v_j \, \beta_j = {\textstyle {1 \over \sqrt{N^N}}} 
\sum_{\sigma_1} \prod_{i=2}^N \omega_N^{(\sigma_1 (i)-1)(i-1)} = \beta_1 \, .~~ 
\end{equation} 
From this and the fact that $1+2+...+(N-1)={1 \over 2} N(N-1)$ we finally arrive 
at the relation 
\begin{eqnarray} 
\beta_j &=& \left( \prod_{k=0}^{N-1} \omega_N^k \right)^{j-1} \beta_1 \nonumber 
\\ 
&=& \left\{ \begin{array}{cl} \beta_1 \, , & {\rm if}~N~{\rm is~odd} \, , \\  
(-1)^{j-1} \, \beta_1 \, , & {\rm if}~N~{\rm is~even} \, . \end{array} \right. 
\end{eqnarray} 
This shows that the amplitudes $\beta_j$ are all of the same size and the Bell 
multiport can indeed generate $N$-photon {\em W}-states. If one wants the 
coefficients $\beta_j$ to be exactly the same, one can remove unwanted minus 
signs in case of even photon numbers by applying a $\sigma_z$ operation in each 
output port with an even number $j$. 

The logic of the described proof exploits the symmetry of a Bell multiport and 
avoids calculating the coefficients of the constituent states of the output 
photon. Indeed, there exist no known efficient method \cite{Scheel04,Minc78} to 
calculate these coefficients in general. 

In the case $N=2$, the above described state preparation scheme reduces to the 
familiar example, where two photons prepared in the two orthogonal states $|+ 
\rangle$ and $|- \rangle$ pass through a 50:50 beam splitter. The collection of 
one photon in the each output port prepares the system in this case in the state 
${1 \over \sqrt{2}} \, [ \, b_{1+}^\dagger b_{2-}^\dagger - b_{1-}^\dagger 
b_{2+}^\dagger \, ] \, |0 \rangle$, which can be transformed into ${1 \over 
\sqrt{2}} \, [ \, b_{1+}^\dagger b_{2-}^\dagger + b_{1-}^\dagger b_{2+}^\dagger 
\, ] \, |0 \rangle$ by flipping the sign of the state, i.e.~depending on 
whether the photon is in $|+ \rangle$ or $|-\rangle$, in one of the output 
ports. 

\subsection{Success probabilities}

\begin{figure}
\begin{minipage}{\columnwidth}
\begin{center}
\resizebox{\columnwidth}{!}{\rotatebox{0}{\includegraphics{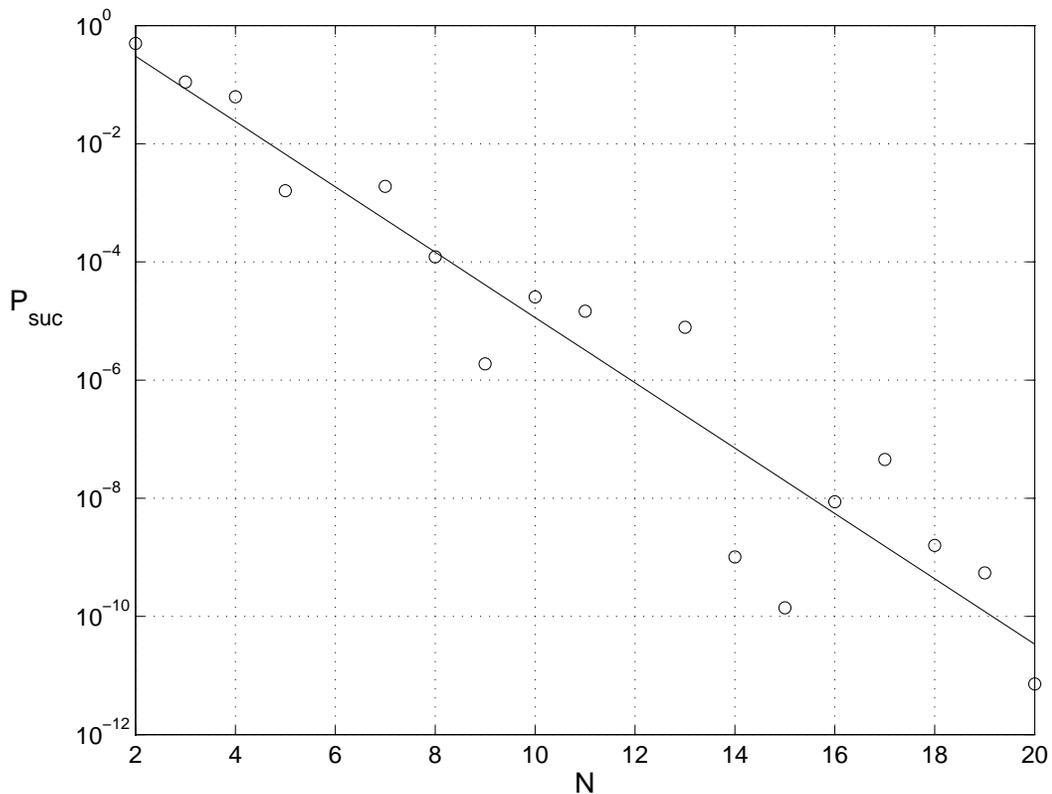}}}  
\end{center}
\caption{The success rate for the generation of $N$-photon {\em W}-states 
$P_{\rm suc}$ as a function of $N$. The solid line approximates the exact 
results via the equation $P_{\rm suc}={\rm e}^{a-bN}$ with $a=1.35 \pm 1.32$ and 
$b=1.27 \pm 0.10$} \label{fireplot}
\end{minipage}
\end{figure}

Let us finally comment on the success rate of the proposed {\em W}-state 
preparation scheme. Computing the probability (\ref{firework:suc}) can be done by finding 
the amplitude $\beta_1$ with the help of Eq.~(\ref{ccc}). Although the 
definition of the permanent of a matrix resembles the definition of the 
determinant, there exist only few theorems that can be used to simplify their 
calculation \cite{Horn85,Scheel04,Minc78}. In fact, the computation of the permanent is an 
NP-complete problem compared to that of a determinant which is only of complexity P. We therefore calculated $P_{\rm suc}$ numerically 
(see 
Fig.~\ref{fireplot}). 

As it applies to linear optics schemes in general, the success probability 
decreases unfavourably as the number of qubits involved increases. Here the 
probability of success drops on average exponentially with $N$. We observe the 
interesting effect of a non-monotonic decreasing success probability as $N$ 
increases. For example, the 
probabilty of success for $N=13$ is higher than for $N=9$. Moreover, for $N=6$ 
and $N=12$, {\em W}-state generation is not permitted due to destructive 
interference. This may lead one to speculate that this is the case for all multiples of $N=6$. Unfortunately, this does not apply to 
$N=18$ and precludes an easy explanation of this effect.

\section{Conclusions} \label{firework:conclusions}

We analysed the generation of multiphoton entanglement with the help of 
interference and postselection in a linear optics network consisting of an $N 
\times N$ Bell multiport beam splitter. Each input port should be entered by a 
single photon prepared in a certain state $|\lambda_i \rangle$. As long as the 
photons are the same with respect to all other degrees of freedom and it can be 
guaranteed that photons prepared in the same state overlap within their 
coherence time inside the linear optics network, the described scheme can be 
implemented using only a single photon source  
\cite{Kuhn02,Mckeever04,Lange04,Benson00,Pelton02,Kurtsiefer00,Beveratos02}. We 
believe that the described approach allows one to entangle much higher photon 
numbers than what can be achieved in parametric down conversion experiments. 

In general, a highly entangled output state is obtained under the condition of 
the collection of one photon per output port. The motivation for this 
postselection criteria is that distinguishing this state from other output 
states does not require photon number resolving detectors, and can also accommodate lossy photon production. Ideally, the detectors should 
have negligible dark counts which is possible with current technology \cite{Rosenberg05}. For 
simplicity of discussion, we would take this to be the assumption in the rest of the thesis.  Moreover, the photons 
can easily be processed further and provide a resource for linear optics quantum 
computing and quantum cryptographic protocols. 

Firstly, we analysed the case $N=4$ and showed that the $4 \times 4$ Bell multiport 
allows for the creation of a variety of highly-symmetric entangled states 
including the {\em W}-state, the GHZ-state and double singlet states. It was 
found that some states are easier to prepare than others. A straightforward 
generalisation of the 4-photon case yields a scheme for the creation of 
$N$-photon {\em W}-states. We calculated the rates for successful state 
preparations and showed that they decrease in a non-monotonic fashion and on 
average exponentially with $N$.

The motivation for considering a Bell multiport beam splitter was that it only 
redirects the photons without affecting their inner degrees of freedom. The 
proposed setup can therefore be used to produce polarisation, time-bin and 
frequency entanglement, respectively. To generate, for example, polarisation 
entangled photons, the initial photon states may differ in polarisation but 
should otherwise be exactly the same. The high symmetry of the Bell multiport 
beam splitter allows for the generation of a variety of highly entangled 
symmetric states.  Furthermore, except for interferometric stability being required for the multiport, the scheme is highly robust to 
slow external and unknown phase fluctutation as this contributes to only a trivial global phase in the scheme.

The results in this chapter need not be limited only to postselected photon entanglement generation. As a foretaste, we will highlight an even more important application based on this chapter in  
Chapter \ref{demand}. We continue our study on multiports in 
the next chapter with the aim of studying multiparticle interference, which is the crucial underlying mechanism in much of this thesis.

\chapter{Generalised Hong-Ou-Mandel Effect for \\ Bosons and Fermions} 
\label{fusion}

\section{Introduction}

The 2-photon Hong-Ou-Mandel (HOM) dip has been demonstrated first in 1987 
\cite{Hong87}. In their experiment,  Hong, Ou and Mandel sent two identical 
photons  through the separate input ports of a $50:50$ beam 
splitter. Each output port contained a photon detector.  No 
coincidence detections within the temporal coherence length of the photons, 
i.e.~no simultaneous clicks in both detectors, were recorded when there is no 
relative delay of the input photons\footnote{The term ``HOM 
dip" refers to the ``dip" of the coincidence counts in both detectors under zero 
relative time delay of the input photons or photon 
detection.}. Crucial for the 
observation of this effect was  the indistinguishability of the pure quantum 
states of 
the input photons, which differed only in the directions of their wave vectors. 
This allowed the photons  to interfere within the 
setup. The detectors could not resolve the origin of each observed photon.

Due to the nature of this experiment, the HOM dip was soon employed for quantum 
mechanical tests of local realism  and for the generation of 
postselected entanglement between two photons \cite{Shih88}. Linear optics Bell 
measurements on photon pairs rely intrinsically on the HOM dip 
\cite{Braunstein95,Mattle96}, which has also been a building block for the 
implementation of linear optics gates for quantum information processing with 
photonic qubits \cite{Knill01}. Shor's factorisation algorithm \cite{Shor94}, 
for 
example, relies on multiple path interference to achieve massive parallelism 
\cite{Ou99} and multiphoton interference has to play a crucial role in any 
implementation of this algorithm using linear optics.  

Since it requires temporal and spatial mode-matched photons, observing the HOM 
dip for two photons is also a good test of their indistinguishability. HOM 
interference has been applied to characterise recently introduced sources for 
the 
generation of single photons on demand by testing the identicalness of 
successively generated photons \cite{Fattal04,Legero04,Kiraz05}. Another 
interesting test based on the HOM dip has been studied by Bose and Home, who 
showed that it can reveal whether the statistics of two identical particles 
passing through a $50: 50$ beam splitter is fermionic or bosonic \cite{Bose02}. 

Motivated by the variety of possible applications of the 2-photon HOM dip, 
this 
chapter investigates generalised HOM experiments. We consider a straightforward 
generalisation of the scattering of two particles through a $50:50$ beam 
splitter, namely the scattering of $N$ particles through a symmetric $N \times 
N$ 
Bell multiport beam splitter. While numerous studies on $N$ photon interference 
in the {\em constructive} sense, i.e.~resulting in the enhancement of a certain 
photon detection syndrome, have been made (see e.g.~Refs.~\cite{Ou99}), not much 
attention has been paid to multiple path interference in the {\em destructive} 
sense. Mattle {\em et al.} \cite{Mattle95} has studied both constructive and 
destructive detection syndromes for two photons scattering 
through an $N \times N$ Bell multiport. Recently, Walborn {\em et al.} studied 
so-called multimode HOM effects for 
photon pairs with several inner degrees of freedom, including the spatial and 
the 
polarisation degrees of freedom \cite{Walborn03}. A notable example for 
destructive HOM interference has been given by Campos \cite{Campos00}, who 
studied certain triple coincidences in the output ports of an asymmetric $3 
\times 3$ multiport beam splitter, which is also known as a tritter.

We consider {\em bosons} as well as the simultaneous scattering of {\em 
fermions}. The difference between both classes of particles is most elegantly 
summarised in the following commutation rules. While the annihilation and 
creation operators $a_i$ and $a_i^\dagger$ for a boson in mode $i$ obey the 
relation 
\begin{equation} \label{boson}
[a_i,a_j^\dagger] \equiv a_i a_j^\dagger - a_j^\dagger a_i = \delta_{ij} ~~~ 
{\rm 
and} ~~~ [a_i^\dagger,a_j^\dagger]=[a_i,a_j]=0 ~~ \forall ~ i, \, j 
\end{equation}
with $\delta_{ij}=0$ for $i \neq j$ and $\delta_{ii} = 1$, the annihilation and 
creation operators $a_i$ and $a_i^\dagger$ of fermionic particles obey the 
anticommutation relation
\begin{equation} \label{fermion}
\{a_i,a_j^\dagger\} \equiv a_i a_j^\dagger + a_j^\dagger a_i =\delta_{ij} ~~~ 
{\rm and} ~~~ \{a_i^\dagger,a_j^\dagger\}=\{a_i,a_j\}=0 ~~ \forall ~ i, \, j \, 
.
\end{equation} 
Here $i$ and $j$ refer to the inner degrees of freedom of the particles, like 
their respective path, polarisation, spin, frequency or energy.

Multiport beam splitters exist in general for a wide variety of fermionic 
and bosonic particles. Possible realisations of a  photonic multiport have been 
discussed in Chapter \ref{firework}.  For example, 
multiports for bosonic or fermionic atoms can 
consist of a network of electrode wave guide beam splitters on an atom chip 
\cite{Cassettari00}. Multiports for electrons, which behave  like fermions, can be 
realised by fabricating a network of quantum point contacts acting as 
2-electron beam splitters \cite{Samuelsson04}. Specially doped optical fibres 
have recently been introduced in the literature and are expected to constitute 
beam splitters for ``fermion-like" photons \cite{Franson04}.

As in the original HOM experiment \cite{Hong87}, we assume in the following 
that 
a particle detector is placed in each output port of  the scattering beam 
splitter array. The incoming particles should enter the different input ports 
more or less simultaneously and in such a way that there is one particle per input port. 
Moreover, we assume that the particles are identical. We will show that it is 
impossible to observe a particle in each output port for even numbers $N$ of 
bosons. We denote this effect of zero coincidence detection as the {\em 
generalised HOM dip}. We will also show that fermions always leave the setup 
separately exhibiting perfect coincidence detection. Since the interference 
behaviour of both types of particles is very different, the Bell multiport can 
be 
used to reveal their quantum statistics. 

This chapter is organised as follows. In Section \ref{scatterfusion}, we introduce the 
theoretical description of particle scattering through a symmetric Bell 
multiport. Section \ref{twofusion} describes the scattering of two particles through a 
$50 : 50$ beam splitter as an example. In Section \ref{many}, we derive the 
condition for the generalised HOM dip for bosons and analyse the scattering of 
fermions through the same setup for comparison. Finally we conclude our results 
in Section \ref{conclusions}.

\section{Scattering through a Bell multiport beam splitter} \label{scatterfusion}

The description of particle scattering through a multiport is essentially the 
same as the previous Chapter \ref{firework}. Suppose each 
input port $i$ is entered by a 
particle with creation operator $a_i^\dagger$. Then the input state of the 
system 
equals 
\begin{eqnarray} \label{infusion}
|\phi_{\rm in} \rangle &=& \prod_{i=1}^N  a_i^\dagger  \, |0 \rangle \, ,
\end{eqnarray}
where $|0 \rangle$ is the vacuum state with no particles in the setup.

If $b_j^\dagger$ denotes the creation operator for a single particle in output 
port $j$,  similar to Chapter \ref{firework}, we obtain for the output state of 
the photons given the input state (\ref{infusion}),
\begin{eqnarray} \label{output1}
|\phi_{\rm out} \rangle &=& \prod_{i=1}^N \, \Bigg( \, \sum_{j=1}^N \, U_{ji} \, 
b_j^\dagger \, \Bigg) \, |0 \rangle \, . 
\end{eqnarray} Again, $U_{ji}$ denotes the matrix element representing the 
transition amplitude of the $i$th input port to the $j$th output port of the matrix 
$U$ defining the multiport. Specially for a Bell multiport, $U$ is a discrete 
fourier transform matrix defined in Chapter \ref{firework}.
Note that up to now, we 
have not invoked any assumptions about the nature of the particles. The 
formalism 
in this section applies to bosons and fermions equally.

\section{HOM interference of two particles} \label{twofusion}

\begin{figure}
\begin{minipage}{\columnwidth}
\begin{center}
\vspace*{0cm}
\resizebox{\columnwidth}{!}{\rotatebox{0}{\includegraphics{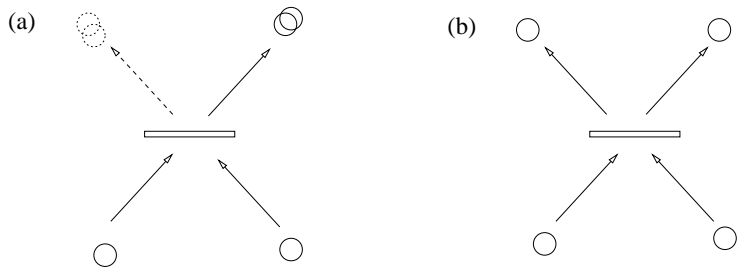}}}  
\end{center}
\vspace*{-1.5cm}
\caption{(a) HOM dip for two bosons scattering through a $50:50$ beam splitter. 
(b) Perfect coincidence in the output ports for fermion scattering.} \label{hom}
\end{minipage}
\end{figure}

Before analysing the general case, we motivate our discussion by considering two 
identical particles entering  the different input ports of a $50 : 50$ beam 
splitter. For $N=2$, the transition matrix (\ref{fourier}) becomes the Hadamard 
matrix\footnote{We remind the reader that the transition matrix chosen here is not unique. It represents rather an equivalence class of $50:50$ beam splitters with can be  transformed to each other via phase shifts. Our discussion on the bunching or antibunching of particles apply to this equivalence class.}
\begin{eqnarray}
U &=& {\textstyle {1 \over \sqrt{2}}} \left( \begin{array}{rr} 1 & 1 \\ 1 & -1 
\end{array} \right)  
\end{eqnarray}
and the input state (\ref{in}) becomes $|\phi_{\rm in} \rangle=  a_1^\dagger 
a_2^\dagger \, |0 \rangle $. Note that local measurements on this input state 
cannot reveal any information about the bosonic or fermionic nature of the two 
particles. However, using Eq.~(\ref{output1}), we find that the beam splitter 
prepares the system in the state 
\begin{equation} \label{n2}
|\phi_{\rm out} \rangle = {\textstyle {1 \over 2}} \, 
\big(b_1^\dagger+b_2^\dagger \big) \big(b_1^\dagger-b_2^\dagger \big) \, |0 
\rangle = {\textstyle {1 \over 2}} \left[ (b_1^\dagger)^2 - b_1^\dagger 
b_2^\dagger + b_2^\dagger b_1^\dagger - (b_2^\dagger)^2 \right] |0 \rangle \, .
\end{equation}
This state no longer contains any information about the origin of the particles, 
since each incoming one is equally likely                        
transferred to any of the two output ports. Passing through the setup, the input 
particles become indistinguishable by detection (see Fig.~ \ref{hom}). Their 
quantum statistics can now be revealed using local measurements.

Bosons obey the commutation law (\ref{boson}). Using this, the output state 
(\ref{n2}) becomes
\begin{equation} 
|\phi_{\rm out} \rangle =  {\textstyle {1 \over 2}} \left[ (b_1^{\dagger})^2 - 
(b_2^{\dagger})^2 \right] |0 \rangle \, ,
\end{equation} 
which implies a zero-coincidence count rate at the output ports. The particles 
bunch together in the same output port and exhibit the famous HOM dip (see Fig.~ 
\ref{hom}(a)). In contrast, fermions obey the anticommutation relation 
(\ref{fermion}) and their output state
\begin{equation} 
|\phi_{\rm out} \rangle = b_1^\dagger b_2^\dagger \, |0 \rangle
\end{equation}  
implies perfect particle coincidence. This means that the fermions always arrive 
in separate output ports and never bunch together (see Fig.~\ref{hom}(b)). A 
$50 : 50$ beam splitter can therefore  be used to distinguish bosons and 
fermions indeed \cite{Bose02}. 

\section{Multiparticle HOM interference} \label{many}

We now consider the general case of $N$ particles passing through an $N \times 
N$ 
Bell multiport beam splitter. As in the $N=2$ case, the setup redirects each 
incoming particle with equal probability to any of the possible output ports, 
thereby erasing the information about the origin of each particle and making 
them 
indistinguishable by detection. For even numbers of bosons, this results in the 
generalised HOM dip and zero coincidence detection. In contrast, fermions leave 
the setup always separately, thus demonstrating maximum coincidence detection. 
Observing this extreme behaviour can be used, for example, to verify the quantum 
statistics of {\em many} particles experimentally. 

\subsection{Bosonic particles} \label{bos}

In order to derive the necessary condition for the appearance of the generalised 
HOM dip for even numbers of bosons, we calculate the output state 
(\ref{output1}) 
of the system under the condition of the collection of one particle per output 
port. Each term contributing to the projected conditional output state 
$|\phi_{\rm pro} \rangle$ can be characterised by a certain permutation, which 
maps the particles in the input ports $1, \, 2, \, ..., \, N$ to the output 
ports 
$1, \, 2, \, ..., \, N$. In the following, we denote any of the $N!$ 
permutations 
by $\sigma$ with $\sigma (i)$ being the $i$-th element of the list obtained when 
applying the permutation $\sigma$ onto the list $\{1,\, 2, \, ..., \, N\}$. 
Using 
this notation, $|\phi_{\rm pro} \rangle$ equals up to normalisation 
\begin{equation} \label{pro}
|\phi_{\rm pro} \rangle = \sum_{\sigma} \Bigg[ \prod_{i=1}^N U_{\sigma (i)i} \, 
b_{\sigma(i)}^{\dagger} \Bigg] \, |0 \rangle \, .
\end{equation} 
The norm of this state has been chosen such that 
\begin{equation} \label{suc}
P_{\rm coinc} = \| \, |\phi_{\rm pro} \rangle \, \|^2  
\end{equation}
is the probability to detect one particle per output port. It is therefore also 
the probability for observing coincidence counts in all $N$ detectors as in Chapter \ref{firework}.

Up to now, the nature of the particles has not yet been taken into account. 
Using 
the commutation relation (\ref{boson}) for bosons, the conditional output state 
(\ref{pro}) becomes  
\begin{equation} \label{properm}
|\phi_{\rm pro} \rangle =  {\rm perm} \, U \cdot \prod_{i=1}^N b_i^{\dagger} \, 
\ket{0} 
\end{equation} 
with the permanent of the square matrix $U$ defined as 
\cite{Scheel04,Horn85,Minc78}
\begin{equation} \label{perm}
{\rm perm} \, U \equiv {\rm perm} \, U^T \equiv \sum_{\sigma} \prod_{i=1}^N 
U_{\sigma (i) \, i} \, .
\end{equation} 
The permanent of a matrix is superficially similar to the determinant. However, 
there exist hardly any mathematical theorems that can simplify the calculation 
of 
the permanent of an arbitrary matrix. 
 
To derive a condition for the impossibility of coincidence detections, we have 
to 
see when the probability (\ref{suc}) equals zero. Using Eq.~(\ref{properm}), we 
find 
\begin{equation} \label{almost}
P_{\rm coinc} = | \, {\rm perm} \, U \, |^2 \, .
\end{equation} 
The key to the following proof is to show that the transition matrix $U$ of the 
Bell multiport possesses a certain symmetry such that its permanent vanishes in 
certain cases. Suppose the matrix $U$ is multiplied by a diagonal matrix 
$\Lambda$ with matrix elements 
\begin{equation}
\Lambda_{jk}  \equiv \omega_N^{j-1} \, \delta_{jk} \, .
\end{equation} 
This generates a matrix $\Lambda U$ with 
\begin{eqnarray} \label{ma}
(\Lambda U)_{ji} = \sum_{k=1}^N \Lambda_{jk}U_{ki} = \Lambda_{jj}U_{ji} 
=  {\textstyle{1 \over \sqrt{N}}} \, \omega_N^{(j-1)i} \, .
\end{eqnarray} 
We now introduce the modulus function defined as ${\rm mod}_N (x)=j$, if $x-j$ 
is 
dividable by $N$ and $0\leq j<N$. Since  $\omega_N^ N = \omega_N^0 =1$, the 
matrix 
elements (\ref{ma}) can be expressed as
\begin{eqnarray}
(\Lambda U)_{ji}  =  {\textstyle{1 \over \sqrt{N}}} \, \omega_N^{(j-1)({\rm 
mod}_N (i)+1-1)} \, .
\end{eqnarray}
Note that the function $\tilde \sigma(i)={\rm mod}_N (i)+1$ maps each element of  
the list $\{1,2,...N-1,N \}$ respectively to the list $\{2,3,...N,1\}$. A 
comparison with Eq.~(\ref{fourier}) therefore shows that 
\begin{equation}
(\Lambda U)_{ji} = U_{j \, \tilde \sigma(i)} \, .
\end{equation} 
In other words, the multiplication with $\Lambda$ amounts to nothing more than a 
cyclic permutation of the columns of the matrix $U$. Taking the cyclic 
permutation symmetry of the permanent of a matrix (see definition (\ref{perm})) 
into account, we obtain  
\begin{equation} \label{proof1}
{\rm perm} \, U ={\rm perm} \, (\Lambda U) \, .
\end{equation} 
However, we also have the relation
\begin{equation} \label{proof2}
{\rm perm} \, (\Lambda U) ={\rm perm} \, \Lambda \cdot {\rm perm} \, U 
\end{equation} 
with the permanent of the diagonal matrix $\Lambda$ given by
\begin{eqnarray} \label{relative} 
{\rm perm} \, \Lambda =  \prod_{k=1}^N \omega_N^{k-1} = \omega_N^{\sum_{k=1}^N 
k} 
= \omega_N^{N(N+1)/2} = {\rm e}^{ {\rm i} \pi (N+1)} = \left\{ \begin{array}{rl} 
1 \, , & {\rm if}~N~{\rm is~odd} \, , \\  -1  \, , & {\rm if}~N~{\rm is~even} \, 
. \end{array} \right.
\end{eqnarray}  
For $N$ being even, a comparison of Eqs.~(\ref{proof1}) - (\ref{relative}) 
reveals that 
\begin{equation} \label{last}
{\rm perm} \, U = - {\rm perm} \, U  = 0 \, . 
\end{equation} 
As a consequence, Eq.~(\ref{almost}) implies that $P_{\rm coinc} =0$. 
Coincidence 
detection in all output ports of the setup is impossible for even numbers of 
bosons. This is not necessarily so, if the number of particles is odd. For 
example, for $N=3$ one can check that there is no HOM dip by calculating ${\rm 
perm} \, U$ explicitly. Campos showed that observing a HOM dip for $N=3$ is 
nevertheless possible with the help of a specially designed asymmetric multiport 
beam splitter \cite{Campos00}. 

Furthermore, even if the number of particles is even, the HOM dip does not 
appear to hold for all symmetric multiports. For example, it 
is known that all symmetric $4 \times 4$ multiport can be represented generally 
by the transition matrix $U$ given in the form as \cite{Zukowski97}
\begin{eqnarray} \label{fiber}
U &=& {\textstyle {1 \over 2}} \left( \begin{array}{rrrr} 1 & 1 & 1 & 1 \\ 1 & 
{\rm e}^{{\rm i}\phi} & -1 & -{\rm e}^{{\rm i}\phi} \\ 1 & 
-1 & 1 & -1 \\ 1 & -{\rm e}^{{\rm i}\phi} & -1 & {\rm e}^{{\rm i}\phi}  
\end{array} \right) \, ,  
\end{eqnarray} where each choice of $\phi$ in the range between $0$ and $\pi$ 
parameterize an equivalence class. Note that the Bell 
multiport coincides with the choice of $\phi=\frac{\pi}{2}$. As before, one can 
compute the probability of coincidence detection and it 
is given by
\begin{equation}
P_{\rm coinc} =\frac{1}{8}(1+\cos (2 \phi)) \, .
\end{equation} This suggest that by performing a HOM experiment on coincidence 
detection, one can characterise an  unknown symmetric $4 
\times 4$ multiport. This may find new application in symmetric multiports made 
by fiber splicing \cite{Pryde03}\footnote{For example, 
it was communicated to me by Geoff Pryde that the phase factor $\phi$ in 
Eq.~(\ref{fiber}) is not a parameter easily controllable in 
fiber splicing. }. In the case of a Bell multiport, one 
recovers the HOM dip.

\subsection{Fermionic particles}

Fermions scattering through a Bell multiport  show another extreme 
behaviour. Independent of the number $N$ of particles involved, they always 
leave 
the setup via different output ports, thereby guaranteeing perfect coincidence 
detection. As expected, particles obeying the quantum statistics of fermions 
cannot populate the same mode. 

Again, we assume that each input port is simultaneously entered by one particle 
and denote the creation operator of a fermion in output port $i$ by 
$b_i^\dagger$. Proceeding as in Section \ref{bos}, one finds again that the 
output state of the system under the condition of the collection of one particle 
per output port is given by Eq.~(\ref{pro}). To simplify this equation, we now 
introduce the sign function of a permutation with $\rm{sgn}(\sigma) = \pm 1$, 
depending on whether the permutation $\sigma$ is even or odd. An even (odd) 
permutation is one, that can be decomposed into an even (odd) number of 
interchanges. Using this notation and taking the anticommutation relation for 
fermions (\ref{fermion}) into account, we find 
\begin{equation}
|\phi_{\rm pro} \rangle = \sum_{\sigma}{\rm sgn}(\sigma) \Bigg( \prod_{i=1}^N 
U_{\sigma (i) \, i} \, b_i^{\dagger} \Bigg) \, |0 \rangle \, .
\end{equation} 
A closer look at this equation shows that the amplitude of this state relates to 
the determinant of the transformation matrix given by
\begin{equation}
{\rm det} \, U =\sum_{\sigma}{\rm sgn}(\sigma)  \prod_{i=1}^N U_{\sigma (i) \, 
i} 
\, .
\end{equation} 
Since $U$ is unitary, one has $|{\rm det} \, U| =1$ and therefore also, as 
Eq.~(\ref{suc}) shows, 
\begin{equation}
P_{\rm coinc} =  | \, {\rm det} \, U \, |^2 = 1 \, .
\end{equation} 
This means that fermions  leave the system separately indeed, i.e.~with one 
particle per output port. In the above, we only used the unitarity of the 
transition matrix $U$ but not its concrete form.  Perfect coincidence detection 
therefore applies to any situation where fermions pass through an $N \times N$ 
multiport, i.e.~independent of its realisation.

\section{Conclusions} \label{conclusions}

We analysed a situation, where $N$ particles enter the $N$ different input ports 
of a symmetric Bell multiport beam splitter simultaneously. If these particles 
obey fermionic quantum statistics, they always leave the setup independently 
with one particle per output port. This results in perfect coincidence 
detection, 
if detectors are placed in the output ports of the setup. In contrast to this, 
even numbers $N$ of bosons have been shown to never leave the setup with one 
particle per output port. This constitutes a generalisation of the 2-photon 
HOM 
dip to the case of arbitrary even numbers $N$ of bosons. The generalised HOM dip 
is in general not observable when $N$ is odd. 

The proof exploits the cyclic symmetry of the setup. We related the coincidence 
detection in the output ports to the permanent or the determinant of the 
transition matrix $U$ describing the multiport, depending on the bosonic or 
fermionic nature of the scattered particles. The NP complexity of computing the 
permanent compared to the determinant has been discussed 
in Chapter~\ref{firework}.
Experimental setups involving the scattering of bosons through a multiport 
therefore have important applications in quantum information processing.

For example, part of the linear optics quantum computing scheme by Knill, 
Laflamme and Milburn \cite{Knill01} is based on photon scattering through a Bell 
multiport beam splitter. In contrast to this, the scattering of non-interacting 
fermions through the same corresponding circuit, can be efficiently simulated on 
a classical computer \cite{Terhal02,Knill01a}. Moreover, the quantum statistics 
of particles has been used for a variety of quantum information processing tasks 
such as entanglement concentration \cite{Paunkovic02} and entanglement transfer 
\cite{Omar02}. Completely new perspectives might open when using setups that can 
change the quantum behaviour of particles and convert, for example, photons into 
fermions \cite{Franson04}.

Finally, we remark that observing HOM interference of many particles is 
experimentally very robust. Our results can therefore also be used to verify the 
quantum statistics of particles experimentally as well as to characterise or 
align an experimental setup. Testing the predicted results does not require 
phase 
stability in the input or output ports nor detectors with maximum efficiency. 
The 
reason is that any phase factor that a particle accumulates in any of the input 
or output ports contributes at most to an overall phase factor of the output 
state $|\phi_{\rm out} \rangle$. However, the coincidence statistics are 
sensitive 
to the phase factors accumulated inside the multiport beam splitter as they 
affect the form of the transition matrix $U$. 

In the next chapter, we propose a scheme for an  entanglement assisted 
photon manipulation. The required entangled photon ancillas 
can be either generated on demand (see Chapter \ref{demand}) or postselectively 
(see Chapter \ref{firework}).

\chapter{An Efficient Quantum Filter for Multiphoton States} \label{hummingbird}

\section{Introduction}

Much effort has been made to find efficient schemes for the realisation of useful operations between photons contributing to quantum information processing. For example, we have discussed the process of entangling photons in Chapter \ref{firework}. In this chapter, we discuss a very useful operation, namely  the {\em parity} or {\em 
quantum filter} \cite{Pan98a,Franson01,Hofmann02,Grudka02,Zou02a}. The application of parity filters is diverse, ranging from quantum non demolition 
measurements of entanglement to the generation of multiphoton quantum codes \cite{Hofmann02} and the generation of multipartite 
entanglement \cite{Zou02a}. Moreover, it has been shown that the parity filter can constitute a crucial component for the generation of 
cluster states for one-way quantum computing \cite{Verstraete04,Browne05}. Furthermore, Nemoto and Munro\cite{Nemoto04} applied the parity filter based on weak nonlinearity to achieve nearly deterministic linear optics quantum computing. Together with single qubit rotations and measurements, the 
parity 
filter constitutes a universal set of gate operations \cite{Browne05}. 

Applied to two photons, the parity filter projects their state onto the 2-dimensional subspace of states where the photons have 
identical polarisation in the $\ket{H}$ and $\ket{V}$ basis\footnote{This is also known as the states of even parity}. We denote the corresponding operator as $P_2$ and define 
\begin{equation} \label{par}
P_2 = \sqrt{p_2} \, \Big( \, \ket{HH}\bra{HH}+\ket{VV}\bra{VV} \, \Big) ~,~
\end{equation} 
where $H$ and $V$ describe a horizontally and a vertically polarised photon, respectively. Besides, $p_2$ is the success probability for 
the performance of the parity projection on an arbitrary input state. This means, even when applied to a parity eigenstate, the photons 
only pass through the filter with probability $p_2$. Here, the term {\em success probability} denotes the projection efficiency 
of a given setup.

In the original proposal of a linear optics implementation of the 2-photon parity filter \cite{Hofmann02}, Hofmann and Takeuchi obtained a success 
probability of $p_2={1 \over 16}$ after passing the photons through several beam splitters and performing postselective measurements. 
Two other proposals yield a higher success probability of $p_2 = {1 \over 4}$ \cite{Grudka02,Zou02a}. Grudka and Wojcik achieve this by 
using 
the idea of teleportation \cite{Knill01} and by employing ancilla states containing six photons. Zou and Pahlke use a 
single 
mode quantum filter that separates the 1-photon state from the vacuum and the 2-photon state. By combining two such single mode filters, 
a parity filter can be realised that requires a 4-photon ancilla state as a resource \cite{Zou02a}. 

In direct analogy to the 2-photon parity filter (\ref{par}), a quantum filter for $N$ photons can be defined
by the operator
\begin{equation} \label{PN}
P_N=\sqrt{p_N} \, \Big( \, \ket{HH \, . \, . \, . \, H}\bra{HH \, . \, . \, . \, H}+\ket{VV \, . \, . \, . \, V}\bra{VV \, . \, . \, .\, 
V} \, \Big) ~.~
\end{equation} 
Applied to an arbitrary input state with $N$ photons, this filter projects the system with probability $p_N$ onto the 2-dimensional 
subspace where all photons have the same polarisation in the $\ket{H}$ and $\ket{V}$ basis. One way to implement this gate is to pass 
the input state through $(N-1)$ 2-photon parity filters, which succeeds with overall probability $p_N= p_2^{N-1}$. This approach 
presents a steep challenge for large photon number $N$, given the above mentioned success probabilities of a single 2-photon parity 
check.

In this chapter, we describe a potential implementation of the $N$-photon quantum filter (\ref{PN}) with a success rate as high as $p_N 
= {1\over 2}$, which is much more effective than performing operation (\ref{PN}) with the previously proposed 2-photon parity filters. 
As a resource we require the presence of the $N$-photon GHZ-state
\begin{equation} \label{GHZ}
|A^{(N)} \rangle = {\textstyle {1 \over \sqrt{2}}} \, \Big( \, \ket{HH \, . \, . \, . \, H} + \ket{VV \, . \, . \, . \, V} \, \Big) ~.~ 
\end{equation}
In principle, this photon state can be prepared on demand \cite{Gheri98,Lange00,LimSPIE04}. Furthermore, we require a photon-number 
resolving 
detector that can distinguish between 0, 1 and 2 photons. To implement the quantum filter (\ref{PN}), we use ideas that have been 
inspired by a recently performed entanglement purification protocol \cite{Pan03}. Indeed, the same setup can be reconfigured and interpreted  as a 2-photon parity filter. It should also be emphasised, with some changes in the definition of the photon basis in our setup, our quantum filter also maps to the CNOT gate proposed by Pittman {\em et al.} \cite{Pittman01} with a $\frac{1}{4}$ probability of success. 

\section{A multipartite quantum filter} \label{humfilter}

The most important component of our scheme is the polarising beam splitter, which redirects a photon depending on its polarisation to 
one of the output modes. In the following, $\ket{\lambda_i}$ describes a photon with polarisation $\lambda$ travelling in mode $i$. 
Besides, we denote the input modes $i=1$ and $2$ and the output modes $i=1'$ and $2'$ such that a $V$ polarised photon entering input 
mode $1$ and an $H$ polarised photon entering input mode $2$ leave the setup through output $1'$. Suppose two photons enter the setup 
in different modes. Then the effect of the beam splitter can  be summarised in the transformation 
\begin{equation} \label{comp}
\ket{\lambda_1 \mu_2} \otimes \ket{0_{1'} 0_{2'}} \longrightarrow 
\left\{ \begin{array}{ll} \ket{0_1 0_2} \otimes \ket{H_{1'}H_{2'}}  ~,~ & {\rm if} ~~ \lambda = \mu = H ~,~ \\[0.1cm]
\ket{0_1 0_2} \otimes \ket{V_{1'}V_{2'}}  ~,~ & {\rm if} ~~ \lambda = \mu =V  ~,~ \\[0.1cm]
\ket{0_1 0_2} \otimes \ket{(HV)_{1'} 0_{2'}}  ~,~ & {\rm if} ~~ \lambda = V ~~ {\rm and} ~~ \mu = H ~, \\[0.1cm] 
\ket{0_1 0_2} \otimes \ket{0_{1'} (HV)_{2'}}  ~,~ & {\rm if} ~~ \lambda = H ~~ {\rm and} ~~ \mu = V ~.~ \end{array} \right.
\end{equation} 
We show now that this operation can be used to realise a filter which compares the polarisation $\lambda$ of a target photon with the 
polarisation of an ancilla photon prepared in $\ket{\mu_2}$. With $\mu$ being either $V$ or $H$, the filter operation corresponds to the 
projector $\ket{\mu} \bra{\mu}$ and can be implemented with {\em unit} efficiency.

Suppose a photon number resolving detector is placed in one of the output modes, say output $2'$, and the target photon enters the 
system prepared in $\ket {\lambda_1} =\alpha \, \ket{H_1} + \beta \, \ket{V_1}$. Using Eq.~(\ref{comp}), one can calculate the 
 unnormalised output state after a click in the detector corresponding to polarisation $\mu$. It is either $\alpha \, \ket{H_{1'}}$ or $\beta \, \ket{V_{1'}}$, depending on whether $\mu$ 
equals $H$ or $V$. Note that the probability for a 1-photon detection ($|\alpha|^2$ or $|\beta|^2$, respectively) is exactly what one 
would expect after applying the filter operation $\ket{\mu} \bra{\mu}$ with efficiency 1 to the incoming photon. Remarkably, the target 
photon is effectively not destroyed in the process. The reason is that it does not matter whether the detector absorbs the target photon 
or the ancilla photon, if both have the same polarisation and are anyway indistinguishable.
 
\subsection{The 2-photon case} \label{two}

\begin{figure}
\begin{center}
\includegraphics{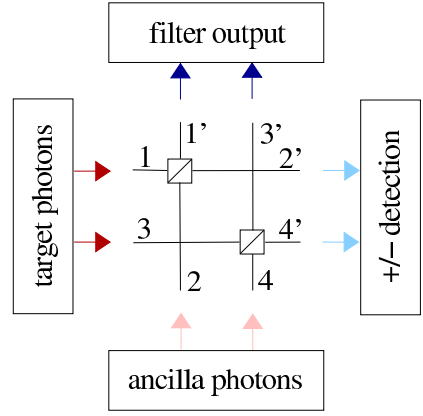}  
\end{center}
\caption{Experimental setup for the realisation of a 2-photon parity filter. The target photons enter the setup via the input 
modes 1 and 3, while the ancilla photons enter the setup via inputs 2 and 4. Within the setup, each photon has to pass one polarising 
beam splitter. Under the condition of the detection of one photon in each of the outputs $2'$ and $4'$, the filter succeeded and the 
projected output state leaves the system via the modes $1'$ and $3'$.} \label{fig1}  
\end{figure}

Let us now describe how the polarising beam splitter (\ref{comp}) can be used for the implementation of a 2-photon parity filter. The 
setup we consider here contains two polarising beam splitters and two polarisation sensitive detectors (see Fig.~\ref{fig1}). The target 
state 
enters the setup via the input modes $1$ and $3$. We further require the presence of the 2-photon ancilla state $\ket{A^{(2)}}$, which 
is a 2-photon Bell state. The ancilla photons should enter the setup via the input modes $2$ and $4$. The two detectors are placed in 
the output modes $2'$ and $4'$. If they both receive a photon each, the filter operation is deemed a success.  Output modes $1'$ 
and $3'$ are designated the filter output.

In the following, we consider the general input pure state 
\begin{equation} \label{huminput}
\ket{\psi_{\rm in}^{(2)}} = \alpha \, \ket{H_1 H_3} +\beta \, \ket{V_1 V_3} +\gamma \, \ket{H_1 V_3} + \delta \, \ket{V_1 H_3}  ~.~
\end{equation}
Our aim is to eliminate the components, where the photons are of different polarisation. Together with the ancilla state $|A^{(2)} 
\rangle$, the setup in Fig.~\ref{fig1} is entered by the 4-photon state
\begin{eqnarray} \label{in}
\ket{\tilde \psi_{\rm in}^{(2)}} &=& \ket{\psi_{\rm in}^{(2)}} \otimes \ket{A^{(2)}} \nonumber \\
&=& {\textstyle {1 \over \sqrt{2}}} \, \Big( \, \alpha \, \ket{H_1H_2H_3H_4} +  \alpha \, \ket{H_1V_2H_3V_4} + \beta \, 
\ket{V_1V_2V_3V_4} +  \beta \, \ket{V_1H_2V_3H_4} \nonumber \\
&& + \gamma \, \ket{H_1H_2V_3H_4} +  \gamma \, \ket{H_1V_2V_3V_4} + \delta \, \ket{V_1V_2H_3V_4} +  \delta \, \ket{V_1H_2H_3H_4} \, 
\Big) ~.~ \nonumber \\
\end{eqnarray}
We now show that the system can act like a parity filter, if one photon is collected in output mode $2'$ and another one is collected in 
output mode $4'$. Using Eq.~(\ref{comp}), one can show that the 4-photon states (\ref{in}) becomes in this case, the unnormalised state
\begin{equation} \label{bus}
\ket{\tilde \psi_{\rm out}^{(2)}} = {\textstyle {1 \over \sqrt{2}}} \, \Big( \, \alpha \, \ket{H_{1'}H_{2'}H_{3'}H_{4'}} + 
\beta \, \ket{V_{1'}V_{2'}V_{3'}V_{4'}} \, \Big) ~.~
\end{equation} 
We further assume that the detectors measure the polarisation of the incoming photons in the rotated basis defined by the 1-photon 
states 
\begin{equation} \label{tractor}
\ket{\pm} \equiv {\textstyle {1 \over \sqrt{2}}} \, \Big( \, \ket{H} \pm \ket{V} \, \Big) ~.~ 
\end{equation}
It is important that the detectors distinguish the polarisation of each incoming photon in this basis (opposed to just absorbing the 
photon), since this approach guarantees that the output becomes the expected pure state. Using the definition (\ref{tractor}), we can 
rewrite the state (\ref{bus}) as 
\begin{eqnarray} 
\ket{\tilde \psi_{\rm out}^{(2)}} &=& {\textstyle {1 \over 2}} \, \Big( \, \alpha \, \ket{H_{1'}H_{3'}} + \beta \, 
\ket{V_{1'}V_{3'}} \, \Big) \otimes {\textstyle {1 \over \sqrt{2}}} \, \Big( \, \ket{+_{2'}+_{4'}} + \ket{-_{2'}-_{4'}} \, \Big) \nonumber \\
&& + {\textstyle {1 \over 2}} \, \Big( \, \alpha \, \ket{H_{1'}H_{3'}} - \beta \, \ket{V_{1'}V_{3'}} \, \Big) \otimes {\textstyle 
{1 \over \sqrt{2}}} \, \Big( \, \ket{+_{2'}-_{4'}} + \ket{-_{2'}+_{4'}} \, \Big) ~.~ 
\end{eqnarray} 
Suppose the photons in output ports $2'$ and $4'$ are absorbed in the measurement process. Then the output state of the system equals in 
case of a single click in each of the detectors
\begin{equation}
\ket{\psi_{\rm out}^{(2)}}  = {\textstyle {1 \over 2}} \, \Big( \, \alpha \, \ket{H_{1'} H_{3'}} \pm \beta \, \ket{V_{1'} V_{3'}} 
\, \Big) ~.~
\end{equation} 
The ``$+$" sign applies when both detectors measure the same polarisation (which happens with probability ${1 \over 2}$); the ``$-$" 
sign applies when both detectors measure different polarisations (which also happens with probability ${1\over 2}$). More generally, every measurement of the state $\ket{-}$ yields a phase flip error on the output state. Therefore, measuring even numbers of $\ket{-}$(or in this case, the same polarisations) yield no phase flip error or identity operation on the output state. The implementation 
of the parity filter only needs a correction of this phase flip error in the event of measuring odd numbers of $\ket{-}$ (in this case,  different polarisations) which can be implemented with the help of a Pauli $\sigma_z$ operation on any  of the output photons. In any case, the 4-photon state (\ref{in}) can be reduced by measurement in the $\ket{\pm}$ with appropriate  $\sigma_z$ 
correction to the following 2-photon state,  
\begin{equation} \label{car}
\ket{\psi_{\rm out}^{(2)}} = {\textstyle {1 \over \sqrt{2}}} \Big( \, \alpha \, \ket{H_{1'} H_{3'}} + \beta \, \ket{V_{1'} V_{3'}} \, 
\Big) ~,~
\end{equation}  with unit efficiency. We have taken into account  all appropriate measurement syndromes which explains the normalisation.
A closer look at the normalisation of this state  tells us that the parity filter shown in Fig.~\ref{fig1} works with 
efficiency $p_2 = {1 \over 2}$. If the success probability of the scheme would be 1, the output state (\ref{car}) would be $\alpha \, 
\ket{H_{1'} H_{3'}} 
+ \beta \, \ket{V_{1'} V_{3'}}$. It can be shown that the filter can also be operated with mixed states as inputs.

\subsection{The $N$-photon case}

\begin{figure}
\begin{center}
\includegraphics{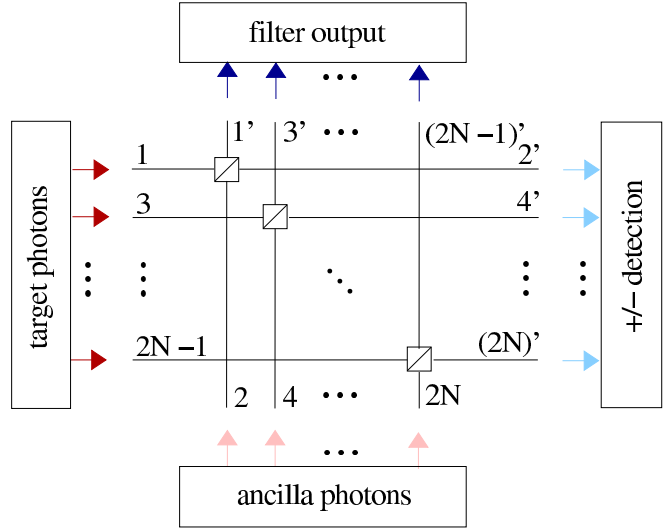} 
\end{center}
\caption{Experimental setup for the realisation of the $N$-photon quantum filter (\ref{PN}). The $N$ polarising beam splitters 
each compare the state of one of the target photons with the state of one of the ancilla photons, which are initially prepared in the 
GHZ state $|A^{(N)} \rangle$. Besides, $N$ detectors perform photon measurements in the polarisation basis (\ref{tractor}). The output 
photons leave the system via the odd numbered output ports.} \label{fig2}  
\end{figure}  

The generalisation of the above described 2-photon parity filter to the $N$-photon quantum filter (\ref{PN}) is straightforward and 
requires $N$ polarising beam splitters and $N$ polarisation sensitive detectors (see Fig.~\ref{fig2}). One side of the setup is 
entered by the $N$-photon input state $|\psi_{\rm in}^{(N)} \rangle$ with one photon in each odd-numbered input mode, while the other 
side is entered by an $N$-photon ancilla state $|A^{(N)} \rangle$ with one photon in each even numbered input mode. In the following we 
denote the modes containing the detectors by $2'$, $4'$, ..., $(2N)'$, while the modes $1'$, $3'$, ..., $(2N-1)'$ contain the output 
state.

Again, the successful operation of the quantum filter is indicated by a single click in each of the detectors. Suppose $\alpha$ denotes the amplitude 
of the state $|H_1H_3 \, . \, . \, . \, H_{2N-1} \rangle$ while $\beta$ is the amplitude of the state $\ket{V_1V_3 \, . \, . \, . \, 
V_{2N-1}}$ with respect to the target state $\ket{\psi_{\rm in}^{(N)}}$. Then we find, using Eq.~(\ref{comp}) and in analogy to 
Eq.~(\ref{bus}), that the collection of one photon in each of the detector output ports transforms the total input state $\ket{\tilde 
\psi_{\rm in}^{(N)}} = \ket{\psi_{\rm in}^{(N)}} \otimes \ket{A^{(N)}}$ into 
\begin{equation} \label{bicycle}
\ket{\tilde \psi_{\rm out}^{(N)}} = {\textstyle {1 \over \sqrt{2}}} \, \Big( \, \alpha \, \ket{H_{1'}H_{2'} \, . \, . \, . \, H_{(2N)'}} 
+ \beta \, \ket{V_{1'}V_{2'} \, . \, . \, . \, V_{(2N)'}} \, \Big) ~.~
\end{equation} 
For the same reason as in the 2-photon case, we assume that the detectors measure the polarisation of the incoming photons in the 
polarisation basis (\ref{tractor}) by absorption. Suppose $J$ is the number of photons found in the $\ket{-}$ state, then one can show 
using Eq.~(\ref{bicycle}) and proceeding as in Section \ref{two} that the output state of the remaining $N$ photons equals 
\begin{equation}
\ket{\psi_{\rm out}^{(N)}} = {\textstyle {1 \over 2}} \, \Big( \, \alpha \, \ket{H_{1'}H_{3'} \, . \, . \, . \, H_{(2N-1)'}} + 
(-1)^{J} \, \beta \, \ket{V_{1'}V_{3'} \, . \, . \, . \, V_{(2N-1)'}} \, \Big) ~.~
\end{equation} Note that the probability of $J$ being an odd number, which incurs a phase flip error on the output state in analogy to the 2-photon filter, is $1 \over 2$. 
As before, we can transform with unit efficiency the state (\ref{bicycle}) with the help of a phase flip correction to  the final state  given by 
\begin{equation} \label{tram}
\ket{\psi_{\rm out}^{(N)}} = {\textstyle {1 \over \sqrt{2}}} \, \Big( \, \alpha \, \ket{H_{1'}H_{3'}...H_{(2N-1)'}} + \beta \, 
\ket{V_{1'}V_{3'}...V_{(2N-1)'}} \, \Big) ~.~
\end{equation} 
This is exactly the output state that one expects after the application of the quantum filter (\ref{PN}) to the input state $|\psi_{\rm 
in}^{(N)} \rangle$ with success probability $p_N = {1 \over 2}$, which is the highest that has been predicted so far without the use of 
 universal two-qubit quantum gate operation such as the CNOT or CZ gates\footnote{Alternatively, a straightforward way of implementing the quantum filter (\ref{PN}) is to replace each polarising beam splitter in the setup (see 
Fig.~\ref{fig2}) by a CNOT gate(which is difficult to realise with linear optics alone). Furthermore, the detectors in all even numbered output modes should perform a polarisation 
sensitive measurement in the $H/V$ basis. The projection efficiency of such a scheme would only be limited by the success probability 
$p$ of a single controlled-NOT operation and would scale like $p^N$. For sufficiently large photon numbers $N$, this might decrease 
below ${1 \over 2}$. Therefore the use of polarising beam splitters, which can operate with a very high fidelity, should be favoured 
\cite{Pan03}.}.
Naively, one might expect that the efficiency of the filter decreases with the number of photons in the setup. However, this is not the case here. 
Furthermore, we remark that the described quantum filter also works for mixed $N$-photon input states. 

Taking into account real detector efficiencies and dark count rates will diminish both the success probability and fidelity of the 
above described filter. In general, the success probability and fidelity depend on the nature of the input state as well as the ancilla. 
Here, we focus  on a simple example of analysing the 
error probability of a 2-photon parity filter by assuming imperfect photon detectors but perfect ancilla state. It is shown in 
\cite{Saavedra00} that the ancilla states considered here can be prepared with high fidelity and success probability. As in 
\cite{Hofmann02}, we assume that the dark count rate can be reduced by time gating and consider the effect of detector inefficiencies causing 
an error due to a mistake of registering a 2-photon detection event as a single photon event. This is  known as preselective error. If 
we also postselect the output state, then such an error can in principle be eliminated. Without loss of generality, we analyse the 
case where the detectors each register a click for an alleged photon in the state $\ket{+}$. This can be represented by a POVM(Positive 
operator valued measure) element $E_{i'}$ given by \cite{Lee04}
\begin{equation} 
E_{i'}=p_d\ket{+_{i'}}\bra{+_{i'}}+2p_d(1-p_d)\ket{(++)_{i'}}\bra{(++)_{i'}} \, ,
\end{equation} where $p_d$ is the single photon detection efficiency. From Kok and Braunstein \cite{Kok01,Barnett98}, we know that the reduced projected state is $\rho_{1'3'}=\frac{{\rm Tr}_{2'4'}(E_{2'}E_{4'}\rho_{1'2'3'4'})}{{\rm Tr}_{1'3'}(\cdot)}$ where $\rho_{1'2'3'4'}$ is the state after passing  $\ket{\tilde \psi_{\rm in}^{(2)}}$ through the 2 polarising beam splitters. We also fix 
$|\alpha|^2=|\beta|^2=|\gamma|^2=|\delta|^2=\frac{1}{4}$ to compute  for the most typical input state to obtain the average fidelity. 
One can show that the fidelity\footnote{This is analogous to the true QND or preselective fidelity discussed in Ref \cite{Kok05a}. Clearly, if we define our fidelity based on coincidence counting, where a photon is detected in all outputs 1', 2' 3' 4', the postselective fidelity  can be much higher.} of the quantum filter is given by $F=\bra{\psi_{\rm out}^{(2)}}\rho_{1'3'}\ket{\psi_{\rm out}^{(2)}}/\langle \psi_{\rm out}^{(2)}\ket{\psi_{\rm out}^{(2)}}=(5-6p_d+2p_d^2)^{-1}$. For example, given a $p_{d}$ of 0.88 
(\cite{Takeuchi99,Rosenberg05}), the maximum error rate $1-F$ would be 0.19 in the light of current  technology. Especially for the 
recent work by  Rosenberg {\em et al.} \cite{Rosenberg05},  superconducting transition-edge sensors are expected to have photon-number 
resolution with negligible dark counts at arbitrary high efficiency in the future.

\section{Conclusions}

We described the realisation of a 2-photon parity filter that requires only two polarising beam splitters, two photons prepared in a 
maximally entangled Bell state and two polarisation sensitive detectors. The success rate of the scheme $p_2 = {1 \over 2}$ is the 
highest that has been predicted so far without the help of universal two-qubit quantum gate operations and is reached here due to employing an 
entangled ancilla state as a resource. A generalisation of the proposed scheme to the $N$-photon case is straightforward. We showed that 
the quantum filter (\ref{PN}) can be implemented with the help of $N$ polarising beam splitters and an $N$-photon GHZ state as a 
resource. Remarkably, the success rate of the filter remains ${1 \over 2}$, irregardless of the size of the input state.

To implement the quantum filter (\ref{PN}), the $N$ polarising beam splitters compare the state of the incoming photons pairwise with 
the state of the ancilla photons. In Section \ref{humfilter}, we showed that a single polarising beam splitter can be used to realise a 
filter, which measures polarisation $H$ or $V$, respectively, with unit efficiency. Preparing the ancilla photons, for example, in the 
state $\ket{HH \, . \, . \, . \, H}$, would result in a filter that measures whether all target photons are prepared in $\ket{H}$. 
However, since we compare the input state with a GHZ state, which contains two terms, namely $\ket{HH \, . \, . \, . \, H}$ and $\ket{VV 
\, . \, . \, . \, V}$, the probability of the described filter is only as high as ${1 \over2}$. Indeed, the highly entangled $N$-photon 
GHZ state acts as a ``mask" for the filter. 

A straightforward extension of the ideas of this chapter is to consider a different form of the ``mask" or ancilla state $\ket{A^{CZ}}$ 
given by $\frac{1}{2}(\ket{H_2H_4}+\ket{H_2V_4}+\ket{V_2H_4}-\ket{V_2V_4})$. Under the condition that the photons pass the filter, 
heralded by single photon detection in both output detectors in  $\ket{\pm}$,  the output state, with correction of sign errors, would 
instead be given by 
\begin{equation}
\ket{\psi^{(CZ)}_{\rm out}}=\frac{1}{2}(\alpha \ket{H_{1'}H_{3'}}+\beta \ket{V_{1'}V_{3'}}+\gamma \ket{H_{1'}V_{3'}}-\delta 
\ket{V_{1'}H_{3'}}) \, . 
\end{equation} This is the same as the application of a CZ filter or gate $P_{CZ}$
\begin{equation}
P_{CZ}=\frac{1}{2}(\ket{H_{1'}H_{3'}}\bra{H_1H_3}-\ket{V_{1'}V_{3'}}\bra{V_1V_3}+\ket{H_{1'}V_{3'}}\bra{H_1V_3}+\ket{V_{1'}H_{3'}}\bra{V_
1H_3})
\end{equation} with efficiency $\frac{1}{4}$ to the input state (\ref{huminput}). This is analogous to the CNOT gate proposed by Pittman 
{\em et al.} \cite{Pittman01} with success probability $\frac{1}{4}$.

We have seen an example of how quantum computing with photons assisted with entangled ancillas can result in a more efficient 
implementation. However, the scheme is still necessarily probabilistic as are  all known linear optics based schemes where the input 
state is not already 
necessarily encoded offline. We move to the next chapter where we add just one more ingredient, a special single 
photon source with encoding 
ability, and show how quantum computing with linear optics can become effectively deterministic.

\chapter{Distributed Quantum Computing with Distant Single Photon 
Sources}\label{minsk}

\section{Introduction}
Practical implementations of quantum computing to solve non-trivial problems 
require a scalable architecture, i.e. the ability to process, 
address and store many qubits. This is particularly challenging if all the 
interactions between qubits are controlled locally and coherently. Particular 
advances in this aspect have been made in ion traps \cite{Kielpinski02} and atoms 
trapped in optical lattices \cite{Jaksch99}. Even with optical 
lattices, with the inherent capability to store many qubits,  controlled 
addressibility and manipulation of individual qubits still 
remains an experimental challenge despite  advances to alleviate these 
requirements \cite{You00,Kay04,Calarco04} through the help of  marker atoms. In 
ion traps, while addressibility is not an issue, interaction between distant 
qubits still requires some form of ion transport to the range 
where coherent interaction is possible between two ions 
\cite{Duan04,Kielpinski02}.

An attractive alternative approach is the concept of distributed quantum 
computing \cite{Eisert00,Grover96,Cirac99}. This consists of a network of 
nodes with each node processing and storing a small number of qubits, which is 
comparatively easy to realise. The qubits in each node are 
stationary qubits, i.e. qubits that are not transported, with long decoherence 
times and serve as a quantum memory. The stationary qubits 
in each node communicate with distant nodes through the means of flying qubits, 
i.e. qubits that are transported. Distributed quantum 
computing can lead to a more efficient implementation of the phase estimation 
problem compared to a classical computer in the presence of 
decoherence \cite{Grover96,Cirac99}. Furthermore, distributed quantum computing 
allows distant users to share quantum resources.

Traditionally, the stationary qubit of a certain node maps its state to a flying 
qubit which leaves the node. On the arrival at the 
target node, the flying qubit maps its state to a stationary qubit in the target 
node. It is thus assumed that interconvertability of 
stationary and flying qubits are required. Schemes related to this have been 
proposed based on atom-cavity as stationary qubits and 
photons as flying 
qubits \cite{Enk97,Cirac97,Sorensen98,Xiao04,Cho04b,Zhou05,Duan05b}.  All these 
schemes involve single photon sources with direct transmisions of photons 
through cavities. In all these cases, such transmissions occur one or several 
times to complete the gate operation protocol. Another 
scheme by Mancini {\em et al.} involves engineering a direct interaction between 2 
distant coupled cavities via fibers \cite{Mancini04}.  All these schemes 
demand a high level of precision and might pose a great experimental challenge 
\cite{Browne03} if one requires a high success probability.

In contrast to this,  we avoid all these challenges by not requiring any form of 
photon transmission through cavities. Note that we do 
not really require the interconversion of stationary qubits and flying qubits 
for quantum computation in a network. The unidirectional 
encoding of stationary to flying qubits is already sufficient for distributed 
quantum computation.  Schemes along these lines 
\cite{Protsenko02,Schlosser03,Zou05,Barrett04} have already been 
proposed\footnote{Very recently, after this work has been submitted for 
publication in August 2004, Benjamin {\em et al.} 
\cite{Benjamin05} reported a scheme on creating graph states by optical 
excitation in stationary qubits. They also obtain the insurance 
scenario reported in this chapter with a $4 \times 4$ multiport at the  cost of 
having two distant stationary qubits encoding a qubit. 
Furthermore, interferometric stability is not inherent in their scheme.}. To 
implement a two-qubit universal gate between two distant 
stationary 
qubits, the basic idea is to redundantly encode the pair of stationary qubits to 
a pair of flying qubits. Following that, a maximally 
entangling or Bell measurement, which is normally accomplished with linear 
optics, is performed on the pair of flying qubits. A universal 
two-qubit gate is accomplished between the stationary qubits if the entangling 
measurement is successful. Another related scheme based on 
trapped ions has been proposed \cite{Duan04a} which uses ancilla ions in which they 
have to be pre-entangled.
 
In this chapter, we will demonstrate that 
scalable quantum computing between distant stationary qubits, where the 
stationary qubits are single photon sources which generate the photons naturally 
as flying qubits, can be made deterministic even if the 
entangling measurement does not succeed.\footnote{We remind the reader here that a never-failing complete Bell measurement on two  photons is not possible with linear optics \cite{Lutkenhaus99}.} We {\em do not} require any ancilla 
stationary qubits nor any photon transmission through 
cavities to achieve this. The ability to encode the state of the atom unto the 
photon is all that is required. We use linear optics to 
perform the Bell measurements on the photon. Generally, the measurement basis we 
choose  does not yield any information about the 
stationary qubits and therefore cannot destroy the qubits in any case. As above, 
for a successful Bell measurement, a two-qubit gate is 
accomplished. If not,  the state of the stationary qubit is not destroyed and 
this allows us to repeat the encoding and subsequent 
measurement until it succeeds. We have shown that this can be done by carefully 
choosing the measurement basis in the entangling 
measurement with linear optics. A related idea to protect a  photon state 
against 
gate failure has been proposed in the past by Knill {\em et al.} \cite{Knill01} in the context 
of 
photon gate implementation by a two-qubit quantum code in their 
teleportation-based gate. Such ideas are closely related to quantum error 
correction \cite{Shor95,Calderbank96,Steane96}. We however use a form of 
redundant encoding 
natural to diverse kinds of  single photon sources and show that distributed 
quantum 
computation between stationary qubits can require  similar experimental 
resources 
as  linear optics computation, i.e. single photon sources, optical elements and 
photon detectors. At the same time, it can be performed much more 
efficiently\footnote{We require no prepared ancillas nor any photon storage 
and feedforward 
operations.} as compared to conventional linear optics computation \cite{Knill01}.

The single photon sources that we use can take the form of atom-cavity systems 
\cite{Law97,Kuhn99}, quantum dots, diamond NV colour centers or 
even atomic ensembles \cite{Matsukevich04}. In principle, any photon source that 
allows redundant encoding of the state of the source to the 
photon it generates is a viable candidate for our scheme. 

Having generated the photons, the photons must subsequently travel to the linear 
optics apparatus that performs the entangling or partial Bell 
measurement. Finally, the partial Bell measurements 
on the encoded photons are performed and based on measurement results, we either
halt the scheme upon a heralded success or repeat the 
scheme until success. 

This chapter is organised as follows. The next section details the general 
principle of a remote two-qubit gate 
implementation with our scheme. We also show how teleportation with insurance 
can be accomplished with minimal change to the setup. 
Following that, in Section \ref{photonencoder} and \ref{photonmeasurement} we 
describe the two ingredients of the scheme, photon encoding 
and measurements.  Finally, we conclude in the last section with a short 
discussion on possible applications to cluster state buildup for 
robust computing.

\section{Basic Idea of a remote two-qubit phase gate} 

One of the requirements for universal quantum computing is the ability to 
perform a universal two-qubit gate 
operation, like a controlled 
phase gate. Here we describe the general concept for the implementation of such 
an entangling two-qubit phase 
gate between two distant 
single photon sources. Note that our method of distributed quantum computing 
only allows the realisation of non-local 
phase gates, since the 
measurement on a photon pair can imprint a phase on the state of the 
corresponding sources but cannot change 
the distribution of their 
populations. This is however sufficient for universal quantum computation. The 
first step for the implementation of a two-qubit gate is 
the generation of a photon within 
each respective source, which 
encodes the information of the stationary qubit.

\subsection{Encoding}

Let us denote the states of the photon sources, which encode the logical qubits 
$|0 \rangle_{\rm L}$ and $|1 
\rangle_{\rm L}$ as $|0 
\rangle$ and $|1 \rangle$, respectively. For example, for atom-like single photon sources, the stable ground states can be chosen as the logical qubits. An arbitrary pure state of two 
stationary qubits can  be written 
as
\begin{equation} \label{original}
\ket{\psi_{\rm in}}=\alpha \, \ket{00} + \beta\, \ket{01} + \gamma \, \ket{10} + 
\delta \, \ket{11} \, ,
\end{equation} 
where $\alpha$, $\beta$, $\gamma$ and $\delta$ are the corresponding complex 
coefficients with $|\alpha|^2 + 
|\beta|^2 + |\gamma|^2 + 
|\delta|^2=1$. Suppose a photon is now generated in each of the two sources, 
whose state (i.e. ~polarisation, 
frequency or generation 
time) depends on the state of the source.  As we see below, it is helpful to 
assume that the encoding is for 
both sources different. In 
the following, we assume that source 1 prepared in $|i \rangle$ leads to the 
creation of one photon in state 
$|{\sf x}_i \rangle$, while 
source 2  prepared in $|i \rangle$ leads to the creation of one photon in state 
$|{\sf y}_i \rangle$, such 
that 
\begin{equation} \label{enc}
\ket{i}_1 \rightarrow \ket{i;{\sf x}_i}_1 \, ,
~~ \ket{i}_2 \rightarrow \ket{i;{\sf y}_i}_2 \, . 
\end{equation} 
The simultaneous creation of a photon in both sources then transfers the initial 
state (\ref{original}) into 
\begin{equation} \label{theencoding}
\ket{\psi_{\rm enc}} = \alpha \, \ket{00;{\sf x}_0{\sf y}_0} + \beta \, 
\ket{01;{\sf x}_0{\sf y}_1} + \gamma 
\, 
\ket{10;{\sf x}_1{\sf y}_0} +\delta \, \ket{11;{\sf x}_1{\sf y}_1} \, . 
\end{equation}
The way this encoding step can be realised experimentally using either emission 
time or 
polarisation degrees of freedom to 
encode the stationary qubits is discussed in Section \ref{photonencoder}.

\subsection{Mutually Unbiased Basis}

Once the photons have been created, an entangling phase gate can be implemented 
by performing an absorbing 
measurement on the photon 
pair. Therefore, it is important to choose the photon measurement such that none 
of the possible outcomes 
reveals any information about the 
coefficients $\alpha$, $\beta$, $\gamma$ and $\delta$. That such measurements 
exists is well 
known \cite{Wootters89}. The corresponding 
measurement basis forms a so-called {\em mutually unbiased basis} (MUB) with 
respect to the computational basis. 
Here we are interested in 
photon pair measurements in a MUB\footnote{In this chapter, our MUB basis is 
always defined with respect to the computational basis.}  
given  the computational basis $\{\ket{{\sf x}_0{\sf 
y}_0}, \, \ket{{\sf x}_0{\sf y}_1}, \, \ket{{\sf x}_1{\sf y}_0}, \, \ket{{\sf 
x}_1{\sf y}_1} \}$.  

More concretely, the potential outcomes of the photon measurement should all be of the 
form
\begin{equation} \label{unbiased}
\ket{\Phi} = {\textstyle {1 \over 2}} \big[ \ket{{\sf x}_0{\sf y}_0} + {\rm 
e}^{{\rm i} \varphi_1} \, 
\ket{{\sf x}_0{\sf y}_1} + {\rm 
e}^{{\rm i} \varphi_2} \, \ket{{\sf x}_1{\sf y}_0} + {\rm e}^{{\rm i} \varphi_3} 
\, \ket{{\sf x}_1{\sf y}_1} 
\big] \, . 
\end{equation} Indeed, this is possible with linear optics as we will show in this thesis.
Detecting this state and absorbing the two photons in the process transfers the 
encoded state 
(\ref{theencoding}) into
\begin{equation} \label{out}
\ket{\psi_{\rm out}} = \alpha \, \ket{00} + {\rm e}^{-{\rm i} \varphi_1} \, 
\beta \, \ket{01} + {\rm e}^{-{\rm 
i} \varphi_2} \, 
\gamma \, \ket{10} + {\rm e}^{-{\rm i} \varphi_3} \, \delta \, \ket{11} \, ,
\end{equation}
and therefore does  not reveal any information about the input state 
(\ref{original}) indeed. It is thus 
equivalent to a phase gate implementation on the stationary qubit.

Here we are especially interested in the implementation of an entangling phase 
gate with maximum entangling 
power. This requires 
detecting the photons in one of the four Bell states.  If 
\begin{equation}
\varphi_3 = \varphi_1 + \varphi_2 \, , 
\end{equation}
the state $\ket{\Phi}$ is a product state and the output (\ref{out}) differs 
from the initial state 
(\ref{original}) only by local 
operations. However, the state (\ref{unbiased}) becomes a maximally entangled one if and only if
\begin{equation}
\varphi_3 = \varphi_1 + \varphi_2 \pm (2n-1)\pi \, .
\end{equation}
. 

\subsection{A deterministic entangling gate}

Let us denote the states of the measurement basis, i.e. the mutually unbiased 
basis, in the following by $\{ 
\ket{\Phi_i} \}$. In order to find a complete Bell basis with all 
states of the form (\ref{unbiased}), we introduce the following notation,
\begin{eqnarray} \label{base}
\ket{\Phi_1} &\equiv & {\textstyle {1 \over \sqrt{2}}} \big[ \ket{{\sf a}_1 {\sf 
b}_1}+\ket{{\sf a}_2 {\sf 
b}_2} \big] \, \, , \,
\ket{\Phi_2} \equiv  {\textstyle {1 \over \sqrt{2}}} \big[ \ket{{\sf a}_1 {\sf 
b}_1}-\ket{{\sf a}_2 {\sf b}_2} 
\big] \, , \nonumber \\
\ket{\Phi_3} &\equiv & {\textstyle {1 \over \sqrt{2}}} \big[ \ket{{\sf a}_1 {\sf 
b}_2}+\ket{{\sf a}_2 {\sf 
b}_1} \big] \,  \, , \,
\ket{\Phi_4} \equiv  {\textstyle {1 \over \sqrt{2}}} \big[ \ket{{\sf a}_1 {\sf 
b}_2}-\ket{{\sf a}_2 {\sf b}_1} 
\big] \, ,
\end{eqnarray}
where the states $|{\sf a}_i \rangle$ describe photon 1 and the states $ |{\sf b}_i 
\rangle $ describe photon 2 and
$\langle {\sf a}_1 | {\sf a}_2 \rangle =0$ and $\langle {\sf b}_1 | {\sf b}_2 
\rangle = 0$. One can then write 
the photon states on the 
right hand side of Eq.~(\ref{base}) without loss of generality as 
\begin{eqnarray} \label{ab}
\ket{{\sf a}_1} &=& \cos{\theta_1} \, \ket{{\sf x}_0} +  {\rm e}^{{\rm i} 
\vartheta_1} \sin{\theta_1} \, 
\ket{{\sf x}_1} \, , \,
\ket{{\sf a}_2} = {\rm e}^{-{\rm i} \xi_1}(  {\rm e}^{-{\rm i} \vartheta_1} 
\sin{\theta_1}\, \ket{{\sf x}_0} - 
 \cos{\theta_1} \, \ket{{\sf x}_1}) \nonumber \\
\ket{{\sf b}_1} &=& \cos{\theta_2} \, \ket{{\sf y}_0} + {\rm e}^{{\rm i} 
\vartheta_2} \sin{\theta_2} 
\,\ket{{\sf y}_1} \, , \,
\ket{{\sf b}_2} = {\rm e}^{-{\rm i} \xi_2}({\rm e}^{-{\rm i} \vartheta_2} 
\sin{\theta_2} \, \ket{{\sf y}_0} - 
\cos{\theta_2} \, \ket{{\sf y}_1}) \, .\nonumber \\
\end{eqnarray} 
Inserting this into Eq.~(\ref{base}), we find 
\begin{eqnarray} \label{base2}
\ket{\Phi_1} &=& {\textstyle {1 \over \sqrt{2}}} \big[ \big 
(\cos{\theta_1}\cos{\theta_2}+{\rm e}^{-{\rm 
i}(\vartheta_1+\vartheta_2)} {\rm e}^{-{\rm i}(\xi_1+\xi_2)} 
\sin{\theta_1}\sin{\theta_2} \big) \ket{{\sf 
x}_0{\sf y}_0} \nonumber \\
&& + \big( {\rm e}^{{\rm i}\vartheta_2}\cos{\theta_1}\sin{\theta_2} - {\rm 
e}^{-{\rm i}\vartheta_1} {\rm 
e}^{-{\rm i}(\xi_1+\xi_2)} \sin{\theta_1}\cos{\theta_2} \big) 
\ket{{\sf x}_0{\sf y}_1} \nonumber \\
&&+ \big( {\rm e}^{{\rm i}\vartheta_1}\sin{\theta_1}\cos{\theta_2}-{\rm 
e}^{-{\rm i}\vartheta_2} {\rm 
e}^{-{\rm i}(\xi_1+\xi_2)} \cos{\theta_1}\sin{\theta_2} \big) \ket{{\sf x}_1{\sf 
y}_0} \nonumber \\
&&+ \big( {\rm e}^{{\rm 
i}(\vartheta_1+\vartheta_2)}\sin{\theta_1}\sin{\theta_2}+ {\rm e}^{-{\rm 
i}(\xi_1+\xi_2)} \cos{\theta_1}\cos{\theta_2} \big) \ket{{\sf x}_1{\sf 
y}_1} \big],  \nonumber \\
\ket{\Phi_2} &=& {\textstyle {1 \over \sqrt{2}}} \big[ 
\big(\cos{\theta_1}\cos{\theta_2}-{\rm e}^{-{\rm 
i}(\vartheta_1+\vartheta_2)} {\rm e}^{-{\rm i}(\xi_1+\xi_2)} 
\sin{\theta_1}\sin{\theta_2} \big)\ket{{\sf 
x}_0{\sf y}_0} \nonumber \\
&& + \big({\rm e}^{{\rm i}\vartheta_2}\cos{\theta_1}\sin{\theta_2}+{\rm 
e}^{-{\rm i}\vartheta_1} {\rm 
e}^{-{\rm i}(\xi_1+\xi_2)} \sin{\theta_1}\cos{\theta_2} \big) \ket{{\sf x}_0{\sf 
y}_1} \nonumber \\
&& + \big( {\rm e}^{{\rm i}\vartheta_1}\sin{\theta_1}\cos{\theta_2}+{\rm 
e}^{-{\rm i}\vartheta_2} {\rm 
e}^{-{\rm i}(\xi_1+\xi_2)} \cos{\theta_1}\sin{\theta_2} \big) \ket{{\sf x}_1{\sf 
y}_0} \nonumber \\
&&+ \big( {\rm e}^{{\rm 
i}(\vartheta_1+\vartheta_2)}\sin{\theta_1}\sin{\theta_2}-{\rm e}^{-{\rm 
i}(\xi_1+\xi_2)} \cos{\theta_1}\cos{\theta_2} \big) \ket{{\sf x}_1{\sf y}_1} 
\big], \nonumber \\
\ket{\Phi_3} &=&{\textstyle {1 \over \sqrt{2}}}\big[ \big({\rm e}^{-{\rm 
i}\vartheta_2} {\rm e}^{-{\rm 
i}\xi_2}\cos{\theta_1}\sin{\theta_2}+{\rm e}^{-{\rm i}\vartheta_1} {\rm 
e}^{-{\rm i}\xi_1} 
\sin{\theta_1}\cos{\theta_2} \big)\ket{{\sf x}_0{\sf y}_0} \nonumber \\
&&- \big({\rm e}^{-{\rm i}\xi_2} \cos{\theta_1}\cos{\theta_2}-{\rm e}^{-{\rm 
i}(\vartheta_1-\vartheta_2)} {\rm 
e}^{-{\rm i}\xi_1}\sin{\theta_1}\sin{\theta_2} \big) \ket{{\sf x}_0{\sf y}_1} 
\nonumber \\
&&+ \big({\rm e}^{{\rm i}(\vartheta_1-\vartheta_2)} {\rm e}^{-{\rm i}\xi_2} 
\sin{\theta_1}\sin{\theta_2}- {\rm 
e}^{-{\rm i}\xi_1} \cos{\theta_1}\cos{\theta_2} \big) \ket{{\sf x}_1{\sf y}_0} 
\nonumber \\
&& - \big({\rm e}^{{\rm i}\vartheta_1} {\rm e}^{-{\rm i}\xi_2} 
\sin{\theta_1}\cos{\theta_2}+{\rm e}^{{\rm 
i}\vartheta_2} {\rm e}^{-{\rm i}\xi_1}\cos{\theta_1}\sin{\theta_2} \big) 
\ket{{\sf x}_1{\sf y}_1} \big] , 
\nonumber \\
\ket{\Phi_4} &=&{\textstyle {1 \over \sqrt{2}}}\big[ \big({\rm e}^{-{\rm 
i}\vartheta_2} {\rm e}^{-{\rm 
i}\xi_2} \cos{\theta_1}\sin{\theta_2}-{\rm e}^{-{\rm i}\vartheta_1} {\rm 
e}^{-{\rm 
i}\xi_1}\sin{\theta_1}\cos{\theta_2} \big) \ket{{\sf x}_0{\sf y}_0} \nonumber \\
&& - \big({\rm e}^{-{\rm i}\xi_2} \cos{\theta_1}\cos{\theta_2}+{\rm e}^{-{\rm 
i}(\vartheta_1-\vartheta_2)} 
{\rm e}^{-{\rm i}\xi_1} \sin{\theta_1}\sin{\theta_2} \big) \ket{{\sf x}_0{\sf 
y}_1} \nonumber \\
&& + \big({\rm e}^{{\rm i}(\vartheta_1-\vartheta_2)} {\rm e}^{-{\rm i}\xi_2} 
\sin{\theta_1}\sin{\theta_2}+ 
{\rm e}^{-{\rm i}\xi_1} \cos{\theta_1}\cos{\theta_2} \big) \ket{{\sf x}_1{\sf 
y}_0} \nonumber \\
&&- \big({\rm e}^{{\rm i}\vartheta_1} {\rm e}^{-{\rm i}\xi_2} 
\sin{\theta_1}\cos{\theta_2}-{\rm e}^{{\rm 
i}\vartheta_2} {\rm e}^{-{\rm i}\xi_1} \cos{\theta_1}\sin{\theta_2} \big) 
\ket{{\sf x}_1{\sf y}_1} \big]  .  
\end{eqnarray}
These states are of the form (\ref{unbiased}), if the amplitudes are all of the 
same size, which yields the 
conditions
\begin{eqnarray} \label{constraint1}
&& \hspace*{-0.6cm} \big| \cos{\theta_1}\cos{\theta_2}\pm{\rm e}^{-{\rm 
i}(\vartheta_1+\vartheta_2+\xi_1+\xi_2)}\sin{\theta_1}\sin{\theta_2} \big| 
\nonumber 
\\
&=& \big|\cos{\theta_1}\sin{\theta_2}\pm{\rm e}^{-{\rm 
i}(\vartheta_1+\vartheta_2+\xi_1+\xi_2)}\sin{\theta_1}\cos{\theta_2} \big| = 
\textstyle{1 \over \sqrt{2}} \, , 
\end{eqnarray}and
\begin{eqnarray} \label{constraint2}
&&  \hspace*{-0.6cm} \big|\cos{\theta_1}\sin{\theta_2}\pm{\rm e}^{-{\rm 
i}(\vartheta_1 - \vartheta_2+\xi_1-\xi_2)}\sin{\theta_1}\cos{\theta_2} \big| 
\nonumber \\
&=& \big|\cos{\theta_1}\cos{\theta_2}\pm{\rm e}^{-{\rm 
i}(\vartheta_1-\vartheta_2+\xi_1-\xi_2)}\sin{\theta_1}\sin{\theta_2}| = 
\textstyle{1 \over \sqrt{2}} \, .
\end{eqnarray}
The constraints (\ref{constraint1}) and (\ref{constraint2}) can be fulfilled by the condition,
 \begin{equation} 
\label{con}
\cos(2\theta_1)\cos(2\theta_2)=\cos(\vartheta_1\pm\vartheta_2+\xi_1 \pm \xi_2)=0 
\, .
\end{equation} 
The $\pm$ sign in Eq.~(\ref{con}) apply to Eq.~(\ref{constraint1}) and 
(\ref{constraint2}), respectively, 
provided that neither 
$\cos(2\theta_1)$  or $\cos(2\theta_2)$ equal 1. In the special case, where 
either $\cos(2\theta_1)=1$  or 
$\cos(2\theta_2)=1$, condition (\ref{con}) 
simplifies to $\cos(2\theta_1)\cos(2\theta_2)=0$ with no restrictions in the 
angles $\vartheta_1$, 
$\vartheta_2$, $\xi_1$ and $\xi_2$\footnote{An example of this special case can be found in Ref. \cite{Zou05}. However, this case does not yield a gate operation with insurance when a partial Bell measurement on the photons is performed.}. One particular way to fulfil 
these restrictions is to set
\begin{equation} \label{angels}
\xi_2=-{\textstyle {1 \over 2}} \pi \, , ~\xi_1=\vartheta_1=\vartheta_2=0  ~~ 
{\rm and} ~~\theta_1 = \theta_2 
= {\textstyle {1 \over 4}} \pi \, ,
\end{equation}
which corresponds to the choice
\begin{eqnarray} \label{angelencoding}
\ket{{\sf a}_{1}} &=& {\textstyle {1 \over \sqrt{2}}} ( \ket{{\sf x}_0} +  \, 
\ket{{\sf x}_1}) \, , 
\ket{{\sf a}_{2}} = {\textstyle {1 \over \sqrt{2}}} ( \ket{{\sf x}_0} -  \, 
\ket{{\sf x}_1}) \, , \nonumber \\
\ket{{\sf b}_{1}} &=& {\textstyle {1 \over \sqrt{2}}} ( \ket{{\sf y}_0} + 
\ket{{\sf y}_1}) \, , 
\ket{{\sf b}_{2}} = {\textstyle {{\rm i} \over \sqrt{2}}} ( \ket{{\sf y}_0} - 
\ket{{\sf y}_1}) \, .
\end{eqnarray}  
Therefore, the Bell states (\ref{base}) have the following form which satisfies 
the form of the mutually 
unbiased basis states (\ref{unbiased}),
\begin{eqnarray}
\ket{\Phi_1} &= & {\textstyle {1 \over 2}}{\rm e}^{{\rm i}\pi/4} \big[ \ket{\sf 
x_0y_0}-{\rm i}\ket{\sf 
x_0y_1}-{\rm i}\ket{\sf x_1y_0}+\ket{\sf x_1y_y} \big] \, , \nonumber \\
\ket{\Phi_2} &= & {\textstyle {1 \over 2}}{\rm e}^{-{\rm i}\pi/4} \big[ \ket{\sf 
x_0y_0}+{\rm i}\ket{\sf 
x_0y_1}+{\rm i}\ket{\sf x_1y_0}+\ket{\sf x_1y_y} \big] \, , \nonumber \\
\ket{\Phi_3} &= & {\textstyle {1 \over 2}}{\rm e}^{{\rm i}\pi/4} \big[ \ket{\sf 
x_0y_0}-{\rm i}\ket{\sf 
x_0y_1}+{\rm i}\ket{\sf x_1y_0}-\ket{\sf x_1y_y} \big] \, , \nonumber \\
\ket{\Phi_4} &= & -{\textstyle {1 \over 2}}{\rm e}^{-{\rm i}\pi/4} \big[ 
\ket{\sf x_0y_0}+{\rm i}\ket{\sf 
x_0y_1}-{\rm i}\ket{\sf x_1y_0}-\ket{\sf x_1y_y} \big] \, . 
\end{eqnarray}
To find out which gate operation the detection of the corresponding maximally 
entangled states (\ref{base}) 
combined with a subsequent 
absorption of the photon pair results into, we  decompose the encoded state 
(\ref{theencoding}) into a state of 
the form 
\begin{equation} \label{bla}
\ket{\psi_{\rm enc}} = {\textstyle {1 \over 2}}  \sum_i^4 \ket{\psi_i,\Phi_i} 
\end{equation} and determine the states $\ket{\psi_i}$ of the stationary qubits.
Using the notation 
\begin{equation} \label{CZ}
U_{\rm CZ} \equiv \ket{00}\bra{00} + \ket{01}\bra{01} + \ket{10}\bra{10} - 
\ket{11}\bra{11} 
\end{equation} 
for the controlled two-qubit phase gate and the notation  
\begin{equation}
Z_i(\phi) \equiv \ket{0}_{\rm ii}\bra{0}+{\rm e}^{-{\rm i}\phi}\ket{1}_{\rm 
ii}\bra{1}
\end{equation}
for the local controlled-Z gate on photon source $i$ \footnote{This gate can be 
 accomplished by 
applying a strongly detuned laser 
field for a certain time $t$.}, we find 
\begin{eqnarray} \label{totalsuccess}
\ket{\psi_1} &=& \exp \big({- \textstyle {1 \over 4}} {\rm i} \pi \big) \, Z_2 
\big(- {\textstyle {1 \over 2}} 
 \pi \big) \, Z_1 \big(- {\textstyle {1 \over 2}}  \pi \big) \, 
U_{\rm CZ} \, \ket{\psi_{\rm in}} \, , \nonumber \\  
\ket{\psi_2} &=& \exp \big({\textstyle {1 \over 4}} {\rm i} \pi \big) \, Z_2 
\big({\textstyle {1 \over 2}}  
\pi \big) \,Z_1 \big( {\textstyle {1 \over 2}}  \pi \big) \, U_{\rm CZ} \, 
\ket{\psi_{\rm in}} \, , \nonumber \\
\ket{\psi_3}&=& \exp \big({- \textstyle {1 \over 4}} {\rm i} \pi \big) \, Z_2 
\big(-{\textstyle {1 \over 2}}  
\pi \big) \, Z_1 \big({\textstyle {1 \over 2}}  \pi \big) \, 
U_{\rm CZ} \, \ket{\psi_{\rm in}} \, , \nonumber \\  
\ket{\psi_4} &=& -\exp \big({\textstyle {1 \over 4}} {\rm i} \pi \big) \, Z_2 
\big({\textstyle {1 \over 2}}  
\pi \big) \, Z_1 \big(-{\textstyle {1 \over 2}}  \pi \big) \, U_{\rm CZ} 
\, \ket{\psi_{\rm in}} \, . 
\end{eqnarray} 
From this we see that one obtains the CZ gate operation (\ref{CZ}) up to local 
unitary operations upon the 
detection of any of the four 
Bell states $\ket{\Phi_{i}}$.

\subsection{Gate implementation with insurance} \label{insurance}

When implementing distributed quantum computing with photons as flying qubits 
and single photon sources as 
stationary qubits, the problem 
arises that it is impossible to perform a complete Bell measurement on the 
photons using only linear optics 
elements. As it has been 
shown in the past \cite{Lutkenhaus99}, in the best case, one can only 
distinguish two of the four Bell states on average. 
The construction of 
efficient non-linear optical elements remains a difficult problem 
experimentally. 
The above described phase 
gate could therefore be operated at most with success rate ${1 \over 2}$. 

Therefore, we  choose the photon pair measurement basis $\{ |\Phi_i \rangle \}$ 
such that two of 
the basis states are maximally entangled 
while the other two basis states are product states. This is also naturally motivated from the fact that such a measurement basis can be 
easily implemented using a linear optics setup \cite{Braunstein95,Mattle96}. In the following, we choose
$\ket{\Phi_3}$ and  $\ket{\Phi_4}$ as in Eq.~(\ref{base}) and $\ket{\Phi_1}$ and 
$\ket{\Phi_2}$ as
\begin{eqnarray} \label{partialproduct}
\ket{\Phi_1} =\ket{{\sf a_1b_1}} \, , ~~ 
\ket{\Phi_2}=\ket{{\sf a_2b_2}} \, .  
\end{eqnarray}
As long as the states $\{ |\Phi_i \rangle \}$ constitute a MUB, the 
implementation of an 
eventually deterministic entangling
phase gate remains possible. In this way, we obtain quantum computing {\em with 
insurance}. In case of the failure of 
the gate implementation, 
a product state is detected and the system remains, up to a local phase gate, in 
the original qubit state. 
This means that the original 
qubit state (\ref{original}) can be restored and the described protocol can be 
repeated, thereby eventually 
resulting in the performance 
of the universal controlled phase gate (\ref{CZ}). The probability for the 
realisation of the gate operation 
within one step equals ${1 
\over 2}$ and the completion of the gate requires, on average, only two steps.

Let us now determine the conditions under which the states $\{ |\Phi_i \rangle 
\}$ constitute a MUB. Proceeding as 
above, we find that $\ket{\Phi_3}$ and $\ket{\Phi_4}$ are of form 
(\ref{unbiased}) if the angles 
$\vartheta_i$, $\xi_i$ and $\theta_i$ in 
Eq.~(\ref{ab}) fulfil, for example, Eq.~(\ref{angels}). In analogy to 
Eqs.~(\ref{constraint1}) and 
(\ref{constraint2}), we find that 
$\ket{\Phi_1}$ and $\ket{\Phi_2}$  belong to a MUB, if 
\begin{equation} \label{magic}
\big|\cos{\theta_1}\cos{\theta_2} \big| = \big |\cos{\theta_1}\sin{\theta_2} \big| 
= \big|\sin{\theta_1}\cos{\theta_2} \big| = \big|\sin{\theta_1}\sin{\theta_2} \big| = \textstyle{1 \over 2} \, ,
\end{equation}
which also holds for the parameter choice in Eq.~(\ref{angels}). Note that 
Eq.~(\ref{magic}) is general and applies 
for any product state detection in a Partial Bell basis measurement. One can 
easily verify with the above 
choice that the product states $\ket{\Phi_1}$ and $\ket{\Phi_2}$ are given by 
\begin{eqnarray}
\ket{\Phi_1}&=& {\textstyle {1 \over 2}} \big[ \ket{\sf x_0y_0}+\ket{\sf 
x_0y_1}+\ket{\sf x_1y_0}+\ket{\sf 
x_1y_1} \big] \, , \nonumber \\
\ket{\Phi_2}&=& {\textstyle {{\rm i} \over 2}} \big[ \ket{\sf x_0y_0}-\ket{\sf 
x_0y_1}-\ket{\sf 
x_1y_0}+\ket{\sf x_1y_1} \big] \, , 
\end{eqnarray} which fulfils (\ref{unbiased}).
This means that choosing the states $|{\sf a}_i \rangle$ and $|{\sf b}_i 
\rangle$ as in Eq.~(\ref{angelencoding}) allows to implement the gate operation 
(\ref{CZ}) with insurance\footnote{The term insurance 
was first coined by Bose {\em et al.} in the context of teleportation between 
atoms in different cavities with the aid of a backup atom 
in one of the cavities \cite{Bose99}.}. 

Finally, we determine the gate operations corresponding to the detection of a 
certain measurement outcome 
$|\Phi_i \rangle$. To do this, 
we decompose the encoded state (\ref{theencoding}) again into a state of the 
form (\ref{bla}). Proceeding as in 
the previous subsection we 
find
\begin{eqnarray}
\ket{\psi_1} &=&  \ket{\psi_{\rm in}} \, , \nonumber \\  
\ket{\psi_2} &=& -{\rm i} \, Z_2 \big( \pi \big) \, Z_2 \big( \pi \big) \, 
\ket{\psi_{\rm in}} \, , \nonumber 
\\
\ket{\psi_3}&=& \exp \big({- \textstyle {1 \over 4}} {\rm i} \pi \big) \, Z_2 
\big(-{\textstyle {1 \over 2}}  
\pi \big) \, Z_1 \big({\textstyle {1 \over 2}}  \pi \big) \, 
U_{\rm CZ} \, \ket{\psi_{\rm in}} \, , \nonumber \\  
\ket{\psi_4} &=& -\exp \big({\textstyle {1 \over 4}} {\rm i} \pi \big) \, Z_2 
\big({\textstyle {1 \over 2}}  
\pi \big) \, Z_1 \big(-{\textstyle {1 \over 2}}  \pi \big) \, U_{\rm CZ} 
\, \ket{\psi_{\rm in}} \, . 
\end{eqnarray} 
From this, we see that one obtains indeed the CZ gate operation (\ref{CZ}) up to 
local unitary operations upon 
the detection of either 
$\ket{\Phi_{3}}$ or $\ket{\Phi_{4}}$ as in (\ref{totalsuccess}). In case of the detection of the product 
states $\ket{\Phi_1}$ or 
$\ket{\Phi_2}$, the initial state 
can be restored with the help of one-qubit phase gates, which then allows to 
repeat the operation. 

It should be emphasized that there are other possible photon pair measurement bases that yield a 
universal two-qubit phase gate 
upon the detection of a Bell-state 
but where the original state is destroyed upon the detection of a product state 
(see e.g.~\cite{Zou05}). The 
reason is that, while the 
detected Bell states might result in a universal gate operation, the 
corresponding product states are not 
mutually unbiased and their 
detection erases the qubit state in the photon sources. To achieve the effect of 
an {\em insurance}, the photon pair
measurement basis should be 
chosen as described in this Section, as an example.

\subsection{Teleportation with insurance}\label{teleportation}
Here, we first show that the setup  can  be directly used to realise a quantum 
filter operation with insurance. 
This would lead us naturally to teleportation. Particularly, we describe a scheme 
for the implementation of the 
parity filter operation(see Chapter \ref{hummingbird})
\begin{equation} \label{filter}
P^{1\pm}_{\rm filter} = |00 \rangle \langle 00| \pm  |11 \rangle \langle 11| \, 
,
\end{equation}
which projects the initial qubit state $\ket{\psi_{\rm in}}$ with probability 
$|\alpha|^2+|\delta|^2$ onto the even-parity state,  
\begin{equation} \label{fin3}
\ket{\psi_{\rm fin}}= \big( \alpha \, \ket{00} \pm \delta \, \ket{11} 
\big)/\sqrt{|\alpha|^2+|\delta|^2} \, .
\end{equation} or 
\begin{equation}
P^{2\pm}_{\rm filter} = |01 \rangle \langle 01| \pm  |10 \rangle \langle 10| \, 
,
\end{equation}
which projects $\ket{\psi_{\rm in}}$ with probability 
$|\beta|^2+|\gamma|^2$ onto the odd-parity state,
\begin{equation} \label{fin4}
\ket{\psi_{\rm fin}}= \big( \beta \, \ket{01} \pm \gamma \, \ket{10} 
\big)/\sqrt{|\beta|^2+|\gamma|^2} \, .
\end{equation}

Again, this can be achieved if the photon states are detected in the
desired form $\frac{1}{\sqrt{2}}(\ket{\sf x_0y_0}\pm{\rm e}^{{\rm 
i}\delta_{1\pm}}\ket{\sf x_1y_1})$
or  $\frac{1}{\sqrt{2}}(\ket{\sf x_0y_1}\pm{\rm e}^{{\rm 
i}\delta_{2\pm}}\ket{\sf x_1y_0})$ where $\delta_{i\pm}$ is any arbitrary phase 
angle. Looking
again at Eq. (\ref{base2}), all the
basis states $\ket{\Psi_i}$ will be the desired form by setting
\begin{equation}\label{teleportationcondition}
\sin(\theta_1\mp\theta_2)=\cos(\theta_1\pm\theta_2)=0 \, ,  
\end{equation}and
\begin{equation}
\sin(\vartheta_1\pm \vartheta_2+\xi_1
\pm \xi_2)=0 \, ,
\end{equation} provided that  $\sin(2\theta_1)\sin(2\theta_2) \neq 0$. In the 
special case where $\sin(2\theta_1)\sin(2\theta_2) =0$, then 
the only constraint would be $\sin(2\theta_1)=\sin(2\theta_2)=0$ with no 
restriction on the angles $\vartheta_1$, $\vartheta_2$, $\xi_1$
and $\xi_2$.
However now, we redefine $\ket{\Psi_1}$ and $\ket{\Psi_2}$ as product states 
defined in Eq.~(\ref{partialproduct}), collectively forming  
a partial Bell-measurement basis. Combining with the condition of insurance in
Eq.~(\ref{magic}), we see that the choice
\begin{equation}
\theta_1=\theta_2=\frac{\pi}{4}, \, \vartheta_1=\vartheta_2=\xi_1=\xi_2=0,
\end{equation} allows us to implement a parity filter with
insurance. This is the choice where
\begin{eqnarray} \label{demonencoding}
\ket{{\sf a}_{1}} &=& {\textstyle {1 \over \sqrt{2}}} ( \ket{{\sf x}_0} +  \, 
\ket{{\sf x}_1}) \, , \,
\ket{{\sf a}_{2}} = {\textstyle {1 \over \sqrt{2}}} ( \ket{{\sf x}_0} -  \, 
\ket{{\sf x}_1}) \, , \nonumber \\
\ket{{\sf b}_{1}} &=& {\textstyle {1 \over \sqrt{2}}} ( \ket{{\sf y}_0} + 
\ket{{\sf y}_1}) \, , \,
\ket{{\sf b}_{2}} = {\textstyle {1 \over \sqrt{2}}} ( \ket{{\sf y}_0} - 
\ket{{\sf y}_1}) \, .
\end{eqnarray}  which yields
\begin{eqnarray} \label{devilencoding}
\ket{\Phi_1} &=& {\textstyle {1 \over 2}} ( \ket{\sf x_0y_0}+\ket{\sf 
x_0y_1}+\ket{\sf x_1y_0}+\ket{\sf x_1y_1}) \, , \nonumber \\
\ket{\Phi_2} &=& {\textstyle {1 \over 2}} ( \ket{\sf x_0y_0}-\ket{\sf 
x_0y_1}-\ket{\sf x_1y_0}+\ket{\sf x_1y_1}) \, , \nonumber \\
\ket{\Phi_3} &=& {\textstyle {1 \over \sqrt{2}}} ( \ket{\sf x_0y_0}-\ket{\sf 
x_1y_1}) \, , \nonumber \\
\ket{\Phi_4} &=& -{\textstyle {1 \over \sqrt{2}}} ( \ket{\sf x_0y_1}-\ket{\sf 
x_1y_0}) \, .
\end{eqnarray}

To see this, we again decompose the input state (\ref{theencoding}) again into a 
state of the form (\ref{bla}). 
Proceedings as in the previous subsection, we 
find
\begin{eqnarray}
\ket{\psi_1} &=&  \ket{\psi_{\rm in}} \, , \,
\ket{\psi_2} = Z_2 \big( \pi \big) \, Z_2 \big( \pi \big) \, \ket{\psi_{\rm in}} 
\, , \nonumber \\ 
\ket{\psi_3}&=& P^{1-}_{\rm filter}  \ket{\psi_{\rm in}} \, , \,
\ket{\psi_4} = -P^{2-}_{\rm filter}  \ket{\psi_{\rm in}} \, .  
\end{eqnarray} 

One application of the quantum parity filter (\ref{filter}) is {\em 
teleportation with insurance}, which now 
requires less resources than previously proposed schemes \cite{Bose99}. Suppose, 
a given state $\alpha \, 
|0\rangle + \beta \, |1 \rangle$ of source A is to be teleported to another 
target source B prepared in ${1 
\over \sqrt{2}} (|0\rangle + |1 \rangle)$. Application of the quantum filter to 
the combined state of the two 
sources, then ultimately transfers this state into $\alpha \, |00 \rangle + \beta \, |11 
\rangle$ or $\alpha \, |01 
\rangle + \beta \, |10 \rangle$. In order to complete the teleportation, the 
state of B should be disentangled 
from the state of source A without revealing the coefficients $\alpha$ and 
$\beta$.  This can be 
achieved by measuring  source A on the basis  given by $|\pm \rangle \equiv {1 
\over \sqrt{2}} (|0\rangle \pm |1 \rangle)$. Depending on 
the outcome of this measurement, a further local operation on the state of B 
might be required. 

We proceed next with the description of possible ways to realise the photon encoding.

\section{Entangled Atom-Photon generation from  Atom-Cavity Systems} 
\label{photonencoder}

\subsection{Introduction}
The purpose of this subsection is to show how a highly efficient encoder of a 
stationary qubit to the flying qubit can be performed in 
the context of an atom-cavity system. We
use an 
atom-cavity photon gun \cite{Law97,Kuhn99,Duan03,Saavedra00, Gheri98,Ciaramicoli01,Maurer04,Keller04} in which either Raman transfer with 
detuning or the 
so-called stimulated Raman adabiatic passage (STIRAP) \cite{Bergmann98,Oreg84}  
is employed. Gheri {\em et al.} \cite{Gheri98} were the first to exploit a limited 
version of encoding with this technique.  Here, we review by 
following closely 
the formalism  of  Duan {\em et al.} \cite{Duan03}, how a 
deterministic single photon encoder with a single or double
$\Lambda$-type level configuration trapped in a cavity can be implemented  in principle. We
illustrate this using the STIRAP method as an example although the Raman 
transfer method using large detuning resulting in adiabatic elimination of the 
excited 
state will yield the same general conclusion.  As in 
\cite{Gheri98,Saavedra00}, we make use of the superposition 
principle and the fact that no-cross couplings can occur between different 
subspaces as described by the system Hamiltonian to show how  efficient encoding to the 
photon 
is possible.

\subsection{Photon gun encoder} \label{encodertrick}
\begin{figure} 
\begin{center}
\psfrag{O}{$\Omega$}
\psfrag{v0}{$v_0$}
\psfrag{v1}{$v_1$}
\psfrag{e0}{$e_0$}
\psfrag{e1}{$e_1$}
\psfrag{u0}{$u_0$}
\psfrag{u1}{$u_1$}
\psfrag{g}{$g$}
\includegraphics{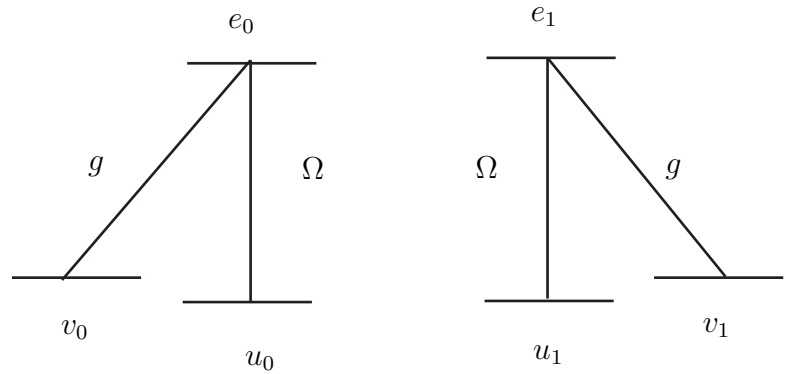}
\end{center}
\caption{Schematic view of a single photon polarisation encoder and level 
configuration of the atomic 
structure containing the qubit.} \label{polencoder} 
\end{figure}

We consider the situation of a degenerate double $\Lambda$-type atom trapped in an optical 
cavity. Specifically, each $\Lambda$ system $i$ where 
$i=0,1$ consists of two ground states $\ket{u_i}$ and $\ket{v_i}$ as well as an 
excited state $\ket{e_i}$ as shown in Fig.~\ref{polencoder}. We assume that the 
atom 
is initially prepared in the state
\begin{equation}
\ket{\psi(t=0)}=(\alpha_0\ket{u_0}+\alpha_1\ket{u_1})\ket{0}_{\rm {\rm 
cav}}\ket{{\rm vac}},
\end{equation} where  $\ket{0}_{\rm {\rm cav}}$ and $\ket{{\rm vac}}$ 
denotes the cavity vacuum field and the 
external photon vacuum field respectively.

Lasers with time-dependent Rabi 
frequency 
$\Omega_i(t)$ with frequency 
$\omega_L$ resonantly couple levels $u_i$ to $e_i$. The cavity resonantly couples levels $v_i$ to $e_i$ with coupling 
strength 
$g_i$ with the cavity resonant frequency $\omega_{{\rm cav}}$. 
In addition, the cavity field associated  with creation(destruction) operator 
$a_i^{\dagger}(a_i)$ corresponding to both $\Lambda$ systems $i$ 
is required to be orthogonal (in this case, in polarisation) with respective decay rates $\kappa_i$. The 
external photon fields can be described by 
creation(destruction) operators $b_i(\omega)^{\dagger}(b_i(\omega))$ with 
mode $i$ and frequency $\omega$. They satisfy the commutation rules 
$[b_i(\omega_1),b^{\dagger}_j(\omega_2)]=\delta_{ij}\delta_{\omega_1\omega_2}$. 

To solve the evolution of the system, it is convenient to borrow the concept of 
quantum jump formalism \cite{Hegerfeldt93,Dalibard92,Carmichael93} with a 
non-event (i.e. no spontaneously emitted photon emitted outside the cavity mode) evolution 
described by a non-Hermitian Hamiltonian. 

The conditional non-Hermitian Hamiltonian ($\hbar=1$), in the case where no 
photon 
is spontaneously emitted outside the cavity mode, in the interaction picture 
with respect to the free evolution is given by 
\cite{Tregenna02,Duan03}
\begin{equation} \label{Hencoder}
H(t)=H_{\rm atom-cavity}(t)+H_{\rm cavity-env}(t)
\end{equation} where 
\begin{equation}
H_{\rm atom-cavity}(t)=\sum_i 
\Omega_i(t)\ket{e_i}\bra{u_i}+g_ia_i\ket{e_i}\bra{v_i}+{\rm H.c.}-{\rm 
i}\frac{\Gamma_i}{2}\ket{e_i}\bra{e_i} \, , 
\end{equation} 
\begin{equation}
H_{\rm cavity-env}(t)=\sum_i {\rm i}\sqrt{\frac{\kappa_i}{2 
\pi}}\int_{-\omega_b}^{\omega_b} d\omega 
a_i^{\dagger}b_i(\bar{\omega}){\rm e}^{-{\rm i}\omega 
t}+{\rm H.c.} \, ,
\end{equation} and $\bar{\omega}=\omega_{{\rm cav}}+\omega$ and $\omega_b$ is 
the bandwidth. We assume $\omega_b$ to be relatively large but it has to be 
much smaller than $\omega_{\rm cav}$. Within this bandwidth, the coupling 
between the free field and the cavity mode is approximately 
constant and given by $\sqrt{\frac{\kappa_i}{2 \pi}}$. The cavity 
coupling $g_i$, which derives from the quantisation of the vacuum field in the 
cavity, is a function of the cavity spatial mode function and the relevant 
dipole transition element \cite{Chen04}. For the purpose of illustration, they 
can be set to be real and equal without loss of generality. We also see 
the effect of the damping factor $\Gamma_i$ on the excited 
state $\ket{e_i}$ due to the possibility of spontaneous emission.

We now define  the states $\ket{D_i(t)}$ and 
$\ket{B_i(t)}$ such that
\begin{eqnarray}\label{dark}
\ket{D_i(t)}&=&\cos\vartheta_i(t) \ket{u_i}\ket{0}_{\rm {\rm 
cav}}-\sin\vartheta_i(t)\ket{v_i}a_i^{\dagger}\ket{0}_{\rm {\rm 
cav}} \nonumber \, , \\
\ket{B_i(t)}&=&\sin\vartheta_i(t) \ket{u_i}\ket{0}_{\rm {\rm 
cav}}+\cos\vartheta_i(t)\ket{v_i}a_i^{\dagger}\ket{0}_{\rm {\rm 
cav}}\, ,
\end{eqnarray} where
\begin{equation}
\cos\vartheta_i(t)=\frac{g_i}{\sqrt{g_i^2+\Omega_i^2(t)}}.
\end{equation} One can easily see that $\ket{D_i(t)}$ and $\ket{B_i(t)}$ are 
orthogonal to each other. An important point, as 
pointed out by Duan {\em et. al.} is that the spatial dependence of 
$\cos\vartheta_i(t)$ can be made to vanish provided that 
$\Omega_i(t)$ and $g_i$ share the same cavity spatial mode structure. This can 
be accomplished by 
collinear pumping where the external pumping laser couples to a similar spatial 
cavity mode of a different polarisation relative to the 
one used in generating the cavity photon that subsequently leaks out and is 
encoded to the atomic state. This suggests that the atom need 
not really be cooled to the 
Lamb-Dicke limit for operation which removes a huge experimental challenge. 
However, the same authors point out that cooling is still important to maintain 
a long trap lifetime of the atom in the cavity.

When $\Gamma_i$ and $\kappa_i$ vanish, $\ket{D_i(t)}$ is an exact eigenstate 
with 
zero eigenvalue of the Hamiltonian $H(t)$ and hence, 
known as a dark state. A system initially in the dark state 
always stays in the dark state provided that 
the adiabatic following condition is fulfilled. This is the essence of the 
STIRAP process, which allows for robust coherent state transfer by remaining always in a
dark state. In the subspace defined by the 
Hamiltonian $H(t)$ and making the assumption that 
only one photon excitation can be put to the external field, the general state 
after time $t$ is given by
\begin{eqnarray}\label{conditionalstate}
\ket{\psi(t)}&=&\sum_i \alpha_i ((c_{D_i}(t)\ket{D_i(t)}+c_{B_i}(t)\ket{B_i(t)}+ 
\nonumber \\
&&c_{e_i}(t)\ket{e_i}\ket{0}_{\rm cav})\ket{{\rm vac}}+ \ket{v_i}\ket{0}_{\rm cav} 
\int_{-\omega_b}^{\omega_b} d\omega s(\omega,t)_i\ket{\bar{\omega}}_i)\, ,
\end{eqnarray} where $\ket{\bar{\omega}}_i=b_i^\dagger(\bar{\omega})\ket{\rm 
vac}$. Note that the last term with the external photon excitation is 
associated with the state $\ket{v_i}\ket{0}_{\rm 
cav}$. This can be inferred by looking at the Hamiltonian given in Eq. 
(\ref{Hencoder}). An external photon 
excitation of mode $i$ comes only through the 
annihilation of a cavity photon $a_i^{\dagger}\ket{0}_{\rm cav}$ from $H_{\rm 
cavity-env}(t)$ which was created accompanying the 
projection of the atomic state to $\ket{v_i}$ from 
$H_{\rm atom-cavity}(t)$.
Up to now, we have not made any adiabatic approximations. We can calculate the 
probability $P_{\rm cond}(t)=|| \,\ket{\psi(t)}||^2$ of the system  
evolving according to $H(t)$. It is this evolution that yields the photon in 
the cavity mode which leaks out subsequently. Otherwise, spontaneous emission 
occurs and this takes place 
with the probability $P_{\rm spon}(t)=1-P_{\rm 
cond}(t)$. 
Explictly, this is given by
\begin{equation}
P_{\rm spon}(t)=1-\sum_i |\alpha_i|^2(|c_{D_i}(t)|^2+|c_{B_i}(t)|^2+|c_{e_i}(t))|^2+\int_{-\omega_b}^{\omega_b} d\omega 
|s(\omega,t)_i|^2) 
\end{equation}
When $P_{\rm spon}(t)$ is small or close to $0$, the 
system can yield an effective photon source on demand. 
Now, the adiabatic condition, which we take as an ansatz (see further discussion by Duan {\em et al.} \cite{Duan03}), results in a very slow change of 
$\cos\vartheta_i(t)$ which implies that the time derivatives of 
$\ket{D_i(t)}$ and $\ket{B_i(t)}$ vanish. This condition also implies that the 
population of 
$\ket{B_i(t)}$ and $\ket{e_i}$ is virtually zero and can be effectively 
neglected.  This allows us to 
simplify the calculation of the evolution of $\ket{\psi(t)}$ with the Schrodinger's equation by calculating the 
two time-dependent 
coefficients given by
\begin{eqnarray}
\dot{c}_{D_i}(t)&=&-\sqrt{\frac{\kappa_i}{2\pi}}\sin\vartheta_i(t)\int_{-\omega_
b}^{\omega_b} d\omega 
s(\omega,t)_i {\rm e}^{-{\rm i}\omega t} \, , \nonumber \\
\dot{s}(\omega,t)_i&=& \sqrt{\frac{\kappa_i}{2\pi}} 
c_{D_i}(t) \sin\vartheta_i(t) {\rm e}^{{\rm i}\omega t} \, .
\end{eqnarray}
The solutions of the two coefficients are given approximately by
\begin{eqnarray}\label{cgsolution}
c_{D_i}(t)&=& \exp(-\frac{\kappa_i}{2} \int_0^{t} dt' \sin^2 \vartheta_i(t'))  
\, ,
\nonumber \\
s(\omega,t)_i&=& \sqrt{\frac{\kappa_i}{2\pi}}\int_0^t dt' {\rm e}^{{\rm 
i}\omega t'} c_{D_i}(t') \sin\vartheta_i(t') \, .  
\end{eqnarray} We have used the Markovian approximation in the process where the 
limits of integration of $\omega$ is artificially 
extended to $-\infty$ and $\infty$ due to the large bandwidth $\omega_b$ to 
yield a delta function.
To have generated an external photon with unit probability by time $t=\tau$, we 
should start with 
$c_{D_i}(0)=1$ at time $t=0$  and end up with $c_{D_i}(\tau)=0$  by looking 
at  Eq. (\ref{conditionalstate})  where $\tau$ is 
chosen to be a 
characteristic time in the tail end of the pumping laser pulse when the 
amplitude is near zero.
This is fulfilled by choosing a large $\tau$ and(or) increasing the laser Rabi 
frequency $\Omega_i(t)$ by looking at Eq. (\ref{cgsolution}). Note that  $\kappa_i^{-1}$ 
must be smaller than $\tau$. Otherwise, adiabatic following will imply coherent 
return to the initial state with $c_{D_i}(\tau)=1$ with no external photon 
generated\cite{Kuhn99} if $\kappa_i^{-1} \gg \tau$.
We can define the pulse shape by the Fourier transform of the spectral evelope
\begin{equation}
f(t,\tau)_i=\frac{1}{\sqrt{2\pi}}\int_{-\infty}^{\infty} d\omega 
s(\omega,\tau)_i{\rm e}^{-{\rm 
i}\omega t}.
\end{equation}
From Eq. (\ref{cgsolution}), we see that the pulse shape is  given by 
\begin{equation}\label{magicpulse}
f(t,\tau)_i=\sqrt{\kappa_i} \sin\vartheta_i(t)c_{D_i}(t)
\end{equation}

For simplicity, we assume all parameters related to different 
$i$ to be the same i.e. $(\Omega_i(t)=\Omega(t), 
g_i=g,\Gamma_i=\Gamma)$. We also assume that after a photon is generated, we 
recycle the state of the system from $\ket{v_i}$ to 
$\ket{u_i}$. In the ideal limit of strong cavity 
coupling $g^2\gg \Gamma \kappa$ and adiabatic following, we are able to perform 
a deterministic 
mapping of the form
\begin{equation} \label{effectiveencoding}
\sum_i \alpha_i\ket{u_i}\ket{0}_{\rm {\rm cav}}\ket{{\rm vac}} 
\to \sum_i 
\alpha_i\ket{u_i}\ket{0}_{\rm {\rm cav}}\int_{-\omega_b}^{\omega_b} d\omega 
s(\omega,t)_i b^{\dagger}_i(\bar{\omega})\ket{\rm vac}
\end{equation}
This implies that we can encode the state of the atoms to the 
externally generated photons with a STIRAP process and was first demonstrated by 
Gheri {\em et. al.} \cite{Gheri98} in the regime of Raman transfer with a large detuning.

Now, all these above calculations invoke the crucial assumption of adiabatic 
following. The adiabatic following condition is 
well defined in the limit where $\Gamma_i$ and $\kappa_i$ vanish. 
Specifically, the evolution time $\tau$ must be longer than 
the inverse of the frequency splitting gap between the dark state $\ket{D_i(t)}$ 
and the rest of 
the eigenstates \cite{Kuhn99}. In our case, the gap $\delta$ is 
$\sqrt{g_i^2+\Omega_i^2(t)}$ and the adiabatic condition is given by 
\begin{equation}
(g_i^2+\Omega_i^2(t))\tau^2 \gg 1.
\end{equation} Note that this condition does not imply adiabatic following  when 
$\Gamma_i$ and $\kappa_i$ is non-zero due to the 
atom-cavity system coupling to an infinite continum of modes. Therefore, a 
fuller description necessitates numerical simulation. 
Duan {\em et. al.} \cite{Duan03} have already performed such simulations and showed that 
spontaneous emission loss is negligible in the limit of 
strong coupling where $g^2/\kappa \Gamma \gg 1$. Empirically, $P_{\rm 
spon}$ scales as $\kappa \Gamma /4 g^2$. Furthermore, it was found 
that for strong coupling, the analytically calculated pulse shape $f(t,\tau)_i$ 
based on the adiabatic following ansatz agrees very well with 
numerical simulations. If we restrict ourselves to the less general case of a 
single $\Lambda$ subsystem characterised for example by $\alpha_i=0$, we recover 
exactly the usual single photon gun \cite{Law97,Kuhn99}.

Finally, we consider the case where the same laser driving pulse is offset by a 
time $T_i$ and we operate in the single $\Lambda$ subsystem, dropping all 
subscripts corresponding to subsystem for readability. For convenience, we can 
define the effective photon creation operator as $B^\dagger (t_i, \tau)$
\begin{equation}
B^\dagger (t, \tau)=\int_{-\omega_b}^{\omega_b} d\omega {\rm e}^{{\rm i} \omega 
t}s(\omega,\tau)b^\dagger(\bar{\omega}).
\end{equation}
 We find that when $|t_i-t_j|\gg \tau$ \cite{Gheri98,Saavedra00},
\begin{equation}
\left[B^\dagger (t_i, \tau),B^\dagger (t_j, \tau) \right]\to 0.
\end{equation}
This implies that photons created by laser driving pulse offset by a time 
separation much larger than the driving pulse duration $\tau$ can be considered 
to be in different modes and hence orthogonal. In fact,$ \left[B^\dagger (t_i, 
\tau),B^\dagger (t_j, \tau) 
\right] $ which depends essentially on the temporal overlap of two identical 
pulses offset by $|t_i-t_j|$ 
approximately decay exponentially with the ratio $\frac{|t_i-t_j|}{\tau}$ and 
thus typically, a 
time-separation $|t_i-t_j|$ of the order of $\tau$ might already be sufficient 
to achieve the photon 
orthogonality condition \cite{Saavedra00}.   This leads us to the possibility 
of time-bin encoding in which photons are created by an early or late driving pulse 
with the appropriate time 
separation 
dependent on the initial ground state of the atom. 

\begin{figure}
\psfrag{O}{$\Omega$}
\psfrag{v1}{$v_1$}
\psfrag{e1}{$e_1$}
\psfrag{u0}{$u_0$}
\psfrag{u1}{$u_1$}
\psfrag{g}{$g$}
\begin{center}
\includegraphics{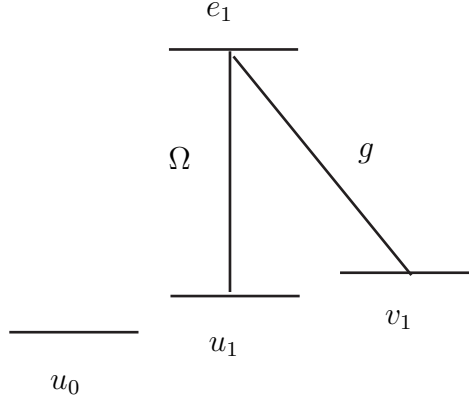} 
\end{center}
\caption{Schematic view of a single photon time-bin encoder and level 
configuration of the atomic structure 
containing the qubit.} \label{binencoder}
\label{photongun2}
\end{figure}

One way to implement such an encoding is to 
first swap the atomic states $\ket{u_0}$ and $\ket{u_1}$. Then a laser pulse 
with 
increasing Rabi frequency should excite the $u_1$-$e_1$ transition (see 
Fig.~\ref{binencoder}) at time $t_0$ for 
example. This transfers the atom into the state $\ket{v_1}$ 
and places one excitation into the field of the strongly coupled optical cavity, 
if the atom was initially 
prepared in $|u_0 \rangle$. The 
photon then leaks out through the outcoupling mirror of the resonator. The 
encoding operation, which is 
feasible with present technology \cite{Duan03}, is completed by transfering 
$|v_1 \rangle$ back into $|u_1 
\rangle$, swapping again the states 
$|u_0 \rangle$ and $|u_1 \rangle$ and repeating the above described photon 
generation process at a later time 
$t_1$. The above process therefore describes the mapping of a form
\begin{equation}
\sum_i \alpha_i\ket{u_i}\ket{0}_{\rm cav}\ket{{\rm vac}} \to \sum_i 
\alpha_i\ket{u_i} \ket{0}_{\rm cav} 
B^\dagger(t_i,\tau)\ket{{\rm vac}}
\end{equation} In general, time-bin encoding which requires a simpler energy level structure compared to polarisation encoding may find 
realisations in systems such as quantum dots and NV color centers where a double $\Lambda$-type configuration may  not be easily found.  
We now proceed 
to the description of photon pair measurement.

\section{Measurements on photon pairs}\label{photonmeasurement}

We give 2 examples of measurements on the photon pairs based on the concrete 
choice given in 
(\ref{angelencoding}). The first method is suitable for polarisation encoded 
photon pairs. The second is 
suitable for dual-rail encoded photon pairs. In general, depending on the 
initial choice of encoding, 
conversion between encodings may be required. For example, one might need to 
convert time-bin encoding to 
either polarisation or dual-rail encoding and method 1 or 2 can be used 
respectively for photon pair 
measurements.

It is worth mentioning that our scheme has the same robustness from slow and 
unknown phase fluctutations along 
the photon paths due to the same reason outlined for example in Ref. 
\cite{Simon03}. Due to the fact that we use 
coincidence measurement for our Bell-state detection, any slow phase error on 
the photons contributes only to 
a global phase factor in the stationary qubits. This also implies that our 
scheme does not require 
interferometric stability as is the case of most schemes requiring coincidence 
detection. We first describe a canonical Bell-state 
measurement in polarisation encoding.

\subsection{Canonical Bell-state measurement}

Bell-state measurement on a photon pair is an important tool used widely in 
quantum information processing with photons. It is crucial for quantum 
teleportation \cite{Bennett93}, quantum 
dense coding \cite{Bennett92} as well as entanglement swapping 
\cite{Zukowski93}. Recently, Browne and Ruldoph 
\cite{Browne05} have exploited Bell-state measurement for the efficient 
construction of a photonic cluster 
state. However, a complete Bell 
measurement cannot be realised with unit success probability in a purely linear 
optics based setup \cite{Lutkenhaus99}. This can be thought as a main limitation 
to purely linear optics 
based quantum computation. We show here a canonical example of how a partial 
Bell measurement can be realised with the aid of a beam splitter 
\cite{Braunstein95}. We recall the basis (\ref{completeBellintro})
states of a complete Bell basis as
\begin{eqnarray} 
&&\ket{\Phi^\pm}=\frac{1}{\sqrt{2}}(a_{1,h}^\dagger a_{2,v}^\dagger \pm 
a_{1,v}^\dagger a_{2,h}^\dagger ) \ket{0}_{\rm vac} \, , \nonumber \\
&&\ket{\Psi^\pm}=\frac{1}{\sqrt{2}}(a_{1,h}^\dagger a_{2,h}^\dagger \pm 
a_{1,v}^\dagger a_{2,v}^\dagger ) \ket{0}_{\rm vac} \, .
\end{eqnarray} Here, $a_{i,\lambda}^\dagger$ refers to a photon creation 
operator for spatial mode $i$ with polarisation $\lambda$. These 2-photon Bell 
states are sent, one in each input arm, into a 50:50 beam splitter which is 
described by the matrix $B(\frac{1}{2},1)$ in Chapter \ref{firework}. We fix the 
convention that the spatial modes defined by the two 
input ports are defined as spatial modes 1 and 2 and that defined by the two 
output ports are 
defined as spatial mode 3 and 4. It can be shown that the basis Bell states at 
the input will transform to the output ports as
\begin{eqnarray} 
&&\ket{\Phi^+} \to \frac{1}{\sqrt{2}}(a_{4,v}^\dagger 
a_{4,h}^\dagger-a_{3,v}^\dagger a_{3,h}^\dagger)\ket{0}_{\rm vac} \, , \nonumber 
\\
&&\ket{\Phi^-} \to \frac{1}{\sqrt{2}}(a_{3,v}^\dagger 
a_{4,h}^\dagger-a_{4,v}^\dagger a_{3,h}^\dagger)\ket{0}_{\rm vac} \, , \nonumber 
\\
&&\ket{\Psi^+} \to  \frac{1}{2\sqrt{2}}((a_{4,v}^\dagger)^2+ 
(a_{4,h}^\dagger)^2-(a_{3,v}^\dagger)^2- (a_{3,h}^\dagger)^2)\ket{0}_{\rm vac} 
\, ,
\nonumber \\
&&\ket{\Psi^-} \to  \frac{1}{2\sqrt{2}} ((a_{4,v}^\dagger)^2- 
(a_{4,h}^\dagger)^2-(a_{3,v}^\dagger)^2+ (a_{3,h}^\dagger)^2)\ket{0}_{\rm vac} 
\, .
\end{eqnarray} We see that $\ket{\Phi^\pm}$ is indicated by detecting photons of 
different polarisations in the same and different output ports respectively. 
Unfortunately, $\ket{\Psi^\pm}$ cannot be distinguished by simple photon 
detection and hence, the simple beam splitter cannot implement a complete Bell 
measurement with unit efficiency. However, the following product states, 
$\frac{1}{\sqrt{2}}(\ket{\Psi^+}+\ket{\Psi^-})=a_{1,h}^\dagger a_{2,h}^\dagger 
\ket{0}_{\rm vac}$ and 
$\frac{1}{\sqrt{2}}(\ket{\Psi^+}-\ket{\Psi^-})=a_{1,v}^\dagger a_{2,v}^\dagger 
\ket{0}_{\rm vac}$ transform as
$\frac{1}{2}((a_{3,h}^\dagger)^2-(a_{4,h}^\dagger)^2)\ket{0}_{\rm vac}$ and 
$\frac{1}{2}((a_{3,v}^\dagger)^2-(a_{4,v}^\dagger)^2)\ket{0}_{\rm vac}$ 
respectively. The output states in this case are distinguishable. Therefore, the 
measurement performed in this 
case is a partial 
Bell measurement with two Bell states and two product states constituting  the 
measurement basis. 

\subsection{Measurement for polarisation encoded photon pair}

We have shown that sending two polarisation encoded photons through the 
different input ports of a 50:50 beam 
splitter together with 
polarisation sensitive measurements in the $\ket{\sf h}/\ket{\sf v}$-basis in 
the output ports would result in 
a measurement of the 
states ${\textstyle {1 \over {\sqrt 2}}}(\ket{\sf hv}\pm\ket{\sf vh})$, 
$\ket{\sf hh}$ and $\ket{\sf vv }$. To 
measure the states $\ket{\Phi_i}$ defined in Section \ref{insurance} or \ref{teleportation}, we 
therefore  proceed as shown in Fig.~\ref{moon2}(a) and 
perform the mapping 
$U_1 = |{\sf h} \rangle 
\langle {\sf a_1}| + |{\sf v} \rangle \langle {\sf a_2}|$ and $U_2 = |{\sf h} 
\rangle 
\langle {\sf b_1}| + |{\sf v} \rangle \langle {\sf b_2}|$ on the photon coming 
from source $i$.  For the states defined Section \ref{insurance}, using Eq.~(\ref{angelencoding}), we see 
that this 
corresponds to the single qubit rotations 
\begin{eqnarray}
U_1 &=&  {\textstyle {1 \over \sqrt{2}}} \, \big[ \, |{\sf h} \rangle \big( 
\langle {\sf h}| + \langle {\sf 
v}| \big) +  |{\sf v} \rangle 
\big( \langle {\sf h}| - \langle {\sf v}| \big) \, \big] \, , \nonumber \\
U_2 &=&  {\textstyle {1 \over \sqrt{2}}} \, \big[ \, |{\sf h} \rangle \big( 
\langle {\sf h}| + \langle {\sf 
v}| \big) - {\rm i} \, |{\sf 
v} \rangle \big( \langle {\sf h}| -  \langle {\sf v}| \big) \, \big] \, .
\end{eqnarray}
After leaving the beam splitter, the photons should be detected in the $\ket{\sf 
h}/\ket{\sf v}$-basis. A 
detection of two ${\sf h}$ (${\sf v}$) polarised photons indicates a measurement 
of $|\Phi_1 \rangle$ 
($|\Phi_2 \rangle$). Finding two photons of different polarisation in the same 
(different) detectors 
corresponds to a detection of $|\Phi_3 \rangle$ ($|\Phi_4 \rangle$). 

\subsection{Measurement for dual-rail encoded photon pair}

\begin{figure}
\begin{minipage}{\columnwidth}
\begin{center}
\resizebox{\columnwidth}{!}{\rotatebox{0}{\includegraphics{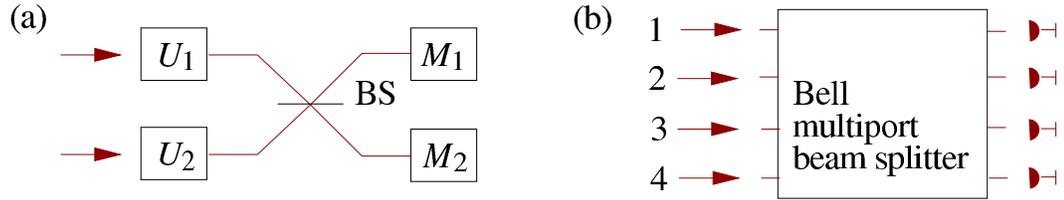}}}
\end{center}
\vspace*{-0.5cm} 
\caption{Linear optics networks for the realisation of a measurement of the 
basis states $\ket{\Phi_i}$ defined in Sections 
(\ref{insurance},\ref{teleportation}) after 
encoding the photonic qubits 
in the polarisation degrees of two photons (a) or into four different spatial 
photon modes (b) involving 
either a beam splitter (BS) or a 
$4 \times 4$ Bell multiport beam splitter.} \label{moon2} 
\end{minipage}
\end{figure}

Alternatively, one can redirect the generated photons ( for example, if the photons are time-bin encoded) to the different input 
ports of a $4 \times 4$ symmetric 
multiport beam splitter as shown in Fig.~\ref{moon2}(b). A symmetric multiport redirects 
each incoming photon with 
equal probability to any of the possible output ports and can therefore be used 
to erase the which-way 
information of the incoming photons as we have mentioned in Chapter \ref{firework} and \ref{fusion}. If $a_n^\dagger$ ($b_n^\dagger$) 
denotes 
the creation operator for a 
photon in input (output) port $n$, 
the effect of the multiport can be summarised as 
\begin{equation} \label{scatter}
a_n^\dagger \to \sum_m U_{mn} b_m^\dagger, 
\end{equation}where  $U_{mn}$ is the probability amplitude to redirect a photon 
from the $n$th input port to the $m$th
output port.  For the implementation of either 
a two qubit universal phase gate or 
a parity filter, one should direct the input
$\ket{\sf x_0}$ ($\ket{\sf x_1}$) photon from source 1 to input port 1 (3) and 
to direct a $\ket{\sf y_0}$  
($\ket{\sf y_1}$) photon from source 2 to input port 2 (4). If 
$\ket{\rm vac}$ denotes the state with no photons in the setup, this results in 
the conversion $|{\sf x_0y_0} 
\rangle \to a_1^\dagger 
a_2^\dagger \, \ket{{\rm vac}}$, $|{\sf x_0y_1} \rangle \to a_1^\dagger 
a_4^\dagger \, \ket{{\rm vac}}$, 
$|{\sf x_1y_0} \rangle \to a_2^\dagger 
a_3^\dagger \, \ket{{\rm vac}}$ and $|{\sf x_1y_1} \rangle \to a_3^\dagger 
a_4^\dagger \, \ket{{\rm vac}}$. 
This conversion should be 
realised such that the photons enter the multiport at the same time. 
For two-qubit universal gate implementation, $U_{mn}$ is given by
\begin{equation}
U_{mn} = {\textstyle {1 \over 2}} \, {\rm i}^{(m-1)(n-1)}.
\end{equation} In such a case, the multiport is also known as a Bell multiport 
which was introduced in Chapter \ref{firework}.
Using Eq.~(\ref{scatter}), one can show that the network transfers 
the basis states $\ket{\Phi_i}$, with the choice Eq. (\ref{angelencoding}) as
\begin{eqnarray} 
|\Phi_1 \rangle & \to &  {\textstyle {1 \over 2}} \, \big( b_1^{\dagger \, 2}  - 
b_3^{\dagger \, 2} \big) \,  
\ket{\rm vac} \, , \,
|\Phi_2 \rangle  \to  - {\textstyle {1 \over 2}} \, \big( b_2^{\dagger \, 2}  - 
b_4^{\dagger \, 2} \big) \, 
\ket{\rm vac} \, , \nonumber \\
|\Phi_3 \rangle & \to & {\textstyle {1 \over {\sqrt 2}}} \, \big(  b_1^\dagger 
b_4^\dagger - b_2^\dagger 
b_3^\dagger \big) \, \ket{\rm vac} \, , \,
|\Phi_4 \rangle  \to  - {\textstyle {1 \over {\sqrt 2}}} \, \big(  b_1^\dagger 
b_2^\dagger - b_3^\dagger 
b_4^\dagger  \big) \, \ket{\rm vac} \, . 
\end{eqnarray}
Finally, detectors measure the presence of photons in each of the possible 
output ports. The detection of two 
photons in the same output port, namely in 1 or 3 and in 2 or 4, corresponds to 
a measurement of the 
state $|\Phi_1 \rangle$ and $|\Phi_2 \rangle$, respectively.
The detection of a photon in ports 1 and 4 or in 
2 and 3 indicates a measurement of the state $\ket{\Phi_3}$, while a photon in 
the ports 1 and 2 or in 3 and 4 
indicates the state 
$\ket{\Phi_4}$. 

On the other hand, to implement a parity filter and hence teleportation with 
insurance, another symmetric 
multiport with $U_{mn}$ given by
\begin{eqnarray*} \label{matrix4}
U &=& {\textstyle {1 \over 2}} \left( \begin{array}{rrrr} 
1 & 1 & 1 & 1\\
1 & 1& -1 & -1\\
1 & -1 & 1 & -1\\
1 & -1 & -1 & 1 \end{array} \right)   
\end{eqnarray*} should be used. One can again show that the network transfers 
the appropriate basis states $\ket{\Phi_i}$ defined in  Eq. (\ref{devilencoding}) as
\begin{eqnarray} 
|\Phi_1 \rangle & \to &  {\textstyle {1 \over 2}} \, \big( b_1^{\dagger \, 2}  
- b_3^{\dagger \, 2} \big) \,  \ket{\rm vac} \, , \,
|\Phi_2 \rangle  \to   {\textstyle {1 \over 2}} \, \big( b_2^{\dagger \, 2}  
- b_4^{\dagger \, 2} \big) \, \ket{\rm vac} \, , \nonumber \\
|\Phi_3 \rangle & \to &  {\textstyle {1 \over {\sqrt 2}}} \, \big(  b_1^\dagger 
b_2^\dagger - b_3^\dagger 
b_4^\dagger  \big) \, \ket{\rm vac} \, , \,
|\Phi_4 \rangle  \to  {\textstyle {1 \over {\sqrt 2}}} \, \big(  b_1^\dagger 
b_4^\dagger - b_2^\dagger 
b_3^\dagger \big) \, \ket{\rm vac} \, .  
\end{eqnarray}
The detection of two photons in the same output port, namely in 1 or 3 and in 2 
or 4, corresponds to a 
measurement of the state $|\Phi_1 \rangle$ and $|\Phi_2 \rangle$, respectively 
as in the case for two-qubit phase
gate implementation.
However now, the detection of a photon in ports 1 and 4 or in 
2 and 3 indicates a measurement of the state $\ket{\Phi_4}$, while a photon in 
the ports 1 and 2 or in 3 and 4 
indicates the state $\ket{\Phi_3}$. 

We  now discuss  an interesting feature of 2-photon coincidence measurement which we can exploit.

\subsection{Time-resolved detection for non-identical photon \\ sources}
 So far, we have assumed identical atom-cavity systems and therefore, photons 
generated from the photon sources 
only differ in their encoded degree of freedom (for example, polarisation, 
time-bin etc.). Non-identical 
atom-cavity systems in general yield different temporal  photon pulse shapes as 
well as spatial modes and thus introduce an additional 
degree 
of freedom which allow  for the origin of the photon pulse to be determined.  
Generally, this leads to errors 
in gate implementation as will be seen in this section.  In principle, this can 
be fixed by 
pulse shape engineering if one has full knowledge of the cavity parameters. 
Otherwise, a time-resolved 
detection technique combined with spatial filters allows us to remove the 
which-way information due to non-identical pulse shapes.  
Recently, it has been shown how the Hong-Ou-Mandel dip, in which 
total photon indistinguishability is normally a necessary requirement, can still 
be observed even 
with two distinguishable photons, provided one performs time-resolved 
postselection \cite{Legero03, Legero04}. Time-resolved postselective detection 
is 
the essential mechanism that wipes away the which-way information as first 
suggested by {\. Z}ukowski {\em et al.} \cite{Zukowski93,Zukowski95}. 

It is convenient to start our discussion using the formalism of Legero {\em et 
al.} \cite{Legero03,Legero04} with the simple situation 
where two 
distant atoms, labelled 1 and 2, are entangled with the aid of a 
beam splitter. Now, we consider a beam splitter where ports 1 and 2 are the input 
ports and ports 3 and 4 define the output ports, each containing  an ideal photon detector. Correspondingly, the photon annihilation and 
creation operator for port 
$j$ and polarisation $i$ in  frequency mode $\omega$ is denoted as $b_{ij}(\omega)$ 
and $b_{ij}^\dagger(\omega)$ respectively. By convention, the $j$th atom is 
placed 
at the $j$th input port. For each $j$th atom, we assume that its initial state is given by 
$c_{0j}\ket{u_{0j}}+c_{1j}\ket{u_{1j}}$. The $j$th atomic state is subsequently 
encoded (see Section \ref{encodertrick}) in the photon state as 
\begin{equation}
\ket{\psi}_j=\sum_{i,k=0,1}c_{ij}\ket{u_{ij}} \int_{-\omega_b}^{\omega_b} 
d\omega 
s_j(\omega)\gamma_{ikj}b_{kj}^\dagger(\bar{\omega})\ket{\rm vac}. 
\end{equation} such that the orthogonality condition $\sum_{k=0,1} 
\gamma_{ikj}\gamma^*_{lkj}=\delta_{il}$ is fulfilled. The coefficients 
$\gamma_{ikj}$ are introduced to allow the generated photons to be 
transformed by arbitrary single qubit rotation. In addition, we have dropped the 
index for photon generation time $\tau$ since it is 
inconsequential to our discussion here. We have assumed perfect redundant 
encoding, where we have set without 
loss of generality, all cavity and laser driving  parameters to be independent 
of polarisation. In principle, photon generation need not be perfect and there 
will generally be terms that do not 
contribute to any photon in the total atom-photon state vector. These terms can 
be neglected for our discussion as we use 2-photon detection to herald 
entanglement, thereby 
allowing us to disregard non-photon contributing terms. In contrast, this is not 
possible in 1-photon detection protocols. We then define 
the total 
input state as $\ket{\Psi}_{\rm in}=\ket{\psi}_1\otimes\ket{\psi}_2$.  We 
further define the unitary transformed total output state, before photon 
detection by $\ket{\Psi}_{\rm out}=S\ket{\Psi}_{\rm in}$ where $S$ is a unitary 
operator that defines the beam splitter transformation. Specifically, we can set
\begin{eqnarray}\label{transforma}
S^{\dagger}b_{i3}(\omega)S&=&U_{31,i}b_{i1}(\omega)+U_{32,i}b_{i2}(\omega) \, , 
\nonumber \\
S^{\dagger}b_{i4}(\omega)S&=&U_{41,i}b_{i1}(\omega)+U_{42,i}b_{i2}(\omega) \, ,
\end{eqnarray} where $U_{mn,i}$ refers to the probability amplitude of 
redirecting a photon of polarisation $i$ from input port $n$ to output port $m$.
Since we are dealing with time-resolved detection, it is convenient to use the 
definition of the time-dependent electric field amplitude operator $E_{\lambda 
j}^+(t)$  for the $j$th port and polarisation $'\lambda'$ where $b_{\lambda 
j}(\omega)=\alpha_{0} b_{0j}(\omega)+\alpha_{1} b_{1j}(\omega)$ and 
$|\alpha_{0}|^2+|\alpha_{1}|^2=1$ given by \cite{Ou99a,Legero03}
\begin{equation}
E_{\lambda j}^+(t)=\frac{1}{\sqrt{2 \pi}} \int_{0}^{\infty} d\omega K(\omega) 
{\rm e}^{-{\rm i}\omega 
t}b_{\lambda j}(\omega).
\end{equation} As in Legero {\em et al.} \cite{Legero03} and Gardiner and Zoller 
\cite{Gardiner04}, we choose $K(\omega) \approx 1$ for reasons of 
normalisation and the fact that $s_j(\omega)$ is 
strongly peaked around $\omega_{\rm cav}$. Accordingly, using Eq. 
(\ref{transforma}), the transformation of 
the electric field operator is thus given by
\begin{eqnarray}\label{transformE}
S^{\dagger}E_{\lambda 3}^+(t)S&=&U_{31,\lambda}E_{\lambda 
1}^+(t)+U_{32,\lambda}E_{\lambda 2}^+(t) \, , \nonumber \\
S^{\dagger}E_{\lambda 4}^+(t)S&=&U_{41,\lambda}E_{\lambda 
1}^+(t)+U_{42,\lambda}E_{\lambda 2}^+(t) \, .
\end{eqnarray}
It is convenient to note that for the input ports 1 and 2,
\begin{eqnarray} \label{pulseshape}
E_{\lambda j}^+(t)\ket{\psi_j}&=&\sum_{i,k,l=0,1}\frac{c_{ij} }{\sqrt{2 
\pi}}\ket{u_{ij}}\int_{-\omega_b}^{\omega_b} d\omega \int_{0}^{\infty} d\tilde 
\omega {\rm e}^{-{\rm i}\tilde \omega t} 
\alpha_{l} b_{lj}(\tilde \omega)s_j(\omega) \gamma_{ikj} b_{kj}^{\dagger}(\bar{
\omega})\ket{\rm vac} \nonumber \\
&=& \sum_{i,k=0,1}\frac{c_{ij} \gamma_{ikj} \alpha_{k}}{\sqrt{2 \pi}}  
\ket{u_{ij}} {\rm e}^{-{\rm i}\omega_{\rm cav} t}
\int_{-\omega_b}^{\omega_b} d\omega {\rm e}^{-{\rm i}\omega t} s_j(\omega) 
\ket{\rm vac} 
\nonumber \\
&=& \sum_{i,k=0,1} c_{ij} \gamma_{ikj} \alpha_{k}{\rm e}^{-{\rm i}\omega_{\rm 
cav} t} f_j(t)\ket{u_{ij}}\ket{\rm vac}
\end{eqnarray} where 
\begin{equation}
f_j(t)=\frac{1}{\sqrt{2 \pi}}\int_{-\omega_b}^{\omega_b} d\omega {\rm e}^{-{\rm 
i}\omega t} s_j(\omega)
\end{equation} is the pulse shape of the photon \cite{Duan03} in the $j$th input 
given by the Fourier transform of its frequency spectrum 
$s_j(\omega)$.
We suppose that a photon is detected at port 3 at time $t_3$ with the 
polarisation $'a'$ and port 4 at time $t_4$ with the polarisation 
$'b'$. The unnormalised conditional state of the system, as in the formalism by 
Legero {\em et al.} \cite{Legero03} and also Gardiner and 
Zoller \cite{Gardiner04} is therefore given by 
$\ket{\Psi}_{\rm cond}$,  
\begin{eqnarray}
\ket{\Psi}_{\rm cond}&=&E_{a3}^+(t_3)E_{b4}^+(t_4)\ket{\Psi}_{\rm out} \nonumber 
\\
&=& SS^{\dagger}E_{a3}^+(t_3)SS^{\dagger}E_{b4}^+(t_4)S\ket{\Psi}_{\rm in} 
\nonumber \\
&=& 
S(U_{31,a}U_{42,b}E_{a1}^+(t_3)E_{b2}^+(t_4)+U_{32,a}U_{41,b}E_{a2}^+(t_3)E_{b1}
^+(t_4))\ket{\Psi}_{\rm in} \, . \nonumber \\
\end{eqnarray} 
For the sake of concreteness, we assume a 50:50 beam splitter so that 
$U_{mn,a}=U_{mn,b}$ is polarisation insensitive and 
$U_{31,k}=U_{41,k}=U_{32,k}=-U_{42,k}=\frac{1}{\sqrt{2}}$. We also set 
$\gamma_{ijk}=\delta_{ij}$ for simplicity. We further set $a='0'$ and $b='1'$  
and require  $c_{ij}=\frac{1}{\sqrt{2}}$. 
Using the relations given by Eq. (\ref{pulseshape}), $\ket{\Psi}_{\rm 
cond}$ can then be simplified to
\begin{equation}
\ket{\Psi}_{\rm 
cond}=\frac{1}{4}(-f_1(t_3)f_2(t_4)\ket{u_{01}}\ket{u_{12}}+f_1(t_4)f_2(t
_3)\ket{u_{11}}\ket{u_{02}})
\end{equation} where we have conveniently dropped all the vacuum terms as well 
as the inconsequential global phase factor for 
readibility.  For $\ket{\Psi}_{\rm cond}$ to describe a maximally entangled singlet state (which we would like to prepare), it is 
necessary that the condition,
\begin{equation}\label{condition}
f_1(t_3)f_2(t_4)=f_1(t_4)f_2(t_3)
\end{equation} holds. This condition can be fulfilled unconditionally if the 
photon pulse shapes originating from both atoms are similar, i.e. 
$f_1(t)={\rm k}f_2(t)$ for some complex constant k.  This means the fidelity of 
the entangled state 
is guaranteed to be unity as long as a photon  is detected in both output 
ports 3 and 4 irrespective of the time of detection. This is easily explained as 
the time of detection does not reveal the origin of the photon given that the 
photon pulse shapes are identical. In the case of distinguishable pulse shapes, 
one can still fulfil the condition given by Eq. (\ref{condition}) by setting 
$t_3=t_4$, thus requiring perfect time-resolved coincidence detection. Note that 
this does 
not require any preknowledge of the pulse shape in either cavity to achieve 
arbitrary high fidelity. This is an illustration of the power of 
measurement-based 
approach to quantum computation in contrast to a fully coherent-based approach. 
Furthermore, since k is an unspecified constant that is 
complex, this implies that the introduction of any unknown slowly varying phase 
factor (with respect to the photon generation and 
detection time) in the path of the photons produces no observable effect on the 
fidelity of operation.

So far, we have assumed that the photon pulses from the 2 cavities have the same 
central frequency $\omega_1=\omega_2=\omega_{\rm cav}$. Further, suppose 
that the 2 cavities 
are detuned relative to each other by $\Delta \omega=\omega_1-\omega_2$, with 
everything else being identical. This is then equivalent to introducing a 
time-dependent phase factor to the pulse shape such that
\begin{equation}
f_1(t)=f_2(t)\exp(-{\rm i \Delta \omega t}).
\end{equation} In this case, pulse similarity can still be obtained for $\Delta 
\omega (t_1-t_2)=2n\pi$. This condition was also previously predicted and used 
in the context of observing Hong-Ou-Mandel dip for photons of 2 different 
central frequencies but identical pulse shape \cite{Legero03,Legero04}.
It is useful to determine analytically, within the 
approximation of adiabatic theorem, how the fidelity of entangled state 
preparation is degraded for 2 different cavity parameters. Firstly, it is 
convenient to recall formula for the pulse shape from Eq. (\ref{magicpulse})
\begin{equation}
f_j(t)=\sqrt{\kappa_j}\sin\theta_j(t)\exp(-\frac{\kappa_j}{2}\int_0^t d\tau 
\sin^2 \theta_j(\tau))
\end{equation} where
\begin{equation}
\sin\theta_j(t)=\frac{\Omega_j(t)}{\sqrt{\Omega^2_{j}(t)+g_j^2}}
\end{equation} and $\kappa_j,\Omega^2_j(t),g_j$ are the cavity decay, Rabi 
frequency of the driving laser and cavity coupling of the  $j$th atom-cavity 
respectively. It is then obvious that the below condition 
with all other parameters being equal, guarantees pulse shape similarity, namely
\begin{equation}
\left|\frac{\Omega_{1}(t)}{g_{1}}\right|=\left|\frac{\Omega_{2}(t)}{g_{2}}\right
|.
\end{equation} Assuming that this condition is fulfilled, we now consider 
$\kappa_1\ne\kappa_2$. The time-dependent fidelity $F(t_3,t_4)=|\bra{\Psi^-}\hat 
\Psi \rangle_{\rm cond}|^2$ 
 given in terms 
of pulse shape functions is then given by\footnote{$\ket{\Psi^-}=\frac{1}{\sqrt{2}}(\ket{u_{01}}\ket{u_{12}}-\ket{u_{11}}
\ket{u_{02}})$ and $\ket{\hat \Psi \rangle_{\rm 
cond}}=\ket{\Psi \rangle_{\rm cond}}/||\,\ket{\cdot}||$}
\begin{equation}
F(t_3,t_4))=\frac{(f_1(t_3)f_2(t_4)+f_1(t_4)f_2(t_3))^2}{2(f_1(t_3)^2f_2(t_4)^2+
f_1(t_4)^2f_2(t_3)^2)} \, .
\end{equation} Following this, it is straightforward to show that the average 
fidelity $F_{\rm av}(b,a)$ with detectors integration time 
from $t=a$ to $t=b$ is given by
\begin{equation}
F_{\rm av}(b,a)=\frac{1}{2}\left(1+\frac{(\int_a^b dt f_1(t)f_2(t))^2}{\int_a^b 
dt f_1(t)^2 \int_a^b dt f_2(t)^2}\right) \, ,
\end{equation} and is related to the overlap between the two pulse shape functions.
It is interesting to also consider the case  where the detected 
polarisation is the same.  In this case, $\ket{\Psi}_{\rm cond}$ is given by
\begin{equation}
\ket{\Psi}_{\rm 
cond}=(-f_1(t_3)f_2(t_4)+f_1(t_4)f_2(t_3))\sum_{i=0,1}\alpha_i\ket{u_{i1}}\otimes\sum_{i=0,1}\alpha_i\ket{u_{i1}}.
\end{equation} For pulse similarity condition, this implies that  
$||\,\ket{\Psi}_{\rm cond}||^2$ vanishes, which is essentially the Hong-Ou-Mandel 
effect. 
As an example, we calculate the fidelity and joint probability density of 
2-photon coincidence detection for the case of a Gaussian driving pulse with 
pulse width $\tau=40/\kappa_2$, being centered in 
$20/\kappa_2$, with  width  $\sqrt{2}\tau/10$, 
${\rm max}(\Omega_j^2(t)/g_j^2)=9$ and $t_3$ and $t_4$  normalised to 
$\kappa_2^{-1}$. 
\begin{figure}
\begin{center}
\psfrag{t3}{$t_3$} \psfrag{t4}{$t_4$}
\includegraphics[scale=0.3]{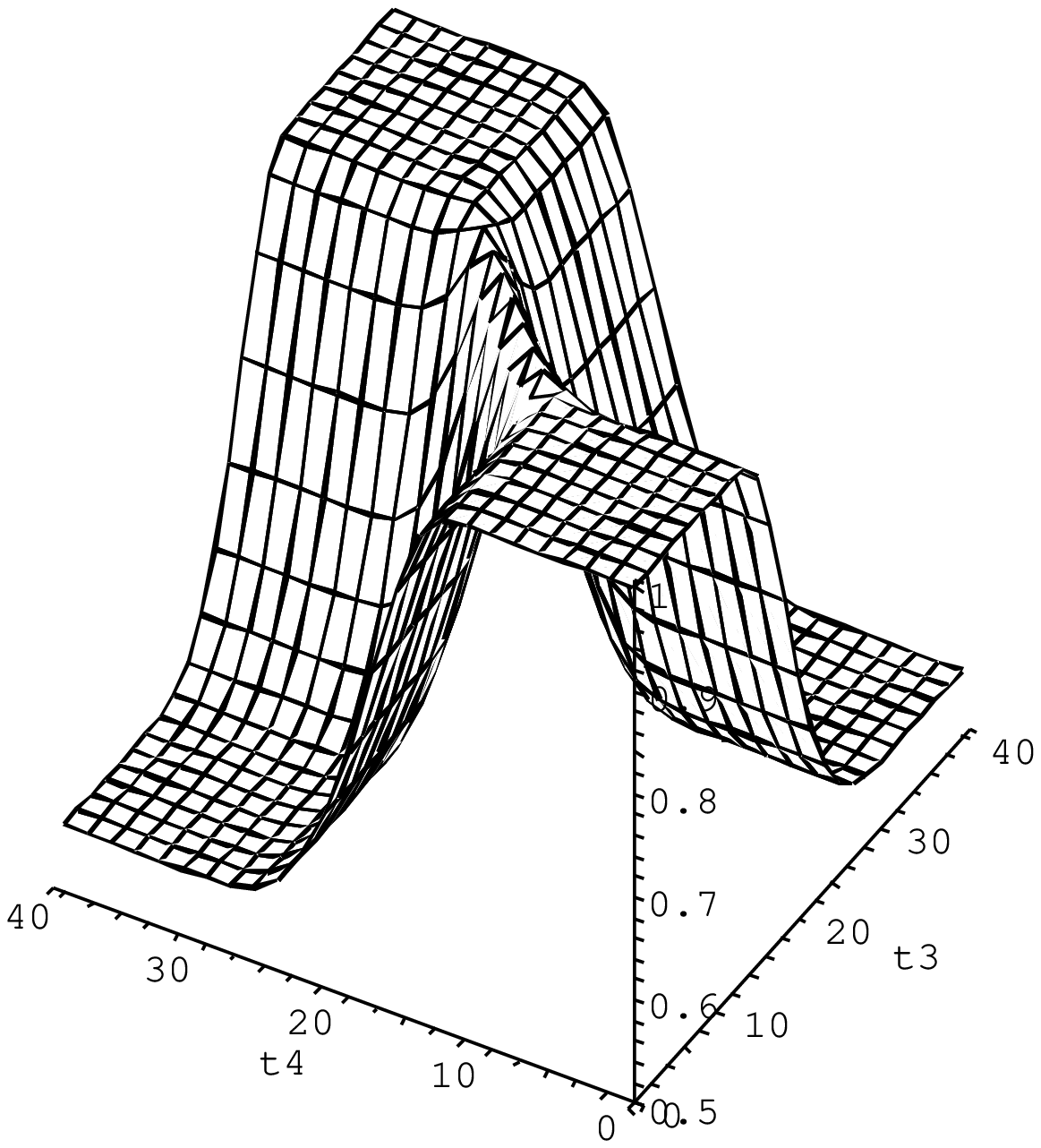} 
\includegraphics[scale=0.3]{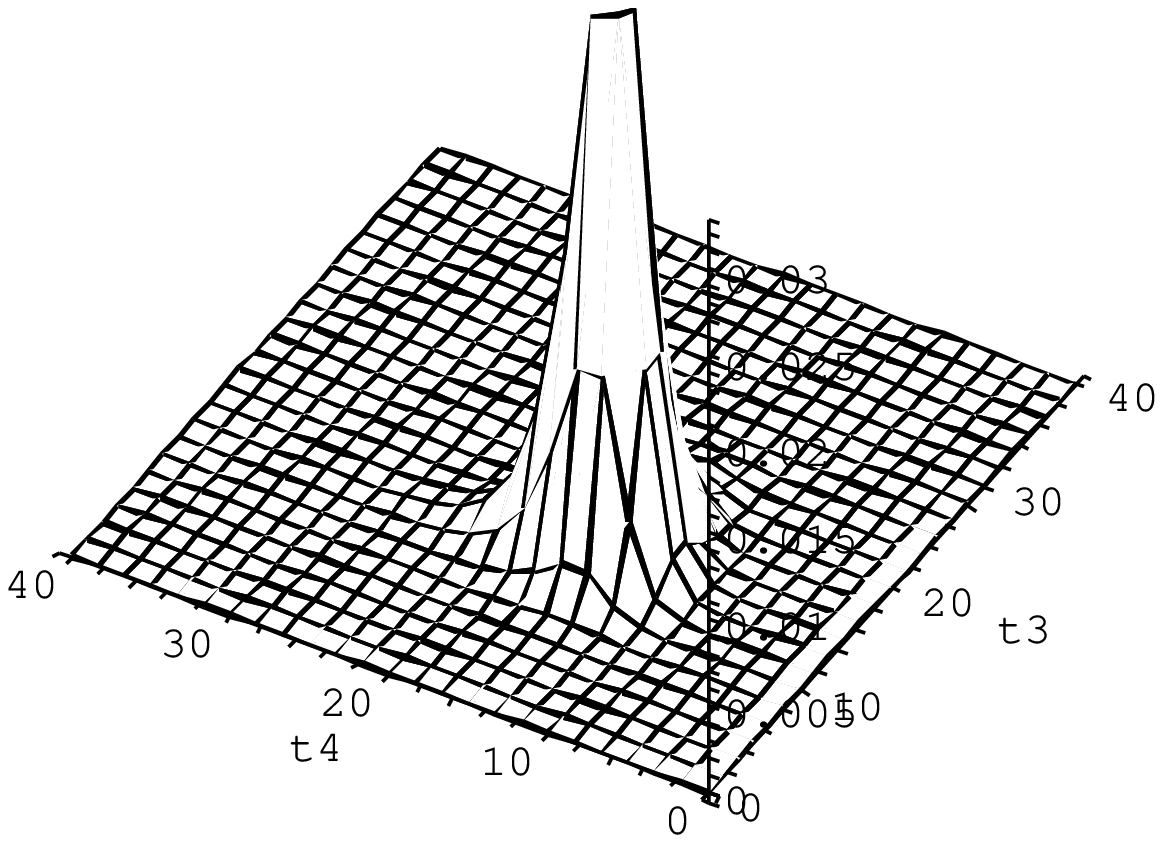} 
\end{center} 
\caption{(a). Fidelity of entangled state and (b). Joint probability density of 
2-photon coincidence detection for $\kappa_1=0.5 \kappa_2$ for a guassian 
driving pulse with pulse width $\tau=40/\kappa_2$,being centered in 
$20/\kappa_2$, with  width   $\sqrt{2}\tau/10$  and fixing 
$max(\Omega_j^2(t)/g_j^2)=9$.}\label{minsk:timeselect5}
\end{figure}

\begin{figure}
\begin{center}
\psfrag{t3}{$t_3$} \psfrag{t4}{$t_4$}
\includegraphics[scale=0.3]{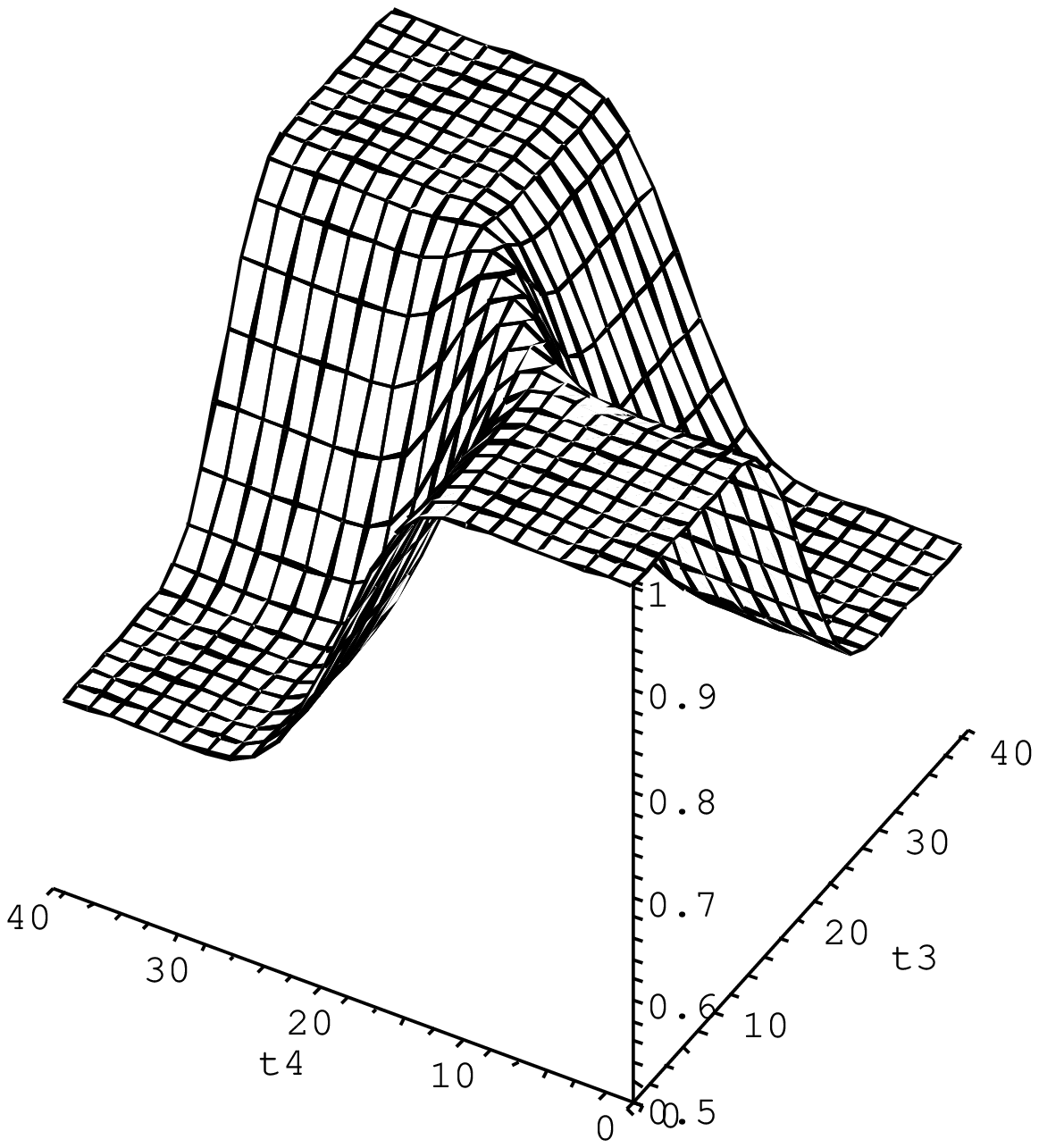} 
\includegraphics[scale=0.3]{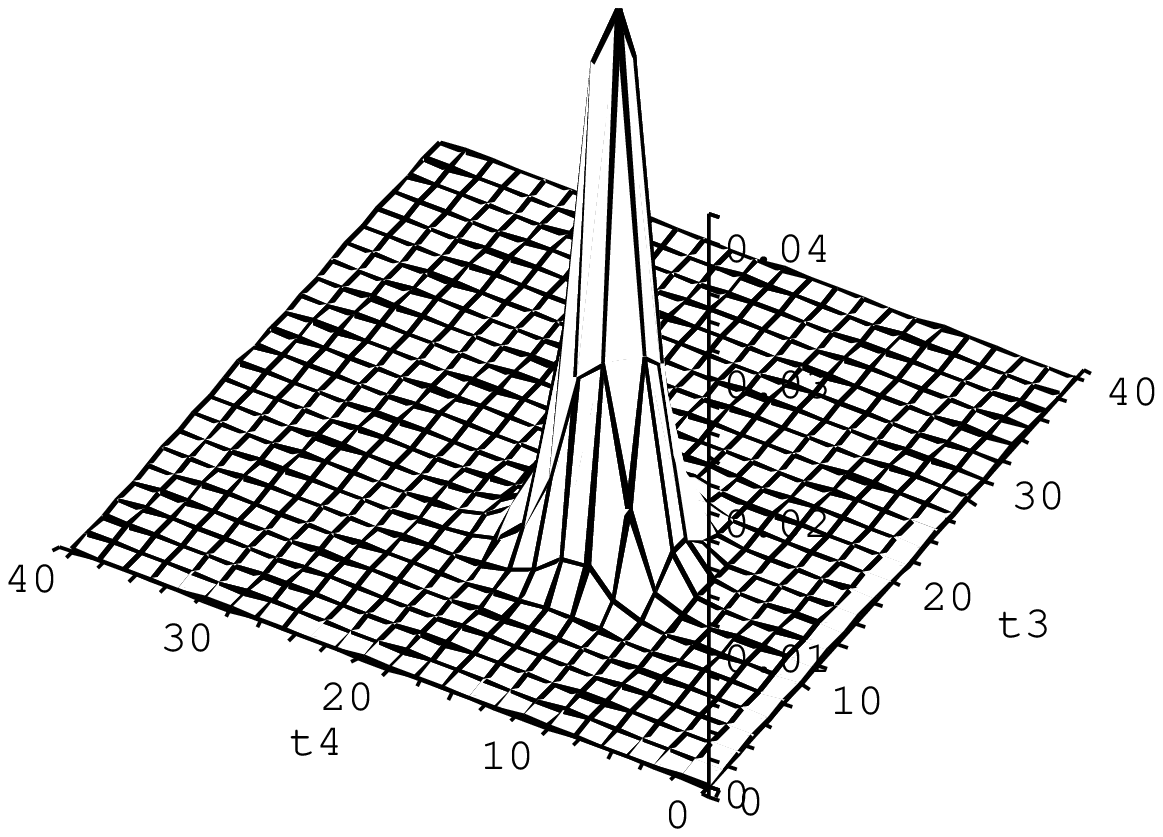} 
\end{center} 
\caption{(a). Fidelity of entangled state and (b). Joint probability density of 
2-photon coincidence detection for $\kappa_1=0.7 \kappa_2$ for the same driving 
condition as above.} \label{minsk:timeselect7}
\end{figure}

\begin{figure}
\begin{center}
\psfrag{t3}{$t_3$} \psfrag{t4}{$t_4$}
\includegraphics[scale=0.3]{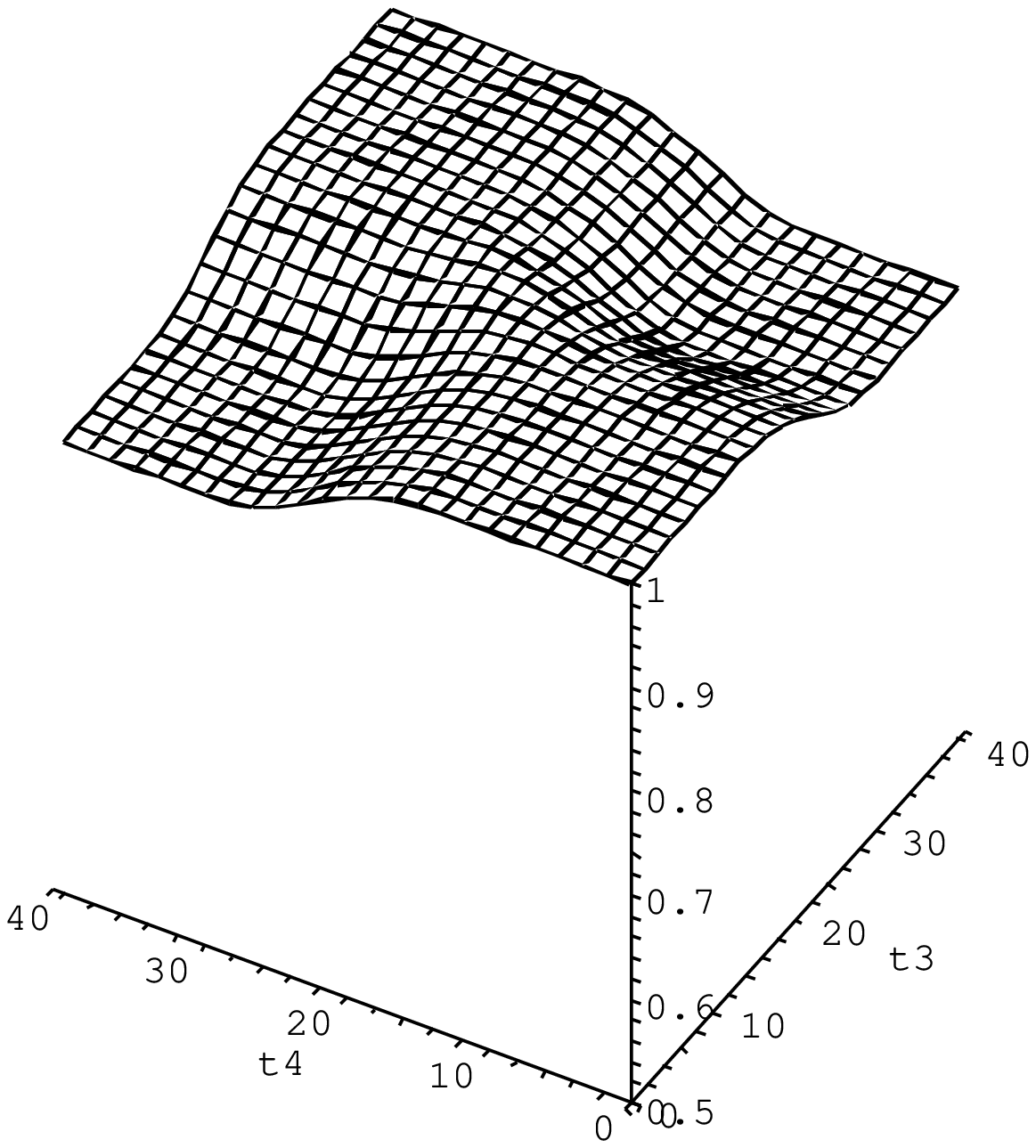} 
\includegraphics[scale=0.3]{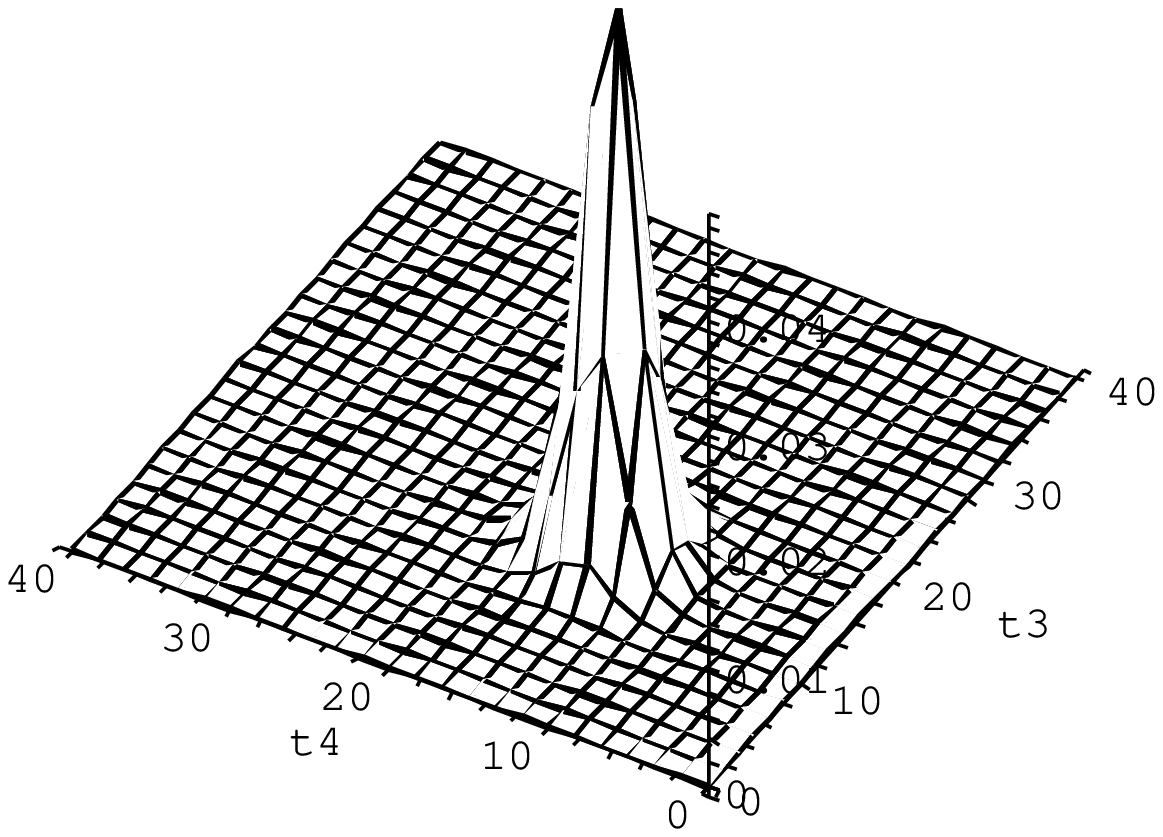} 
\end{center}
\caption{(a). Fidelity of entangled state and (b). Joint probability density of 
2-photon coincidence detection for $\kappa_1=0.9 \kappa_2$ for the same driving 
condition as above} \label{minsk:timeselect9}
\end{figure}
From Fig.~\ref{minsk:timeselect5}, Fig.~\ref{minsk:timeselect7} and 
\ref{minsk:timeselect9} it can 
be seen that the fidelity is always unity for perfect coincidence. The more 
similar the cavities, the greater the tolerance of the 
fidelity. In the limit of identical cavities, coincidence photon pairs can be 
detected at any time interval with no effect on the fidelity. For comparision, the joint probability density 
for 2-photon detection is also calculated and  the more different the 
cavities are, the lower the probability density of obtaining perfect coincidence generally. One 
observes the interesting trend of almost perfect fidelity at photon detection 
times at the leading and tail-end of the pulse even at non-perfect coincidence. 
This is attributed to the fact that the photon pulse shape at these times is 
relatively flat even for 2 non-identical cavities and in no way contributes to 
any which-way information. Mathematically, this corresponds to the condition 
where $f_1(t) \approx {\rm k}f_2(t)$ at the times $t$ during the leading or 
tail-end 
of the photon pulses. Of course, the probability density for such coincidence is 
low as can be seen from the calculation. Without time-resolved detection, the 
average fidelity $F_{\rm av}(\infty,0)$ corresponding to 
Fig.~\ref{minsk:timeselect5}, Fig.~\ref{minsk:timeselect7} and 
\ref{minsk:timeselect9} is  0.94, 0.98 and 0.99 respectively, consistent 
with the fact that dissimilarities of cavities generally yield lower 
fidelity.

We have given an example of how a time postselective 2-photon coincidence 
detection can yield high fidelity of entangled state preparation. It is 
straightforward to extend this to gate operation as described in this chapter.  
For 
example, one can assume the same setup as before except that now the following 
holds
\begin{equation}
\gamma_{001}=\gamma_{011}=\gamma_{101}=-\gamma_{111}=\gamma_{002}=\gamma_{102}={\rm 
i}\gamma_{012}=-{\rm i}\gamma_{112}.
\end{equation}
Repeating the same procedure as before and assuming the same detection syndrome with $a='0'$ and $b='1'$, 
one can show that the fidelity 
of gate operation is maximal when the pulse similarity condition is fulfilled.

\section{Conclusions and discussions }

We have shown that despite the fact that linear optics based Bell-state measurements on 
photons are incomplete, that does not 
prevent the deterministic implementation of a gate between distant qubits. 
Therefore, rather surprising, we 
show that it is not necessary to demonstrate deterministic entanglement 
generation in order to achieve a 
universal two-qubit gate with unit efficiency\footnote{The insurance aspect of this 
scheme allows processing of a single copy of unknown 
input state in principle. In contrast to cluster state implementation of 
two-qubit gates, the non-deterministic teleportation of an 
unknown unencoded input state  into the 
cluster without insurance can destroy the input qubit if the teleportation fails.}. 
Furthermore, this scheme is intrinsically 
interferometrically stable if one uses polarisation 
encoding and measurement due to the same reason as mentioned in 
Chapter~\ref{firework}.
In the real world, nonideal situations of photon loss, inefficient photon detectors and photon 
generation  
would all lower the gate 
probability of success. Indeed, given photon number-resolving detectors (with no 
dark counts) of quantum efficiency $p_d$ each, the 
probability of single-shot gate failure 
$p_F$, when two photons are not detected, is given by $1-p_d^2$. One can also 
account for photon loss in the factor $p_d$.   
The solution towards fault-tolerant quantum computation in this scheme is to 
make a cluster 
state of high fidelity.  High fidelity can be achieved because our scheme is a 
2-photon heralded scheme. In the case 
where only one  or no photon is detected, this is equivalent to tracing out the 
photon degrees of freedom. 
This effect can be removed by simply 
destroying the two qubits by measuring them in the computational basis in which an 
attempted cluster bond is 
required. This does not decrease 
the fidelity of the cluster state. Therefore, in the presence of imperfections 
such as 
photon loss and inefficient photon detectors, distributed quantum computation 
can still be performed with high 
fidelity by building a cluster state of distant qubits.  Further discussions on 
cluster state buildup can be found in the work by Barrett 
and Kok \cite{Barrett04} and Lim {\em et al.}\footnote{Details of efficient cluster state buildup based on the insurance scenario described in this chapter can be found here.} \cite{Lim05}, and they lie out of the scope of this thesis. In 
addition, Benjamin {\em et al.} \cite{Benjamin05} have also 
recently argued how the insurance scenario in our 
scheme can lead to a higher efficiency in building cluster states.  We now proceed to the 
next chapter where we highlight an important application of the results of this chapter on generating photon 
entanglement on demand.

\chapter{Distributed Photon Entanglement on Demand} \label{demand}

In this short chapter, we highlight a useful result from Chapters \ref{firework} 
and \ref{minsk} and show that distributed photon 
entanglement generation is possible on demand. In particular,  we also 
demonstrate a duality relation that arises from our previous study 
of multiports which may give new perspectives in multiport design for quantum 
information processing.

\section{Multiatom entanglement and multiphoton entanglement on demand}

The key to multiphoton entanglement on demand lies in the initial creation of 
multiatom entanglement. We denote this step as the initialisation. Recalling Chapter \ref{firework} 
where we have mainly considered the Bell multiport, we place no restriction 
on the multiports considered in this chapter. We first allow 
each atom to be entangled with a 
photon and study 
what happens if the photons are passed through a 
multiport.

We assume that each $i$th atom is specified by two states $\ket{\pm}$ notated by 
$g^{\dagger}_{i\pm}\ket{0}$ using a second-quantised notation.  
Each of the 
atoms should first be maximally entangled with a photon feeding into each input 
of the $N \times N$ multiport such that we can write the total combined initial state  as
\begin{equation}
\ket{\Psi_{in}}= \frac{1}{\sqrt{2^N}} \prod_{i=1}^N \Big( 
\sum_{\mu=+,-} 
g_{i \mu}^\dagger \, a_{i \mu}^\dagger \Big) \, \ket{0}.
\end{equation}
Such a state preparation can be performed  
deterministically \cite{Gheri98,LimSPIE04} by an atom-cavity system as described 
in Chapter \ref{minsk}. The general  
atom-photon state $\ket{\Psi_{in}}$ after passing the photons through the 
multiport and upon 
collecting one photon per output port is then, up 
to normalisation and by analogy to Eq. (\ref{firework:output2}), given by
\begin{equation} \label{firework:output3}
\ket{\Psi_{out}}=\frac{1}{\sqrt{2^N}}\sum_{\sigma} \Bigg[ \prod_{i=1}^N 
U_{\sigma(i) i} 
\Big( 
\sum_{\mu=+,-} g_{i \mu}^\dagger b_{\sigma (i) \mu}^{\dagger} \Big) 
\Bigg]  \, |0 \rangle \, .~
\end{equation}
Now, the next step is to choose a detection syndrome of the photons. Let the 
detection syndrome be defined by the  postselected state 
\begin{equation}
\ket{{\rm S}}=\prod_{j=1}^N \sum_{\mu}\alpha_{j \mu}^\ast b_{j \mu}^\dagger 
\ket{0}.
\end{equation} For example, $\alpha_{j \mu}^\ast b_{j \mu}^\dagger 
\ket{0}$ defines the state of a photon detected in output port $j$ with 
polarisation $\mu$. Note that we 
see a direct correspondance or analogy of $\ket{{\rm S}}$  to the input photon state 
(\ref{firework:in}).
Applying the relevant projector, the multiatomic state $\ket{\rm A}$ can be 
shown to be projected onto
\begin{equation} \label{dual1}
\ket{{\rm A}}=\frac{1}{\sqrt{2^N}}\sum_{\sigma} \Bigg[ \prod_{i=1}^N 
U_{\sigma(i) i} \Big( 
\sum_{\mu=+,-} \alpha_{\sigma(i) \mu} g_{i \mu}^{\dagger} \Big) \Bigg]  \, |0 
\rangle 
\end{equation} Hence, a multiatomic state can be prepared by choosing an appropriate detection syndrome. Provided that the photons detectors have negligible dark counts, photon loss and inefficient detectors do not decrease the fidelity of state preparation. This is an advantage of choosing a coincidence (one photon per output port) detection syndrome. To get further insight, we can next substitute $i=\sigma^{-1}(j)$ and  obtain
\begin{eqnarray} \label{dual2}
\ket{{\rm A}}&=&\frac{1}{\sqrt{2^N}}\sum_{\sigma^{-1}} \Bigg[ \prod_{j =1}^{N} 
U_{j \sigma^{-1}(j)} \Big( 
\sum_{\mu=+,-} \alpha_{j \mu} g_{\sigma^{-1}(j) \mu}^{\dagger} \Big) \Bigg]  \, 
|0 \rangle \nonumber \\
&=&\frac{1}{\sqrt{2^N}}\sum_{\sigma} \Bigg[ \prod_{i=1}^N U_{i \sigma(i)} \Big( 
\sum_{\mu=+,-} \alpha_{i \mu} g_{\sigma(i) \mu}^{\dagger} \Big) \Bigg]  \, |0 
\rangle \, . 
\end{eqnarray} since $\sigma^{-1}$ is just a dummy index for a permutation. 
A close comparision of  (\ref{dual2})
with (\ref{firework:output2}) shows that this multiatomic state is of a similar\footnote{For simplicity, we have assumed bosonic statistics in the atoms for the discussion of the analogy with (\ref{firework:output2}). In any case, (\ref{dual1}) is the projected multiatomic state.} 
form to 
that of prepared multiphoton state given the input photon 
state (\ref{firework:in}), 
except that the input and output of the multiport are swapped. This suggests a 
duality relation of designing optical 
circuits aimed at preparing entangled atoms by  examining   how entangled 
photons(in analogy to the desired entangled atoms) can be prepared from product states of photons(analogous to the detection syndrome).

An example of producing a three atom W-state was demonstrated using a $3 \times 3$ 
multiport 
\cite{Zou04a}. It was then conjectured, but not proven, 
that an $N \times N$ multiport may be used to prepare an $N$ atom W-state. The 
work in this chapter  qualifies this result rigourously for a 
Bell multiport.

\begin{figure}\label{demand:setup}
\begin{center}
\begin{tabular}{c}
\includegraphics[height=3.5cm]{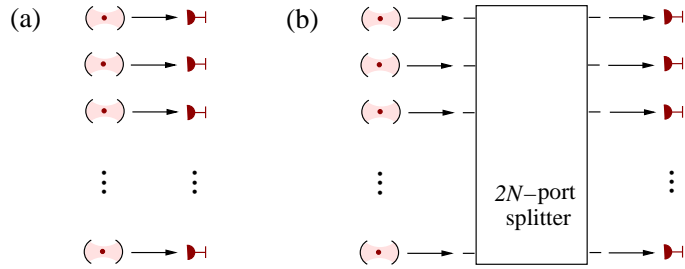}
\end{tabular}
\end{center}
\caption{  
Experimental setup. (a) During the final ``{\em push button}'' step, the 
entanglement of $N$ atom-cavity systems is mapped onto the state of $N$ newly 
generated photons. (b) The initialisation of the system requires postselective 
measurements on the photon emission from the $N$ cavities through a 
multiport beamsplitter.}
\end{figure} 

In general, one can prepare or initialise a wide variety of atomic states with 
this method. One only needs to design the multiport that 
yields the desired mulitphoton state postselectively as described in Chapter 
\ref{firework}. An alternative and more general way to 
create multiatom states by using universal two-qubit gates with insurance is 
described in Chapter \ref{minsk}. 

Once the atomic qubits have been initialised, $N$ photons in 
exactly the same state can be 
created by simply mapping the state of the sources onto the state of $N$ newly 
generated photons whenever required \cite{LimSPIE04}\footnote{ A very recent 
proposal by Kok {\em et al.} 
\cite{Kok05} built on the same point (i.e. initialise and map) to implement a 
multiphoton entanglement on demand source with a slightly 
different physical setup and procedure. The double-heralding step they used in the initialisation process can have the same advantages as  coincidence detection. In addition, a cluster state is prepared offline and arbitrary multiatomic qubits can then be prepared by single qubit measurements to complete the initialisation process. Our scheme employs either the multiport approach for direct preparation of the atomic states or a series of universal two-qubit gates for atomic state initialisation.} (See Fig.~\ref{demand:setup}).  To 
accomplish this, the 
state of each photon source 
should first be encoded as in Chapter \ref{minsk}. Afterwards, the atomic
qubits can be decoupled from the flying qubits by measuring again in a mutually 
unbiased basis 
with respect to the computation basis (i.e. the encoding basis) and 
performing a local operation on the photon whenever necessary. The generation of 
multiphoton entanglement on 
demand superficially resembles a remote state preparation of
the state of $N$ newly 
created photons by the multiatomic state.  This mapping can also be accomplished 
more efficently without measurement by choosing atomic 
levels similar to that in Lim {\em et al.} \cite{LimSPIE04}. For example, one 
can use a 5-level atom with ground states $\ket{0}$ and 
$\ket{1}$ constituting the logical qubit states as well as another ground state 
$\ket{2}$. The excited states are $\ket{e_0}$ and 
$\ket{e_1}$ where the cavity couples the transition $e_i-2$ with a cavity photon 
of polarisation $i$. The exciting laser couples to the 
transition $e_i-i$ and drives an initialised atom similar to the description in 
Chapter~\ref{minsk}, for example in the state 
$(\alpha\ket{0}+\beta\ket{1}) \otimes \ket{\rm vac}$ to the state $\ket{2} \otimes 
(\alpha a_0^\dagger+\beta a_1^\dagger) \ket{\rm vac}$ 
where $a_i^\dagger \ket{\rm vac}$ denotes a photon with polarisation $i$ in the 
external cavity field. 

We can in principle create any arbitrary distributed $N$-photon state on demand 
in comparision with the schemes of Gheri {\em et al.} 
\cite{Gheri98} and Sch{\"o}n {\em et al.}\footnote{I thank Christian Sch{\"o}n for stimulating discussions. } \cite{Schon05}, where a restricted set of 
states on the same spatial mode can be created efficiently from a 
single atom-cavity system.
Before we conclude our thesis, we observe that linear optics resources have been a 
crucial component in most parts of this thesis. We 
remove this resource next and consider entanglement generation with distant 
sources without cavities, i.e. in free space.
\chapter{Photon Polarisation Entanglement from Distant Sources in Free 
Space}\label{photon}

\section{Introduction}\label{photon:intro}
In this chapter, we attempt to develop new perspectives to photon generation through distant sources in free 
space. There exist roughly
two general approaches to create entangled photon pairs. Firstly, entangled photon pairs can  be 
created within the {\em 
same} source as in atomic cascades \cite{Aspect82}, in 
parametric down conversion schemes \cite{Kwiat95} and in the biexciton emission of 
a single quantum dot in a cavity \cite{Stace03}. If 
the entanglement is not created within the same source, single photons can be 
brought together to overlap within their coherence time on 
a beamsplitter where a postselective entangling measurement can be performed on 
the output ports \cite{Shih98}. A more detailed survey of single photon sources and entanglement 
generation is given in Chapter \ref{firework}.

\begin{figure} 
\begin{center}
\includegraphics[scale=1.0]{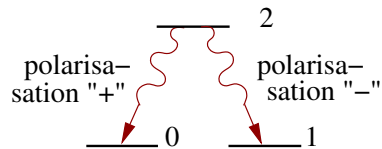} 
\end{center} 
\caption{$\Lambda$-level configuration of the dipole source with the two degenerate ground states $|0\rangle$ and $|1 \rangle$, the excited state 
$|2 \rangle$ and optical transitions corresponding to the 
two orthogonal polarisations ``$+$" and ``$-$"} \label{lambda}
\end{figure}

In contrast to this, we show that polarisation entanglement can also be obtained postselectively
when the photons are created by {\em distant}  sources in 
free space without having to control their photon collection times. As an example, 
we analyse the photon emission from two dipole sources 
that might be realised in the form of trapped atoms, diamond NV color centres, 
quantum dots or by using single atoms doped onto a 
surface. An interaction between the sources is not required. Each source should 
possess a $\Lambda$-type three-level configuration with 
the two degenerate ground states $|0\rangle$ and $|1 \rangle$, the excited state 
$|2 \rangle$ and optical transitions corresponding to the 
two orthogonal polarisations ``$+$" and ``$-$" along a well defined axis (see Fig. \ref{lambda}). Polarisation entanglement arises 
under the condition of the emission of two photons in 
different but carefully chosen directions independent from the initial state of 
the sources. Furthermore in our scheme, this leaves the 
dipoles in a maximally entangled state. Therefore, we can obtain 
both usable postselected 2-photon entanglement\footnote{For example, postselected photon entanglement can be used for quantum 
cryptography} and 
preselected dipole-dipole entanglement. 
 
In order to understand how the scheme works, it is important to note that 
fluorescence from two distant dipole sources can produce an 
interference pattern on a far away screen, if the distance between 
the sources \cite{Scully82,Eichmann93,Schon01} is comparable to the wavelength of the emitted 
photons. This can be understood as both 
sources contributing {\em coherently} to the creation of each photon. 
Consequently, the emission of one photon leaves a trace in the 
states of {\em all} its potential sources, depending on its polarisation and the 
direction of its wave vector \cite{Schon01,Schon02}, and 
can thus affect the state of the subsequently emitted photon. Such a picture is 
seen most directly using the quantum jump formalism 
\cite{Hegerfeldt93,Dalibard92,Carmichael93}. 

The described interference pattern  has already been observed \cite{Eichmann93} 
in the intensity profile due to the flourescence of two 
four-level atoms scattered by laser light. Various attempts 
\cite{Wong97,Itano98,Schon01,Agarwal02} to elucidate this  have been made with the central theme that 
interference can only be 
observed when the which-way information is in principle 
absent. In addition, work aimed at investigating aspects of second-order photon or intensity-intensity correlations at perfect 
photon detection coincidences (i.e. at the same time) has also been made 
\cite{Mandel83,Schon01,Agarwal02}. 
The modulation depth of such intensity-intensity 
correlations of the same polarisation is shown to be reachable to 100 \% even when 
the intensity interference pattern may disappear. In other 
words, there exist a strong spatial antibunching of emitted photons of the same 
polarisation in free space where the detection of one 
photon does not permit the detection of another photon in certain directions at the 
same instant. Here, we exploit this feature for the 
generation of entangled photon pairs.  

In this chapter, the detectors of Alice and Bob are placed such that all wave 
vector amplitudes contributing to the creation of a second 
photon with the same polarisation as the first one interfere destructively. In 
case of the collection of two photons (one by Alice and 
one by Bob) the shared pair has to be in a superposition of the state where Alice 
receives a photon with polarisation ``$+$" and  Bob a 
photon with polarisation ``$-$" and the state where Alice receives a photon  with 
polarisation ``$-$" and  Bob a photon with  
polarisation ``$+$". Both share a maximally entangled pair, if the amplitudes for 
these two states are of the same size. In summary, 
polarisation entanglement is obtained with the help of postselection and 
interference effects. Related mechanisms have been proposed in 
the past to create atom-atom entanglement \cite{Cabrillo99,Plenio99,Protsenko02,Simon03}. 

The pair creation scheme proposed in this paper is feasible with present 
technology and might offer several advantages to quantum 
cryptography. In contrast to parametric down conversion, the setup 
guarantees antibunching between subsequent photon pairs 
since the creation of a new pair is not possible without reexcitation of both 
sources. Furthermore, the scheme is robust with respect to the possible phase fluctuation in the exciting laser\footnote{Axel Kuhn first 
brought this to my attention on discussion of this scheme.}. The final 
photon state does not depend on the initial state of the sources  in case of a 
successful collection.
Finally, the scheme may offer the possibility to generate {\em multiphoton} 
entanglement by incorporating more than two radiators in the 
setup.

\section{Theory}\label{photon:theory}

Let us now discuss the creation of such an entangled photon pair in detail. We describe 
the interaction of the dipole sources with the 
surrounding free radiation field by the Schr\"odinger equation.  The annihilation 
operator for a photon with wave vector ${\bf k}$, 
polarisation $\lambda$ with polarisation vector\footnote{In this thesis, the 
notation is chosen such that $\hat {\bf x} \equiv {\bf x}/\| 
{\bf x} \| $.} defined as ${\bf \epsilon}_{\hat {\bf k}\lambda}$ is $a_{{\bf 
k}\lambda}$. The two dipole sources considered here are placed at 
the fixed positions ${\bf r}_1$ and ${\bf r}_2$ and should be identical in the 
sense that they have the same dipole moment ${\bf D}_{2j}=\bra{2}{\bf D}\ket{j}$ 
for the 2-$j$ transition ($j=0,1$). The energy separation between the degenerate 
ground states and level 2 is $\hbar \omega_0$ while 
$\omega_k=kc$ and $L^3$ is the quantisation volume of the free radiation field.  In 
addition, we define the $i$th atomic lowering and 
raising operator as
\begin{equation}
S^-_{i,j}=\ket{j}_{ii}\bra{2}, \,\,\, S^+_{i,j}=\ket{2}_{ii}\bra{j}.
\end{equation}
Using this notation, the system Hamiltonian becomes 
within the rotating wave approximation and with respect to the interaction-free 
Hamiltonian in the interaction picture,
\begin{eqnarray} \label{21}
H_{\rm I} &=& \sum_{i=1,2} \sum_{j=0,1} \sum_{{\bf k},\lambda}  \hbar g_{{\bf 
k}\lambda}^{(j)} \, 
{\rm e}^{ -{\rm i} (\omega_0 - \omega_k) t} \,  {\rm e}^{ -{\rm i} {\bf k} \cdot 
{\bf r}_i} \, 
a_{{\bf k}\lambda}^\dagger \, S^-_{i,j}+ {\rm H.c.} ~, \nonumber \\ 
&=& \sum_{j=0,1} H_{{\rm I}1}^j+H_{{\rm I}2}^j 
\end{eqnarray}
which can be decomposed into terms $H_{{\rm I}i}^j$ relating only to each of the $i$th 
atom and
\begin{eqnarray}
g_{{\bf k}\lambda}^{(j)} 
&=& {\rm i} e \, \left[ \frac{\omega_k}{2 \epsilon_0 \hbar L^3 } \right]^{1/2}
\! ({\bf D}_{2j} ,  {\bf \epsilon}_{\hat {\bf k}\lambda}) 
\end{eqnarray}
is the coupling constant for the field mode $({\bf k},\lambda)$ to the 2-$j$ 
transition of each source. With $H_{\rm I}$, we can associate the unitary operator describing the evolution of the combined system from 
time $t_1$ to $t_2$ as $U_{\rm I}(t_2,t_1)$. The rotating wave approximation 
corresponds to neglecting the non-energy conserving terms that describe the 
excitation of atoms combined with the creation of a photon or 
the deexcitation of atoms combined with the annihilation of a photon. These 
effects are not unphysical \cite{Knight73} but their 
contribution to the time evolution of the described system can be shown to be very 
small and almost impossible to observe.

\subsection{Entangled photon and entanged dipole generation}
To describe the effect of an emission on the state of the sources, we introduce 
the spontaneous decay rate of the 2-$j$ transition 
$\Gamma_j \equiv (e^2 \omega_0^3 \, |{\bf D}_{2j}|^2)/(3\pi \epsilon_0 \hbar c^3)$ 
and the reset or collapse operator $R_{{\bf \hat k},\lambda}$ which is associated with the quantum jump formalism \cite{Hegerfeldt93, 
Dalibard92, 
Carmichael93}. A good review of the quantum jump approach can be found in Ref. \cite{Plenio98}. For the 
sake of simplicity, we set $\Gamma_0=\Gamma_1=\Gamma$ in this chapter.
The quantum jump formalism is an instance of a type of  unravelling of the master equation describing the evolution of the dipole sources 
in an open environment such as the free radiation  field. The source follows a so-called quantum state trajectory based on knowledge 
obtained from a real or ficticious continuous\footnote{More precisely, the measurement is not truly continuous but coarsed-grained at a 
timescale of $\Delta t$ much larger than the transition optical period but also much smaller than the average timescale of atomic evolution. A 
truly continuous measurement will instead freeze the system due to the quantum Zeno effect.} time-resolved measurement that yields 
generally two types of observables. One of them is the no-photon observation and the other is a photon detection observation that can be 
direction and (or) polarisation  specific.  We first denote the 
free radiation field in the vacuum state $\ket{0_{\rm ph}}$ and define the reduced density operator of the dipole sources at time $t$ as 
$\rho_{a}(t)$. We furthermore denote the 1-photon state of wave vector 
${\bf k}=k{\bf \hat k}$ and 
polarisation $\lambda$ by $\ket{1_{k{\bf \hat k} \lambda}}$. In the theory of quantum evolution of an open system, under the Born-Markovian approximation, the 
evolution of $\rho_{a}(t)$ given that at time $t$, the combined state is $\rho(t)=\ket{0_{\rm ph}} \rho_{a}(t) \bra{0_{\rm ph}}$, can be 
described by a superoperator ${\cal L}(\Delta t)$ that yields a Kraus operator sum representation given as
\begin{equation}
\rho_{a}(t) \to {\cal L}(\Delta t) \rho_{a}(t) = \rho_{a}(t+\Delta t)= \sum_{\mu} M_{\mu} \rho_{a}(t) M^\dagger_{\mu} \, .
\end{equation} Here $M_{\mu}=\bra{\mu}U_{\rm I}\ket{0_{\rm ph}}$  is associated with an observable $\mu$ and is also known as a Kraus operator. The above evolution is valid 
if we have no information of $\mu$ and puts $\rho_{a}(t+\Delta t)$ into a generally mixed state. This can be intepreted as an environment-induced measurement \cite{Schon01} where the results of the measurement is not known. The situation changes if we perform a 
 measurement and have full information on $\mu$. The evolution is now described by
\begin{equation}
\rho_{a}(t) \to {\cal L}(\Delta t) \rho_{a}(t) = \rho_{a}(t+\Delta t)= \frac{M_{\mu} \rho_{a}(t) M^\dagger_{\mu}}{{\rm Tr}(M_{\mu} 
\rho_{a}(t) M^\dagger_{\mu})} \, ,
\end{equation} if the measurement in $\Delta t$ yields an observable $\mu$ with probability  ${\rm Tr}(M_{\mu} \rho_{a}(t) 
M^\dagger_{\mu})$.
The evolution of the source is thus generally stochastic leading to a quantum trajectory and the average of all stochastic evolutions 
yields the density operator obtained on solving the master equation, which gives an  ensemble description. An instance of a stochastic 
evolution of the state is defined as a quantum trajectory of the quantum jump formalism.
If a photon of polarisation  $\lambda$ and a wave 
vector pointing in the $\hat {\bf k}$ direction is detected within a coarse-grained time $\Delta t$ small compared to the average timescale of 
the system evolution and large compared to the optical period, the evolution of the state of the source (see \cite{Wong97,Schon01,Beige97} 
for more details) is given by
\begin{equation}
\rho_a(t+\Delta t)=\sum_{k} M_{\mu_k} \rho_{a}(t) M^\dagger_{\mu_k}/{\rm Tr}(\cdot) \approx R_{{\bf \hat k},\lambda}\rho_{a}(t_0) 
R^{\dagger}_{{\bf \hat k},\lambda} \Delta t/{\rm Tr}(\cdot) \, ,
\end{equation} with  $M_{\mu_k}=\bra{1_{k{\bf \hat k} \lambda}} U_I(\Delta t +t,t) \ket{0_{\rm ph}}$ and\footnote{Our analysis here apply only for degenerate levels $\ket{0}$ and $\ket{1}$ or in the case of non-degeneracies, when their frequency split $\Delta \omega$ is small enough such that $|\Delta \omega|\Delta t \approx 0$. For a more general discussion, see the formulation in Ref. \cite{Hegerfeldt93}. }
\begin{eqnarray} \label{R21}
R_{{\bf \hat k},\lambda} & \equiv & \sum_{i,j}  \left[ {3 \Gamma \over 8 \pi} 
\right]^{1/2}
(\hat{\bf D}_{2j} ,  {\bf \epsilon}_{\hat{\bf k} \lambda})  \, 
{\rm e}^{-{\rm i} k_0 \hat {\bf k} \cdot {\bf r}_i} \,  S^-_{i,j} ~.~~
\end{eqnarray} One can see  that the detection of a photon within a short time $\Delta t$ is always accompanied with a lowering or 
jump of the source within the same time $\Delta t$ hence motivating the name ``quantum jump" approach. Note that the probability density 
for the described emission is given by  ${\rm Tr} 
(R_{{\bf \hat k},\lambda}\rho_{a}(t) R^{\dagger}_{{\bf \hat k},\lambda})$. 

The no-photon time evolution of the system say between $t_2$ and $t_1$ is associated with a Kraus operator $M_0=\bra{0_{\rm ph}} 
U_I(t_2,t_1) \ket{0_{\rm ph}}=U_{\rm cond}(t_2,t_1)$.  More precisely, the state of the sources at $t_2$  after  a 
no-photon event from $t_1$ is given by
\begin{equation}
\rho_{a}(t_2)=U_{\rm cond}(t_2,t_1)\rho_a(t_1)U^{\dagger}_{\rm cond}(t_2,t_1) /{\rm Tr}(\cdot) \, 
\end{equation} where $U_{\rm cond}(t_2,t_1)={\rm e}^{-{\rm i}H_{\rm cond}(t_2-t_1)/\hbar}$.
This approach provides a 
non-Hermitian conditional Hamiltonian $H_{\rm cond}$ 
given by the following 
relation
\begin{equation} \label{Hcondrelation}
I-\frac{\rm i}{\hbar} H_{\rm cond} \Delta t \approx U_{\rm cond}(\Delta t+t_0,t_0)=\bra{0_{\rm ph}} U_I(\Delta t+t_0,t_0) 
\ket{0_{\rm ph}}
\end{equation} where the r.h.s is evaluated by second-order perturbation theory 
for a coarse-grained time $\Delta t$  similar to that of the timescale used in the
derivation of the reset operator. In Eq.~(\ref{Hcondrelation}), $\bra{0_{\rm ph}} U_I(\Delta t+t_0,t_0) \ket{0_{\rm ph}}$ is 
given by
\begin{eqnarray}
&& \bra{0_{\rm ph}} U_I(\Delta t+t_0,t_0) \ket{0_{\rm ph}} \\ \nonumber
&=& \bra{0_{\rm ph}} I-\frac{\rm i}{\hbar} \int_{t_0}^{\Delta t+t_0} dt'H_{\rm I}(t') 
-\frac{1}{\hbar^2}\int_{t_0}^{\Delta t+t_0}dt' \int_{t_0}^{t'} dt'' 
H_{\rm I}(t')H_{\rm I}(t'') +O(\Delta t^2) \ket{0_{\rm ph}} \\ \nonumber
&=&I-\frac{1}{\hbar^2}\int_{t_0}^{\Delta t+t_0}dt' \int_{t_0}^{t'} dt'' \bra{0_{\rm ph}} \Big[
\sum_m H_{{\rm I}1}^m(t')H_{{\rm I}1}^m(t'')+H_{{\rm 
I}2}^m(t')H_{{\rm I}2}^m(t'') \\ \nonumber
&&+\sum_{m \neq n} H_{{\rm I}1}^m(t')H_{{\rm I}1}^n(t'')+H_{{\rm 
I}2}^n(t')H_{{\rm I}2}^m(t'') \\ \nonumber
&&+\sum_{m,n} H_{{\rm I}1}^m(t')H_{{\rm I}2}^n(t'')+H_{{\rm I}2}^n(t')H_{{\rm 
I}1}^m(t'') \Big] \ket{0_{\rm ph}}+O(\Delta t^2)
\end{eqnarray}

We further define the relative position vector ${\bf r}={\bf r}_1-{\bf r}_2$ where 
$r=\|{\bf r}\|$ is the distance between the atoms and denote $k_0=\frac{\omega_0}{c}$. For 
the setup considered here,  one finds in the absence of laser driving \cite{Beige97,Wong97},
\begin{eqnarray} \label{hcondgrand}
H_{\rm cond} &=& \frac{\hbar}{2{\rm i}} \Big[ \Gamma \sum_m 
(S^+_{1,m}S^-_{1,m}+S^+_{2,m}S^-_{2,m}) \\ \nonumber
&&+\sum_{m,n}C_{1m,2n}S^+_{1,m}S^-_{2,n}+C_{1n,2m}S^+_{1,n}S^-_{2,m} \Big]\, ,  
\end{eqnarray} where $C_{in,jm}$ arises from dipole-dipole interaction and is 
given by
\begin{eqnarray} \label{dipoledipole}
C_{in,jm} &=& \frac{3 \Gamma}{2} {\rm e}^{{\rm i}k_0r} \Big[  \frac{1}{{\rm i} 
k_0r} ((\hat{\bf D}_{2m},\hat{\bf D}_{2n})-(\hat{\bf 
D}_{2m},\hat{\bf r})(\hat{\bf r},\hat{\bf D}_{2m})) \nonumber \\ 
&&+ \Big( \frac{1}{(k_0r)^2}-\frac{1}{{\rm i} (k_0r)^3} \Big) ((\hat{\bf 
D}_{2m},\hat{\bf D}_{2n})-3(\hat{\bf 
D}_{2m},\hat{\bf r})(\hat{\bf r},\hat{\bf D}_{2m})) \Big]. 
\end{eqnarray}
We consider only the cases where $C_{in,jm}$ is very small, or in other words, 
where the dipole-dipole interaction is insignificant. Without 
calculating the terms $C_{in,jm}$ explicitly, one can see that relative to the 
rate of spontaneous decay $\Gamma$, $C_{in,jm}$ scales as  
$(k_0r)^{-1}$ in the strongest possible dipole-dipole coupling scenario. This 
occurs when both dipoles are parallel with each other and orthogonal to the 
line joining both atoms. Therefore, in the limit of large $k_0r$, for example, 
$r>25 \lambda_0$ with $\lambda_0=\frac{2 \pi}{k_0}$, 
dipole-dipole coupling becomes insignificant. 

The two-atom double slit experiment 
performed by Eichmann {\em et al.} \cite{Eichmann93} 
also operates at this regime.
We can thus simplify (\ref{hcondgrand}) and get
\begin{equation} \label{hcond}
H_{\rm cond} = \frac{\hbar \Gamma}{2{\rm i}} \sum_m 
(S^+_{1,m}S^-_{1,m}+S^+_{2,m}S^-_{2,m}). 
\end{equation} This Hamiltonian can also be derived by assuming that each atom 
couples to its own separate radiation field.

We now determine the state of the system under the condition of the collection 
of two photons, the first one at $t_1$ in the $\hat{\bf 
k}_{\rm X}$ direction with polarisation ${\bf \epsilon}_{\hat {\bf k}_{\rm X} \lambda}$ and the second one at $t_2$ in the  $\hat{\bf 
k}_{\rm Y}$ 
direction with polarisation ${\bf \epsilon}_{\hat {\bf k}_{\rm Y} \lambda'}$.  If the initial state of the dipole sources at 
$t=0$ is $|\varphi_0 \rangle$, whilst the free radiation field is in its vacuum 
state, the unnormalised state of the dipole sources \cite{Schon01}  after the collection of the second photon is given by
\begin{eqnarray}
\ket{\psi( {\bf \epsilon}_{\hat {\bf k}_{\rm Y} \lambda'}t_2| {\bf 
\epsilon}_{\hat {\bf k}_{\rm X} \lambda}t_1)}&=& R_{\hat{\bf k}_{\rm Y},\lambda'} \, U_{\rm 
cond}(t_2,t_1) \, R_{\hat{\bf k}_{\rm X},\lambda} U_{\rm cond}(t_1,0) |\varphi_0 \rangle \nonumber \\
&=& N(t_1,t_2)\langle 22| \varphi_0 
\rangle  \sum_{i,j=0}^1 ( (\hat{\bf D}_{2i} ,  {\bf \epsilon}_{\hat {\bf k}_{\rm Y} 
\lambda'})(\hat{\bf D}_{2j} , {\bf \epsilon}_{\hat 
{\bf k}_{\rm X} \lambda}){\rm e}^{-{\rm i} k_0 \hat {\bf k}_{\rm Y} \cdot {\bf 
r}_1}{\rm e}^{-{\rm i} k_0 \hat {\bf k}_{\rm X} \cdot {\bf 
r}_2} \nonumber \\
&& +(\hat{\bf D}_{2j} , {\bf \epsilon}_{\hat {\bf k}_{\rm Y} 
\lambda'})(\hat{\bf D}_{2i} ,  {\bf \epsilon}_{\hat {\bf k}_{\rm X} 
\lambda}){\rm e}^{-{\rm i} k_0 \hat {\bf k}_{\rm Y} \cdot {\bf r}_2}{\rm e}^{-{\rm 
i} k_0 \hat {\bf k}_{\rm X} \cdot {\bf r}_1}) \ket{ij} \, ,
\end{eqnarray} with
\begin{equation}
N(t_1,t_2)={3\over 8 \pi} \,  \Gamma {\rm e}^{ -\Gamma (t_1+t_2) } \, .
\end{equation}
Note that $\| \, |\psi (\hat{\bf k}_{\rm Y}, t_2 | \hat{\bf k}_{\rm X}, t_1) 
\rangle \, \|^2$ yields the probability density for the 
corresponding event \cite{Plenio98}. 

We  now calculate the polarisation correlation $C_{\hat {\bf k}_{\rm A}\lambda, 
\hat {\bf k}_{\rm B} \lambda'}$ (i.e. the joint probability where Alice and Bob get a $
\lambda$ and $\lambda'$ polarised photon respectively  if Alice and Bob 
collect a photon each). It is simply given by
\begin{equation} \label{correlation}
C_{\hat {\bf k}_{\rm A} \lambda, \hat {\bf k}_{\rm B} 
\lambda'}=|\ket{\psi({\bf \epsilon}_{\hat {\bf k}_{\rm B} \lambda'}t_2|
{\bf \epsilon}_{\hat {\bf k}_{\rm A} \lambda}t_1})|^2 /\sum_{\lambda_1, 
\lambda_2}|\ket{\psi( {\bf \epsilon}_{\hat {\bf k}_{\rm B} 
\lambda_1}t_2| {\bf \epsilon}_{\hat {\bf k}_{\rm A} \lambda_2}t_1)}|^2 \, .
\end{equation} One can easily check that this is independent of $t_1$ and $t_2$ as all the time 
dependence cancels out in the 
normalising factor $N(t_1,t_2)$. 
We can use (\ref{correlation}) to calculate the probability $C_{\pm}$ $(C_{hv})$ that both Alice and Bob get orthogonal 
polarisation if they each collect a 
photon in the circular (linear) basis. The importance of such a calculation lies in the fact that the circular and linear basis are mutually 
unbiased.  This corresponds closely to the quantum cryptographic BB84 protocol where Alice and 
Bob  perform measurements in a set of
mutually unbiased bases. The existence of polarisation correlations in a set of mutually unbiased bases is a signature of entanglement 
(See Appendix~\ref{sec:appendix_a}.).

To assure that Alice and Bob can receive a polarisation entangled pair, they 
should place their detectors in directions $\hat {\bf 
k}_{\rm A}$ and $\hat {\bf k}_{\rm B}$ with
\begin{equation} \label{hold}
{\rm e}^{ -{\rm i} k_0  (\hat{{\bf k}}_{\rm A}-\hat{{\bf k}}_{\rm B}) \cdot  ({\bf r}_1-{\bf r}_2) } = -1 \, .
\end{equation}

One can see that these positions are in general not unique. They have the physical interpretation of
corresponding to a half-fringe interval 
in the far 
field of a double-slit experiment, in which the two atoms are replaced by pinholes which are symmetrically irradiated by a laser.

With condition (\ref{hold}), one obtains
\begin{eqnarray} \label{hurray}
&& \ket{\psi( {\bf \epsilon}_{\hat {\bf k}_{\rm B} \lambda'}t_2| {\bf 
\epsilon}_{\hat {\bf k}_{\rm A} \lambda}t_1)} =  \ket{\psi( {\bf \epsilon}_{\hat {\bf k}_{\rm A} \lambda}t_2| {\bf 
\epsilon}_{\hat {\bf k}_{\rm B} \lambda'}t_1)} \nonumber \\
&&  = N(t_1,t_2)2^{1/2}
\, {\rm e}^{ -{\rm i} k_0 (\hat {\bf k}_{\rm A} \cdot  {\bf r}_1 + \hat {\bf k}_{\rm B} \cdot  {\bf r}_2 )}  \, \langle 22| \varphi_0 
\rangle \nonumber \\
&&     \left[ \big( \hat{\bf D}_{20} , 
 {\bf \epsilon}_{\hat {\bf k}_{\rm B} \lambda'} \big) 
\big( \hat{\bf D}_{21} , {\bf \epsilon}_{\hat {\bf k}_{\rm A} \lambda} \big) 
- \big( \hat{\bf D}_{21} ,  {\bf 
\epsilon}_{\hat{\bf k}_{\rm B} \lambda'} \big) \big( \hat{\bf D}_{20} ,  {\bf 
\epsilon}_{\hat{\bf k}_{\rm A} \lambda} \big) \right]  \otimes  |a_{01} \rangle 
\end{eqnarray}
with $|a_{01} \rangle \equiv (|01 \rangle - |10 \rangle)/\sqrt{2}$. After two 
emissions, the dipole radiators are left in a maximally 
entangled state which is completely disentangled from the free radiation field.

A coordinate 
system is introduced where the $\hat {\bf z}$-axis points in 
the direction of the line connecting the two sources and the $\hat {\bf x}$-axis 
coincides with the quantisation axis. In addition, we 
choose $\hat {\bf k}_{\rm B} = (1,0,0)^{\rm T}$, $  {\bf \epsilon}_{\hat{\bf 
k}_{\rm B} +} =  \hat{\bf D}_{20}  = (0,1,{\rm i})^{\rm 
T}/\sqrt{2}$ and $ {\bf \epsilon}_{\hat{\bf k}_{\rm A} -} = \hat{\bf D}_{21} = 
\hat{\bf D}^*_{20}$. Using the spherical coordinates 
$(\vartheta,\varphi)$ for Alice's detector position, one can write $  {\bf 
\epsilon}_{\hat{\bf k}_{\rm A} \pm} =\frac{1}{\sqrt{2}}( {\bf 
\epsilon}_{\hat{\bf k}_{\rm A} h}\pm {\rm i} {\bf 
\epsilon}_{\hat{\bf k}_{\rm A} v}) $ with linear polarisations $  {\bf 
\epsilon}_{\hat{\bf k}_{\rm A} h}=( - \sin \varphi,  \cos \varphi  , 0)^{\rm T}$ and $  {\bf 
\epsilon}_{\hat{\bf k}_{\rm A} v}=( - \cos \vartheta \cos \varphi, - \cos \vartheta \sin \varphi  , \sin \vartheta)^{\rm T}$ (see 
\cite{Itano98}).
Using (\ref{hurray}), we have
\begin{equation}
C_{\pm}=C_{\hat {\bf k}_{\rm A}  +, \hat {\bf k}_{\rm B}  -}+C_{\hat {\bf k}_{\rm A}  -, \hat {\bf k}_{\rm B}  +}=\frac{(\cos 
\varphi+\sin \vartheta)^2+(\cos \vartheta \sin \varphi)^2}{2(1+(\sin \vartheta \cos \varphi)^2)} 
\end{equation} and
\begin{equation}
C_{hv}=C_{\hat {\bf k}_{\rm A}  h, \hat {\bf k}_{\rm B} v}+C_{\hat {\bf k}_{\rm A}  v, \hat {\bf k}_{\rm B} h}=\frac{(\cos 
\varphi)^2+(\sin \vartheta)^2}{(1+(\sin \vartheta \cos \varphi)^2)} \, .
\end{equation} A straightforward evaluation of both equations for the range $\frac{\pi}{2}-0.5< \vartheta < \frac{\pi}{2}+0.5$ and 
$-0.5<\varphi<0.5$ shows that\footnote{Note that $\vartheta = \pi/2$ should be excluded and intepreted as a limit point since we demand that condition (\ref{hold}) is fulfilled. At this limit where also $\varphi=0$, then $C_{\pm} = C_{hv} = 1$ and we have a maximally entangled postselected state (See also Appendix~\ref{sec:appendix_a}). }  
\begin{equation}
C_{\pm} \approx C_{hv} \approx 1 \,.
\end{equation}
Therefore,  having $\hat {\bf k}_{\rm A}$ pointing in a direction 
relatively close to the quantisation axis ($\vartheta=\pi/2 
, \varphi=0$) which is the  $\hat {\bf x}$-axis with a tolerance of $\pm 0.5$ radians for both angles $\vartheta 
, \varphi$ together with condition (\ref{hold}) fulfilled will guarantee that Alice and Bob 
obtain an approximate  postselected maximally entangled photon pair state which is maximally entangled in the ideal limit when $\hat {\bf k}_{\rm A} \to \hat {\bf x}$. 

\begin{figure} 
\begin{center}
\psfrag{a}{$\vartheta$}
\includegraphics[scale=0.5]{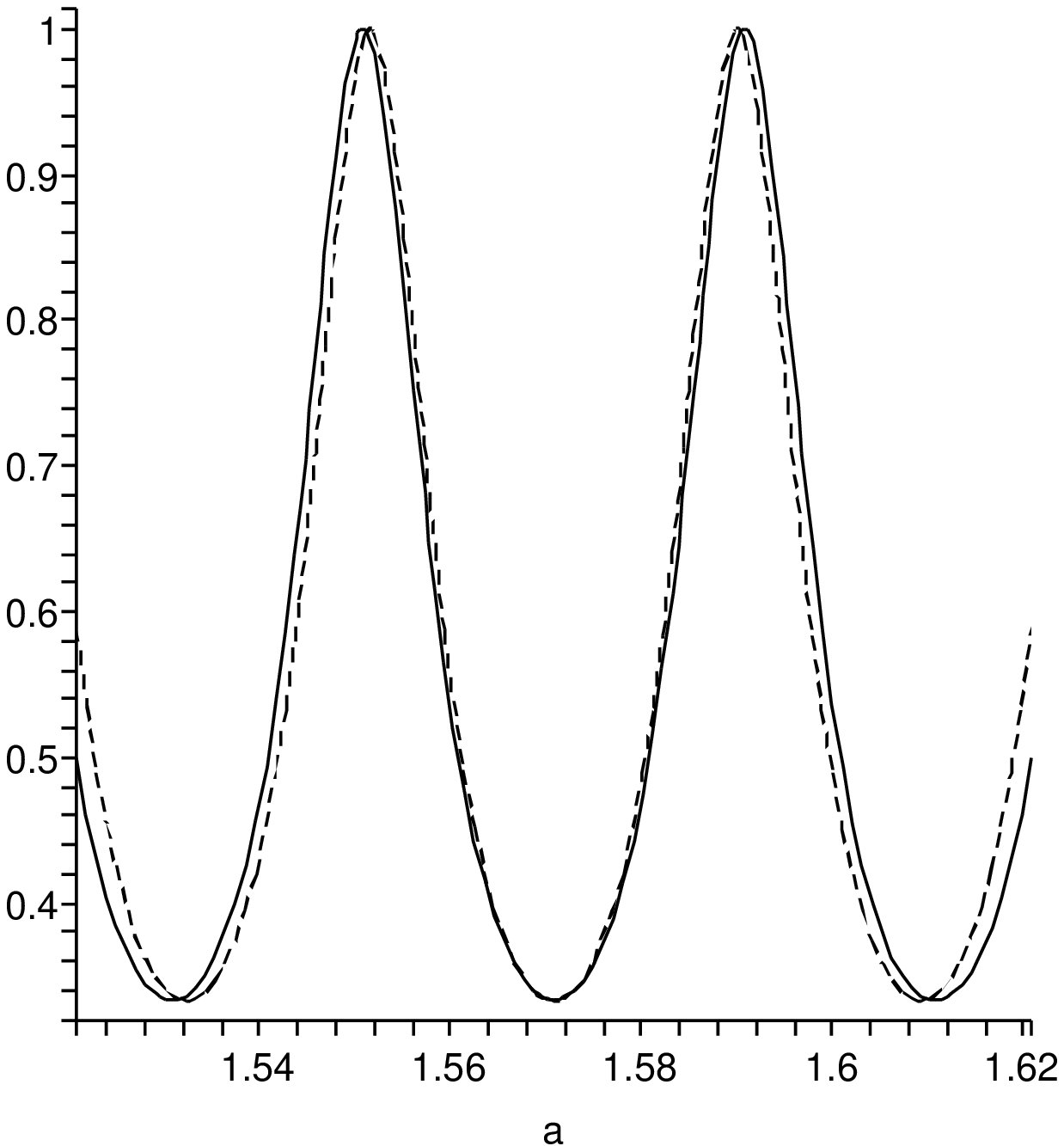} 
\includegraphics[scale=0.5]{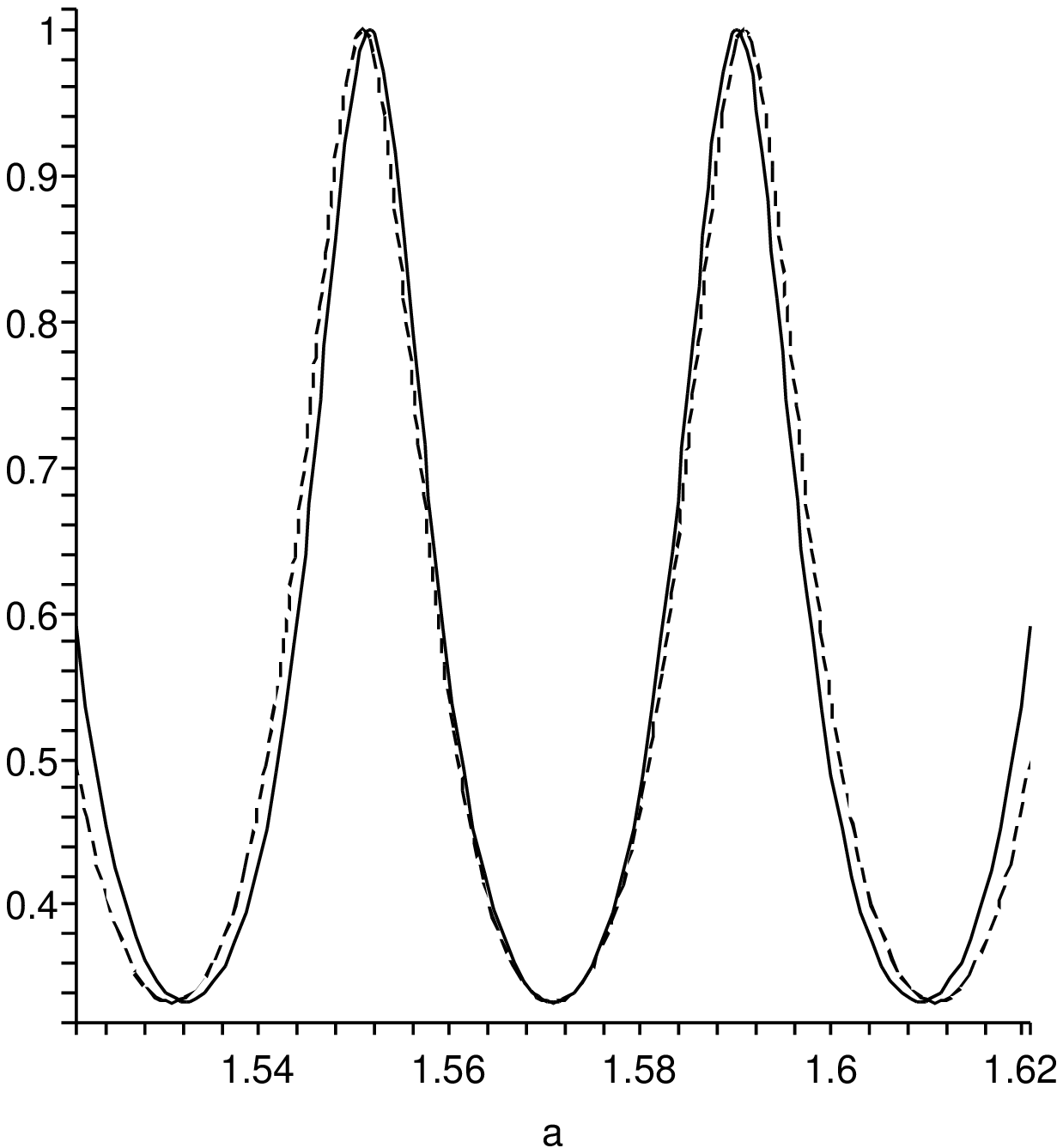} 
\end{center} 
\caption{Photon-photon polarisation correlation for orthogonal polarisation as a function
  of the spherical coordinate $\vartheta$  of Alice's
  detector location while Bob collects photons in the $\hat {\bf
    x}$-direction in the circular(left) 
and vertical(right) basis for $r= 25 
\lambda_0$ (solid curve), and $r= 26 \lambda_0$ (dotted curve). } \label{photonerror1}
\end{figure}

\begin{figure} 
\begin{center}
\psfrag{a}{$\vartheta$}
\includegraphics[scale=0.5]{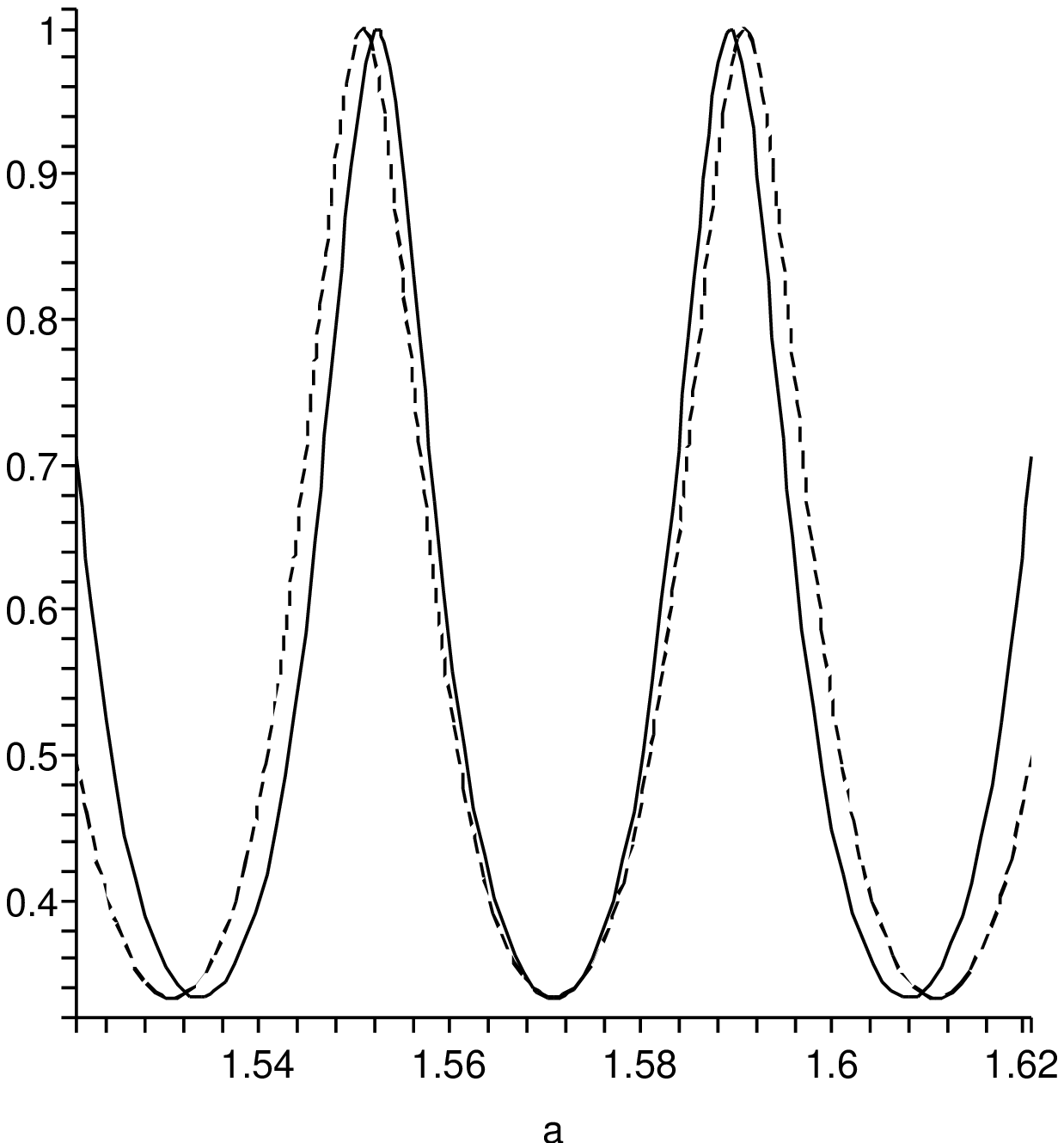} 
\includegraphics[scale=0.5]{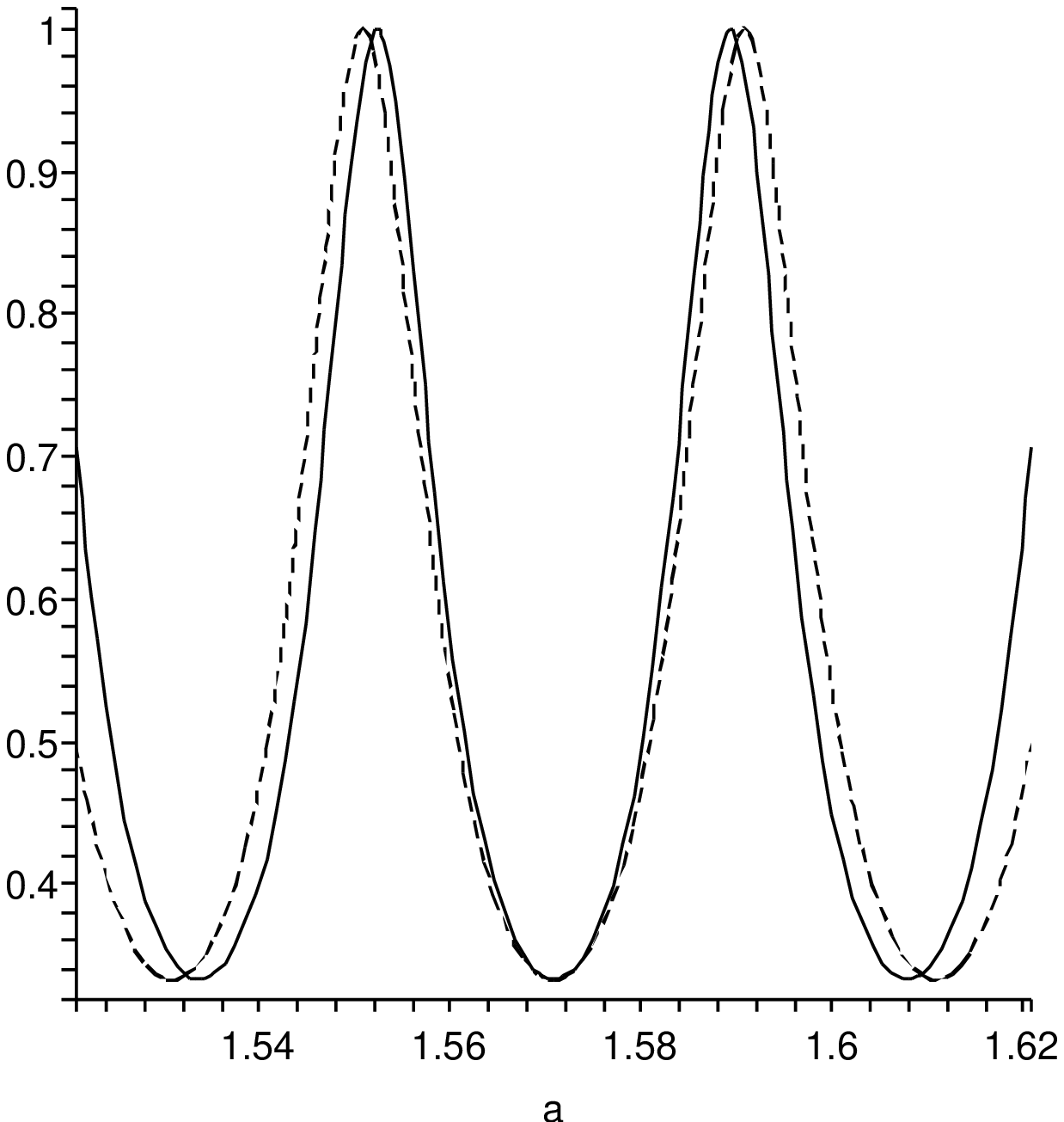} 
\end{center} 
\caption{Photon-photon polarisation correlation for orthogonal polarisation as a function
  of the spherical coordinate $\vartheta$  of Alice's
  detector location while Bob collects photons in the $\hat {\bf
    x}$-direction in the circular(left) 
and vertical(right) basis for $r= 25 
\lambda_0$ (solid curve), and $r= 27 \lambda_0$ (dotted curve).}\label{photonerror2} 
\end{figure}

For illustration, we fix Alice's azimuthal angle $\varphi=0$ and consider the case where $\theta \approx \frac{\pi}{2}$. Fig.~\ref{photonerror1} 
illustrates this case for both dipole separations 
at 25 and 26 wavelengths. As a 
comparision, Fig.~\ref{photonerror2} illustrates the case for both dipole separations at 25 and 27 
wavelengths.

We observe that only when (\ref{hold}) is fufilled, Alice and Bob always collect photons of orthogonal 
polarisation each be it in the circular or linear basis, which therefore agrees with a postselected 2-photon entangled state in the 
singlet form. It can also be seen 
from both figures that even when an error of  the dipole separation occurs within 2 wavelengths, 
the orthogonal polarisation correlation for collecting a photon pair can 
still above $90\%$. Therefore, strict Lamb-Dicke localisation of the dipole source is not essential. One may estimate the order of 
magnitude of the probability of an entangled photon pair collection with the help of Fig.~\ref{photonerror1}.  One can then obtain the 
maximum count rate for detectors with solid angles $\Delta_A$ and  $\Delta_B$ with the approximate formula $\frac{9}{64 \pi^2} \Delta_A 
\Delta_B$ \cite{LimJPA05}.  For example,  for two detectors each of solid 
angular extent of $0.0225$ steradians\footnote{Each detector consist of an array of slit detectors, with 15 slits each of $0.002 \times 
0.75$ steradians at intervals fulfiling (\ref{hold}) for $r=25 \lambda_0$.}  yielding a minimum orthogonal polarisation correlation  of $0.96$, the order of 
magnitude for the collection probability $P_{c}$ is approximately  $10^{-6}$. This is comparable to the scenario considered by Duan {\em 
et al.} \cite{Duan04a} where he estimated the probability (also about $10^{-6}-10^{-7}$) of entangling two distant ions in free space 
with the aid of a beamsplitter based on a similar scheme by Simon and Irvine \cite{Simon03}.

\section{Experimental Implementation} 
As an example\footnote{I acknowledge Phillip Grangier for his kind discussion on experimental issues concerning this scheme during the 2003
summer school in Les Houches, Session 79, {\em Quantum Information and Entanglement}.} we describe now a setup for entangled photon pair creation with two 
trapped $^{87}$Rb atoms that is feasible with present 
technology \cite{Schlosser01}. The ground states $|0 \rangle$ and $|1 \rangle$ are 
obtained from the $5^2S_{1/2}$ levels with $F=1$ and 
have the quantum numbers $m_{\rm F}= -1$ and $m_{\rm F}= 1$. The excited state $|2 
\rangle$ is provided by the $5^2P_{3/2}$ level with 
$F=0$. Suppose the atoms are initially in the $5^2S_{1/2}$ ground state with $F=1$ 
and $m_{\rm F}=0$ and a $\pi$ polarised laser field is 
applied to excite to level 2 by a sharp $\pi$ -pulse. After spontaneous emission into the ground 
states 
$|0\rangle$ and $|1 \rangle$, another $\pi$ polarised laser 
reinitialises the system by coupling these states to the $5^2P_{3/2}$ states with 
$F=1$. From there the atoms return into the initial 
state via spontaneous decay. Due to their differences in polarisation and because 
of the detector locations, ``$+$" ($\sigma^+$) and 
``$-$" ($\sigma^-$) polarised signal photons are distinguishable from the laser 
photons and spontaneously emitted $\pi$ polarised photons. With a typical spontaneous decay  time of order $10^{-8}$ s and assuming a 
rapid excitation with efficiency $90\%$ and recycling time of order $10^{-7}$ with detection efficiency of $0.88$ 
\cite{Takeuchi99,Rosenberg05} and taking the estimate for collection efficiency $P_{c}$, the estimated count rate of entangled photons 
from this setup is expected to be $10^2s^{-1}$. Compared to the yield possible in parametric downconversion being  $10^6s^{-1}$ 
\cite{Kumar04}, this scheme has relatively low yield. However, it does not require frequency filters for entangled photon detection 
and it also offers entanglement of the dipole sources as an attractive byproduct.

\section{Conclusion}
In conclusion, we proposed a scheme for the creation of polarisation entangled 
photon pairs by using two distant dipole radiators in free 
space. The entanglement is obtained by carefully choosing the detector positions 
with respect to the sources and arises under the 
condition of the collection of two photons independent of their emission times and 
the initial state of the sources. This also results in the source being maximally entangled. It is 
important to note that the photon 
entanglement detected can be used for quantum cryptography or Bell's inequality test. The scheme introduced in this chapter  has the advantage of not requiring any linear optics and cavities compared to schemes in the previous chapters. 
Another application of the scheme would be to merely prepare two distant dipole sources 
in the maximally entangled ground state $|a_{01} 
\rangle$. In this case, no degeneracy of the atomic ground levels $\ket{0}$ and $\ket{1}$ is 
required. As in the case of Simon and Irvine \cite{Simon03}, this 2-photon detection protocol 
for preparing an entangled dipole state is robust against random laser phase fluctuations during the 
atomic excitation process as it 
contributes to just a trivial global phase factor. Furthermore, the  2-photon protocol can yield high fidelity of entangled state 
preparation more easily compared to the 1-photon protocol originally proposed by Cabrillo {\em et al.} \cite{Cabrillo99}. This is due to the 
fact that in the 1-photon protocol, photons have to be gathered around the entire solid angle of emission  to rule 
out the possibility of an undetected  2-photon emission which ruins the entanglement. This problem may be solved  at the cost of a 
very weak excitation on the photon sources. The presented idea might find interesting 
applications in quantum 
computing with trapped atoms, diamond NV color centres, 
quantum dots or single atoms doped onto a surface and opens new possibilities for 
the creation of antibunched polarisation entangled 
photon pairs and even multiphoton entanglement by including more than two 
radiators in the setup. 

Finally, we remark that the free-radiation field can be perceived roughly as a type of continuous 
beamsplitter, similar to a discrete multiport with infinite inputs and outputs. This leads generally to low entangled photon pair 
collection efficiency of  $10^{-6}$ if we only gather photons in a 
well-defined directional spatial mode as explained earlier. 

Photon entanglement 
schemes are therefore generally  more realistic in the long-run with linear optics resources and single photon sources emitting on demand in well 
directed spatial modes as demonstrated in the rest of the thesis, owing to the higher success probability that can be obtained.  Linear 
optics also offers flexibility in generating a wider variety of 
entangled states compared to the free space approach described in this chapter. It is now time to conclude this thesis.

\chapter{Summary and Outlook}\label{outlook}

The work of this thesis 
demonstrates various closely related aspects of quantum information processing 
with 
single photons as motivated in Chapter \ref{overview}. Hopefully, it adds new 
perspectives to the relationship between single 
photons and their sources and the implication to quantum information processing 
in general. The summary 
of the main work is as follows.

In Chapter \ref{firework}, we showed that a wide range of highly entangled 
multiphoton states, including {\em W}-states, can be prepared 
by interfering  {\em single} photons inside a Bell multiport beam splitter and 
using postselection. The described setup, being photon 
encoding independent can be used to generate polarisation, time-bin and 
frequency encoded multiphoton entanglement, even when using only a single 
photon source. The success probability has a surprisingly non-monotonic 
decreasing trend as the number of photons increases.  

In Chapter \ref{fusion}, we demonstrated  how the HOM dip can be generalised 
to multiphoton coincidence detection in multiport 
beamsplitters. We considered the canonical symmetric Bell multiport and show 
that the HOM dip can be observed for all $N \times N$  Bell
multiports where $N$ is even but not necessarily when $N$ is odd. Note that this observation 
applies generally to all bosons, of which photons 
are an example, thus having wide applicability. For the sake of completeness, we 
also discussed multifermionic scattering through a 
multiport and showed that identical fermions always leave the  output 
ports of the multiport separately.
 
In Chapter \ref{hummingbird}, we proposed a scheme for implementing a 
multipartite quantum filter that uses entangled photons as a 
resource. It is shown that the success probability for the 2-photon parity 
filter can be as high as ${1 \over 2}$, which is the highest 
that has so far been predicted without the help of universal two-qubit quantum gates. 
Furthermore, the required number of ancilla photons is the 
least of all current parity filter proposals. Remarkably, the quantum filter 
operates with probability ${1 \over 2}$ even in the 
$N$-photon case, regardless of the number of photons in the input state. 

In Chapter \ref{minsk} we described the efficient implementation 
of  eventually deterministic two-qubit gate operations between single photon 
sources, despite the restriction of the no-go theorem 
on deterministic Bell measurement with linear optics. No entangled ancilla 
photons  and photon-feed into cavities are needed. 
The key principle for our approach is based on source encoding to the photon 
that is generated as well as measurements in a mutually 
unbiased basis with respect to the computational basis. The described approach 
is highly general and lends wide implementation to various 
types of single photon sources. Furthermore, the scheme is still robust even in 
the case of dissimilarities of the photon sources, a 
testament to the unique character of a measurement-based approach to quantum 
computing. Our approach also gives fresh perspectives on the 
use of mutually unbiased basis in quantum computation, besides existing 
applications in quantum cryptography and for solving the Mean King's problem.

In Chapter \ref{demand}, we used ideas from Chapters \ref{firework} and 
\ref{minsk} to show how multiphoton entanglement on demand can be 
realised. Generally speaking, any multiphoton qubit state can be generated on 
demand in a distributed manner. At the same time, we also 
relate  a duality relation between preparing photon entanglement and atom 
entanglement  which may lead to new perspectives in multiport designs for 
entanglement generation. 

In Chapter \ref{photon}, we showed, using a setup closely resembling a Young 
double-slit experiment, that dipole-dipole as well as 
2-photon entanglement can be generated with photons emitted from two distant 
dipole sources in free space (i.e. without the aid of 
linear optics setup). The scheme is  highly robust to the dipole excitation 
imperfections.  In the case of two sources, the 
entanglement arises under the condition of two emissions in certain spatial 
directions and  leaves the dipoles in a maximally 
entangled state. This work adds new perspectives to current views on the entanglement 
generation using measurements.

The outlook and possible extensions to the work in this thesis are manifold. For 
example, the work in Chapter \ref{firework} and \ref{fusion}  is mainly 
restricted to Bell multiports due to a cyclic symmetry which 
we exploited. It is interesting to examine a greater variety of multiports 
defined by redirection or transfer matrices of various symmetries and their 
implications on multiphoton scattering and entanglement generation. Here, we have 
confined ourselves to analysing pure states and have not considered 
photon mixed-states for simplicity. This might lead to applications like the
characterisation of photons or multiports.  We also did not consider EPR photon 
pairs as possible inputs to multiports which may yield 
exciting possibilities in multiphoton entanglement generation with higher 
probability of success. One could also extend this 
work to the investigation of POVMs with detectors and multiports.   An extension 
of Chapter \ref{fusion} may  find application in 
experiments with particles with exotic statistics such as anyons\footnote{I thank Vlatko Vedral for stimulating discussions.}. Finally, multiports with weak 
nonlinearities may yield interesting prospects in enhanced multiphoton state 
preparation due to cooperative enhancement\footnote{I thank Jim Franson 
for stimulating discussions.}.

In Chapter \ref{hummingbird}, we have dealt with a simple setup for a 
multiphoton filter. It is interesting to see how this can be 
extended to arbitrary multiphoton gates and if the probability of success could be 
increased by combining approaches using an $N \times N$ 
multiport. With the aid of an arbitrary photon ancilla, generated perhaps by a 
multiphoton source on demand, one might be able to 
implement a programmable multiphoton gate. Further extensions to this work, 
hinted by the duality relation obtained in Chapter 
\ref{demand}, might lead to a multiatom filter implementation.

Chapter \ref{minsk} presents possible extensions to higher dimensional quNit 
operations or 
direct multiatom gate implementation. This might be 
implemented with the aid of a linear optics multiport. For example in this 
chapter, we have 
already used multiports for measurements leading to useful 
gate implementations.  These techniques could be extended  to new and
interesting results. An intriguing observation of the choice of 
mutually unbiased basis used in this chapter yields, on suitable rearragement of 
the coefficents of the computational basis,  a $4 \times 
4$ Fourier transform matrix, which defines a Bell multiport\footnote{I thank Thomas Durt for 
bringing this to my attention.}. The relationship 
between mutually unbiased basis and 
multiports may be worth investigating.

The work of  Chapter \ref{photon} may in 
principle be extended to multiple  dipoles or various energy level structures 
and 
by taking  into account of dipole-dipole 
interaction or various means of dipole excitation. So far, we have restricted 
ourselves to the simplest case of two initially excited 
dipoles which is experimentally reasonable. Spatial polarisation correlations or 
intensity correlations may, for example be exploited 
for certain search tasks as demonstrated by Agarwal {\em et al.} 
\cite{Agarwal04}.

In this thesis, we have confined ourselves to discrete quantum information 
processing with photons and have said nothing about the 
equally rich field of continuous variable processing with photons or even a 
hybrid field of discrete-continuous variables. It may be that 
the work here can be extended to such domains and might lead to analogous applications.

Finally, although we have focused our attention to single photons, many parts of 
this thesis may  find analogous applications in other 
flying qubits such as electrons, which in contrast to photons, have
fermionic statistics. It is noteworthy that many linear optical
operations on photons can also be implemented on electrons. At the same 
time, there exist an intriguing prospect of setups, such as 
doped fibers, that modify the quantum statistics of the photons. One might 
envision new capabilities of quantum information processing 
with single photons using multiport setups consisting of such doped fibers in 
the future. It is hoped that the work in this thesis adds to the overall 
development as well as inspiring new research in quantum information processing.

\appendix
\chapter{Inferring the  singlet state from polarisation statistics} 
\label{sec:appendix_a} 

We assume that Alice and Bob possess a shared reference frame of photon polarisation. They perform random measurements in two basis, one of which is the linear basis $B_1$ spanned by $\ket{h}$ and $\ket{v}$ and the other, the circular basis $B_2$ spanned by $\ket{\pm}=\frac{1}{\sqrt{2}}(\ket{h}\pm\ket{v})$. We assume that the 2-photon state to be measured by Alice and Bob is in the general form $\rho$.
We can write the POVM elements $E_{1s}$ and $E_{1d}$ for $B_1$ as
\begin{equation}
E_{1s}=\ket{hh}\bra{hh}+\ket{vv}\bra{vv} \, \, \, , E_{1d}=\ket{hv}\bra{hv}+\ket{vh}\bra{vh} \, ,
\end{equation} and the POVM elements $E_{2s}$ and $E_{2d}$ for $B_2$ as
\begin{equation}
E_{2s}=\ket{++}\bra{++}+\ket{--}\bra{--} \, \, \, , E_{2d}=\ket{+-}\bra{+-}+\ket{-+}\bra{-+} \, .
\end{equation} Note that $E_{1s}+E_{1d}=E_{2s}+E_{2d}=1$.
If Alice and Bob always detects orthogonal polarisations in $B_1$ and $B_2$, we wish to show that the only state consistent with this observation is the singlet state(which is maximally entangled) given by $\rho=\ket{\Phi_-}\bra{\Phi_-}$\footnote{I thank Jens Eisert for his help in this problem.} where $\ket{\Phi-}=\frac{1}{\sqrt{2}}(\ket{hv}-\ket{vh})$. This approach does not require a full tomographic measurement and may shed new perspectives to the notion of optimal measurements in quantum tomography.
We therefore have 
\begin{equation} \label{Povm1}
{\rm Tr}(E_{1d}\rho)=1
\end{equation} and
\begin{equation}\label{Povm2}
{\rm Tr}(E_{2d}\rho)=1 \, .
\end{equation}
We know that $\rho$ must be positive semi-definite. This is to allow a valid probability interpretation should we choose to arbitrarily reduce the degree of freedom specifying $\rho$. This restriction together with Eq.~(\ref{Povm1}) implies that $\rho$ can only be of the form $\rho=a\ket{hv}\bra{hv}+c\ket{hv}\bra{vh}+c^*\ket{vh}\bra{hv}+b\ket{vh}\bra{vh}$ where $a+b=1$ and $a$,$b$ are positive real numbers.
Adding condition  (\ref{Povm2}) yields the further constraint $a=b=\frac{1}{2}$ as well as $c=-\frac{1}{2}$. This implies that $\rho=\ket{\Phi_-}\bra{\Phi_-}$.

\bibliographystyle{thesbib}
\bibliography{thesis}

\end{document}